\documentclass[aps,prb,twocolumn]{revtex4-1}
\usepackage{longtable}
\usepackage{amsmath,amssymb,amsfonts,bm}
\usepackage{graphicx}
\usepackage{ulem}
\usepackage{color}
\usepackage[colorlinks=true,citecolor=green,linkcolor=red,urlcolor=blue]{hyperref}

\def\KeyWord#1{$\backslash$\IfColor{$\!\!$\textRed{#1}\textBlack}{#1}$\!\!$}
\newcommand{\be}{\begin{equation} }
\newcommand{\ee}{\end{equation} }
\newcommand{\ba}{\begin{eqnarray} }
\newcommand{\ea}{\end{eqnarray} }
\newcommand{\n}{\nonumber \\ }
\newcommand{\m}{\item}
\newcommand{\mac}{\mathcal}

\newcommand{\mz}{\mathbb{Z}}

\newcommand{\bit}{\begin{itemize}}
\newcommand{\eit}{\end{itemize}}

\graphicspath{{Figures/}}

\begin{document}

\title{Anyon condensation and its applications}

\author{F.~J.~Burnell$^1$}
\affiliation{$^1$School of Physics and Astronomy, University of Minnesota, Minneapolis, Minnesota 55455, USA}

\date{\today}

\begin{abstract}

Bose condensation is central to our understanding of quantum phases of matter.
Here we review Bose condensation in topologically ordered phases (also called topological symmetry breaking), where the condensing bosons have non-trivial mutual statistics with other quasiparticles in the system.  We give a non-technical overview of the relationship between the phases before and after condensation, drawing parallels with more familiar symmetry-breaking transitions.  We then review two important applications of this phenomenon.  First, we describe the equivalence between such condensation transitions and pairs of phases with gappable boundaries, as well as examples where multiple types of gapped boundary between the same two phases exist.  Second, we discuss how such transitions can lead to global symmetries which exchange or permute anyon types.  Finally we discuss the nature of the critical point, which can be mapped to a conventional phase transition in some -- but not all -- cases.

\end{abstract}

\maketitle

\section{Introduction}

One of the richest and longest-lived questions in theoretical condensed matter physics has been to understand different phases of matter.  Recent progress on these questions has focussed primarily on phases ``beyond Landau", meaning those which cannot be distinguished by a local order parameter.  
These new families of phases require a re-thinking of much of the machinery developed by Landau and others, 
which relies on a local order parameter not only to distinguish phases, but also to describe phase diagrams and phase transitions of interacting many-body systems.  
What is the analogue of this framework when a local order parameter is absent?

This question is particularly interesting for topologically ordered systems,\cite{WenGSDeg,WenTopOrd} which are intrinsically strongly interacting.  In 2 dimensions (which will be our focus here), topological order is characterized by point-like quasiparticles known as anyons that interact through self- and mutual-  (possibly non-abelian) Aharanov-Bohm phases.\cite{WilczekAnyon}  
How does this topological order dictate which other phases may be attained via a direct (and possibly continuous) phase transition?

In a conventional ordering transition, the order parameter results from condensing a bosonic excitation-- for example, the ordering of a magnet can be viewed as the Bose condensation of spinons. 
Similarly, one way to approach the phase structure of topologically ordered systems is to study the possible ways in which anyons  can condense.
Such {\it anyon condensation} is fundamentally different from the Bose condensation that describes a conventional magnet or superfluid: though an anyon can be a boson in the sense that the phase of the wave function is not affected by exchanging it with another anyon of the same type, by definition there must be at least one other anyon type in the system with which it has a non-trivial exchange phase.
Consequently it always has long-ranged statistical Aharanov-Bohm type interactions with one or more other quasiparticles, and can never be a {\it local} boson.  Thus no local order parameter is produced by condensing such objects.

The study of transitions in which non-local objects condense dates back to Kosterlitz and Thouless,\cite{KT,KosterlitzRG} who described how vortices in a 2D superfluid proliferate at finite temperature and destroy the quasi-long ranged superfluid order.  However, these vortices are quite different from anyons: they are non-local in the sense that they disturb the superfluid everywhere in space, leading to a vortex energy cost that grows at least logarithmically with the system size.  Anyons, in contrast, have a finite energy, but are non-local due to their long-ranged statistical interactions.  This is much more like the situation in a 2D superconductor, where the vortices are (energetically speaking) point defects, because the 
gauge field screens the kinetic energy cost of the superflow, but have long-ranged statistical interactions with the Bogoliubov-de Gennes quasiparticles.\cite{AharonovCasher} Thus vortex proliferation in a 2D superconductor 
can be viewed as a process in which anyons (the vortices)\cite{SondhiSC} condense, destroying the superconducting phase.   Unlike an ordinary Bose condensate, this vortex condensate does not break any global symmetry; consequently it does not lead to Goldstone modes but rather to a gapped phase with trivial topological order.   This is a general feature of anyon condensation transitions: they relate two incompressible phases with distinct topological orders.

The idea that vortex proliferation can be used to relate different topological orders was pursued by a number of authors,\cite{DasguptaHalperin,LeeFisher,FisherLee,WenKMatrix,BarkeshliYaoLaumann} who showed that it can also occur in systems where the vortices are not bosons; this allows a systematic description of the hierarchy of fractional quantum Hall states\cite{HaldaneHierarchy,HalperinHierarchy,ReadHierarchy,FrohlichHierarcy}   
in terms of cascades of anyon condensation transitions.
More recently, this idea has also been applied to study the possible surface states of interacting topological insulators,\cite{WangSenthil,MetlitskiKaneFisher,SenthilVishwanath} which can also be topologically ordered.  
These approaches illustrate the fact that anyon condensation can lead to a rich phenomenology relating different topological orders. 

A general framework to describe how anyon condensation connects different topological orders would clearly reveal interesting structure in the phase space of topologically ordered systems.  However, the field-theoretic techniques used in the  fractional quantum Hall hierarchy states are understood only for a limited (abelian) class of models.
Indeed, for processes in which arbitrary anyons condense, no fully general prescription is known.
However, when the condensing anyons are bosons, 
it is possible to systematically describe the relationship between the condensed and uncondensed topological orders.  \cite{ConformalZoo,MooreSeibergNatural,Gepner89,Schellekens89,GoddardCoset1,GoddardCoset2,TSBPRL,SlingerlandBais,BaisMathy,Eliens,NeupertBernevig,MugerTSB,BruguieresTSB,Muger,KIRILLOV2002183,Davydov13,Hung15,NeupertNoGo} 
This description requires knowledge only of the initial topological order and the anyon(s) to condense, rather than full field theoretic description of the transition.  
This is sufficient to understand the gapped phases related by anyon condensation, much as for conventional symmetry-breaking transitions, the  symmetry of the broken phase is fully determined by the unbroken symmetry group and the order parameter.  
To distinguish this framework from the more general situation where we could contemplate condensing non-bosonic anyons, we refer to it as topological symmetry breaking (TSB).\cite{BaisGaugeTheories,TSBPRL,SlingerlandBais,BaisMathy}

The mathematics of TSB is far from new: it was first developed in 2D conformal field theory,\cite{ConformalZoo,MooreSeibergNatural,Schellekens89,Gepner89,GoddardCoset1,GoddardCoset2,SCHELLEKENSFixed,FUCHSFixed,FuchsCoset}.  (For a review of the connection between topological order and conformal field theory, see Ref. \cite{KitaevHoney}.)  
In mathematics, it has been studied in the context of modular tensor categories,\cite{MugerTSB,BruguieresTSB,Muger,KIRILLOV2002183,Davydov13,Hung15} which are the mathematical structures underpinning topological order.   
Refs. \cite{BaisGaugeTheories,TSBPRL,BaisMathy,SlingerlandBais} described how, in the context of topologically ordered systems, this formalism can be interpreted as condensation of non-local bosons;  
more systematic methods to use this approach to obtain the topological order of the condensed phase have been described by Refs.\cite{Eliens,NeupertBernevig}.  
Here we will primarily use their approach and language.  

Though the framework is not new, TSB's relevance to several important questions about topologically ordered phases has only recently been appreciated.  
Notably, it provides a complete answer to the question of whether the boundary between two topological phases can be gapped, and whether a given pair of phases admits multiple distinct gapped boundary types.\cite{BravyiKitaev,KitaevKong,LevinBdy,KongCond,BaisBoundaries,Beigi11, BaisChiralBoundaries,YanWanCond,KapustinAbelianBdies,WangWenBdy,LanWangWen,GaneshanBdy} It also naturally gives rise to symmetries that interchange different species of anyons.\cite{DrinfeldGauging,GuWan14,HungWan14,GarciaCond,TeoTwist,LongQPaper,ChengAPS,Heinrich16,FradkinTwist,TarantinoTwist}
Both of these are intimately connected to the possibility\cite{FuKane,LindnerPara,ClarkePara,ChengPara,BarkeshliQi,BombinTwist,WenGenon,BarkeshliAbelian1,BarkeshliAbelian2,BarkeshliBraiding,HughesSantos}  of engineering bound states whose statistical interactions emulate those of non-abelian anyons.  
By studying concrete models of TSB,\cite{TSBShort,TSBLong,WenBarkeshli,WenBarkeshliLong,WB2,Bombin,HormoziSlingerland,FengKitaevJW} some progress has also been made towards understanding the (possibly second-order) critical points that arise when non-local bosons condense.\cite{Gilsetal,GilsIsing,BaisMonteCarlo,schulzIsi,VidalMF,SchulzFib,SchulzSU2,VidalBS}
The present work aims to provide a basic introduction to the formalism itself, and give an overview of these recent developments.

\section{ Condensing anyons that are bosons: topological symmetry breaking}

Our first objective is to review how condensing non-local bosons -- i.e. anyons with bosonic self-statistics -- changes a system's topological order.
 This is the analogoue of understanding how condensing (local) bosons, which generates a local order parameter, affects a system's unbroken symmetries.   In the latter case Bose condensation reduces the symmetry group to a (possibly trivial) subgroup.  Here we review how TSB describes an analogous reduction in the topological order resulting from condensing non-local bosons.  

  The term ``topological symmetry breaking" is somewhat misleading: in general we will not be able to identify a symmetry of the uncondensed phase that is broken by the condensate.  The transitions always connect gapped phases, and never produce Goldstone modes. Rather, the name refers to numerous similarites with the Anderson-Higgs mechanism, in which a gauge symmetry is effectively reduced by a Bose condensate.  
Indeed for discrete gauge groups, where Yang-Mills theory describes a fully gapped topologically ordered phase,\cite{BaisDiscrete}  the Anderson-Higgs mechanism is equivalent to TSB.   For more general topological orders the phenomenology remains very similar, as we shall see.

Unlike general anyon condensation processes, which can increase the anyon content or even create topological order from a trivial phase,  TSB necessarily reduces the total number of deconfined anyon states (in technical terms, the total quantum dimension).   Thus while (at least for some topological orders) it can describe a process in which the topological order is completely destroyed, an alternative framework is needed to describe the reverse process, in which the topological order is created from the vacuum.  This is similar to the situation in symmetry-breaking transitions, where Bose condensation necessarily reduces the symmetry; to describe the inverse process a different (dual) description is required.  


\subsection{Topological order, anyons, and bosons} \label{Sec2}

To begin, we will review the key features of topological order that play a role in TSB.
The aspects of a phase captured by topological order are similar in spirit to those described by its symmetry. 
In the latter case, point-like quasiparticles are characterized by the representation of the symmetry group under which they transform. This representation cannot change unless either the symmetry is broken, or the quasiparticle fuses with another quasiparticle.  The action of symmetry on a region $A$ containing multiple quasiparticles is obtained by combining their respective representations.  If the symmetry is non-abelian this combination is not unique, and $A$ may transform in one of several representations.  

In a topologically ordered system, the analogue of a representation is the anyon type, or topological charge.  We use the set of labels $\{ a_i \}$ to denote the possible topological charges; each label represents a fixed set of statistical interactions (i.e. braiding) with other anyon types.   A particle's statistical interactions, and hence the topological charge, must be conserved unless it fuses with another quasiparticle.  If a region $A$ contains multiple anyons, its net statistical interaction with the outside world is fixed by its total topological charge, which is obtained by combining the topological charges of all of its quasiparticles.   The anyons are said to be non-abelian if this combination is not unique, such that $A$ may carry one of several topological charges.

In practise, there is a key difference between topological order and symmetry: the latter is generally present at the level of the microscopic Hamiltonian, whereas in physically realistic situations the former is necessarily emergent.  Thus one might worry that if the energy density in region $A$ is too large, its topological charge is no longer meaningful.  Provided that outside of region $A$ (or more generally, outside of a finite set of bounded spatial regions, which could contain an arbitrary energy density or even holes in the material) the system is in its ground state, however, a quasiparticle encircling the region $A$ from a sufficient distance  will pick up a path-independent Berry phase determined only by the total topological charge in $A$.   In this sense the topological charge in a bounded high-energy region is a meaningful quantity.

To understand TSB, we will require a basic understanding of some of the properties of anyons in 2 spatial dimensions:
\begin{enumerate}
\m
{\it The fusion rules}
\be \label{Fusion}
a_i \times a_j = \sum_{k} N^k_{ij} a_k
\ee
 tell us how anyons brought close together combine to give other anyons. (This is analogous to the way that the Clebsch-Gordon coefficients tell us how particles transforming in representations $r$ and $r'$ of a symmetry group combine to give a particle transforming in one of the possible representations $r\times r'$).    For the examples we will discuss here $N_{ij}^k$ will always be either $1$ or $0$, though in general they need only to be non-negative integers. 
   An anyon $a$ for which $a \times b = c$ (with no sum) for every $b$ is called {\it abelian}; otherwise, if the right-hand side of (\ref{Fusion}) contains multiple terms, $a$ is {\it non-abelian}.
   \m
{\it Unique antiparticle and conserved topological charge}: Every anyon $a_i$ must have a unique anti-particle $\overline{a}_i$, with which it can annihilate to give the vacuum (which we denote by $1$)-- i.e.
 \be
 a_i \times \overline{a}_i = 1 + ...
 \ee
Note that we can -- and in the present work, often will -- have $a = \overline{a}$.
Conversely, the total topological charge is conserved: only particle-antiparticle pairs may be created from the vacuum.  
\m
{\it The topological spin} $s_a =e^{i \theta_a} = s_{\overline{a}}$ of an anyon $a$ is the phase incurred when the anyon is rotated by $2 \pi$.  
Note that $S_1 = 1$.  In the examples we will consider, this spin dictates the anyon $a$'s exchange statistics.  Fig. \ref{ExchangeTwistFig} shows that rotating the anyon's world line by $2 \pi$ is equivalent to exchanging $a$ with $\overline{a}$, and then interchanging a space-like and a time-like direction.
The statistical phase associated with exchanging anyons $a$ and $\overline{a}$ (when fused in the vacuum channel, if they are non-abelian) is therefore given by the anyon $a$'s topological spin. (Specifically, it is the product of $a$'s topological spin and a quantity known as its Frobenius Schur indicator; see e.g. \cite{KitaevHoney,BondersonThesis}.  For simplicity, here we will assume that the Frobenius Schur indicator is $+1$, though this is not necessary for the formalism we discuss to hold.\cite{Eliens})   

If $a = \overline{a}$ (as is the case for the examples we discuss here), then this implies that in the vacuum fusion channel $a$ anyons are bosons (fermions) under exchange if they have 
 topological spin $+1$ ($-1)$.   
 
\m 
 {\it Braiding}
is the process in which two anyons are exchanged twice (see Fig. \ref{BraidFig}).  In general the resulting complex phase depends on the fusion channel of the two anyons, and we call the resulting matrix element $M^{ab}_c$.  
(If $a\neq b$ a single exchange results in a different spatial configuration of anyons, so that only the double exchange $M^{ab}_c$ is a physically meaningful quantity).
Note that braiding with the identity is always a trivial operation, i.e. $M^{a 1}_a = 1$.


\begin{figure}[ht]
\includegraphics[width=0.35\textwidth]{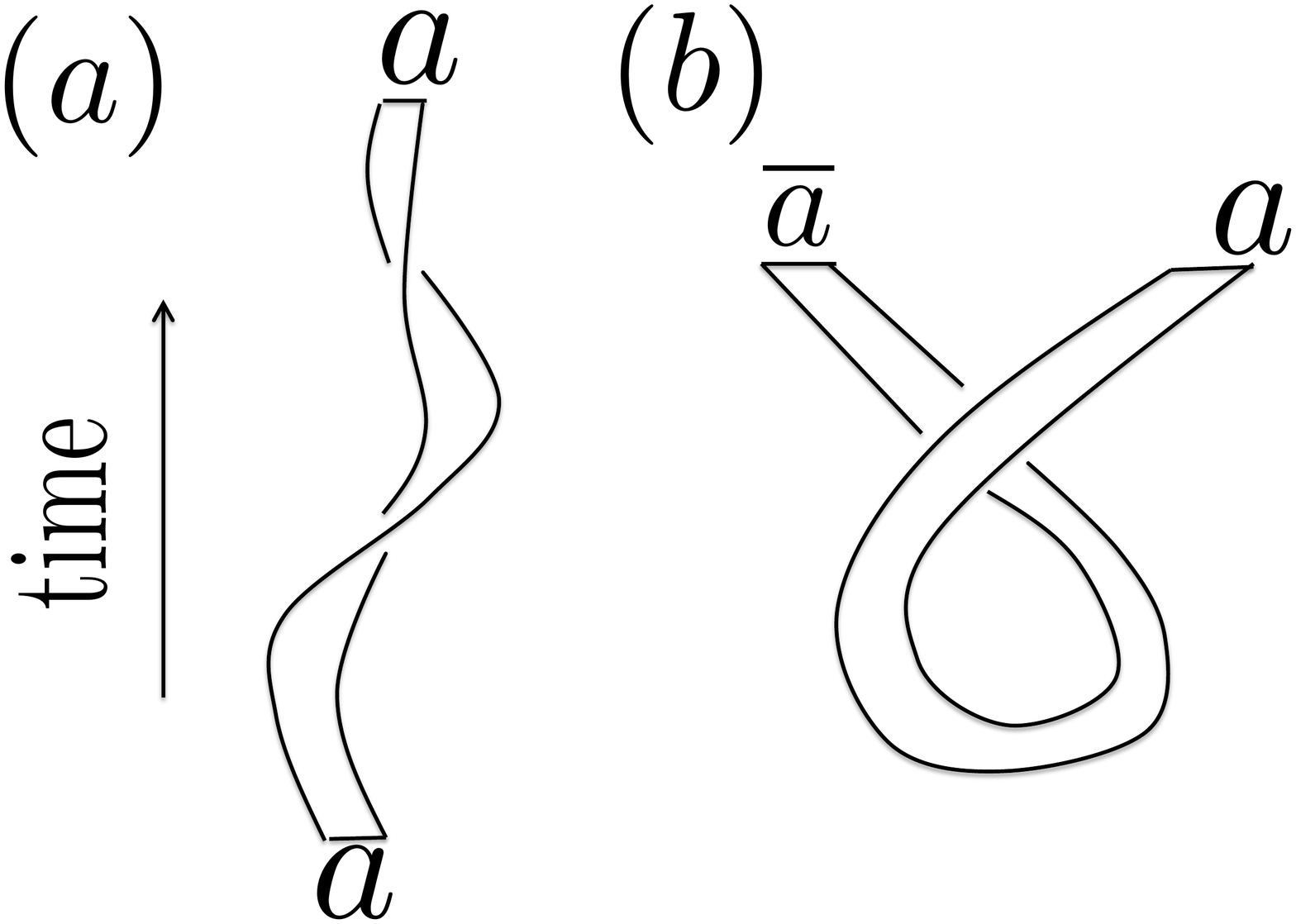}
\caption{\label{ExchangeTwistFig}  A space-time depiction of the relationship between the topological spin of the anyon $a$ and the phase incurred when exchanging $a$ with $\overline{a}$.  (a) In order to represent the effect of rotating an anyon by $2 \pi$, it is convenient to fatten its worldline into a ribbon; the rotation is then depicted by a $2 \pi$ twist in the ribbon.  (b)  A process in which a pair of anyons $a$ and $\overline{a}$ is created from the vacuum and then exchanged.  This worldline can be smoothly deformed into a horizontal ribbon with a $2 \pi$ twist.  (Do try this at home!)
}
\end{figure}

\begin{figure}[ht]
\includegraphics[width=0.45\textwidth]{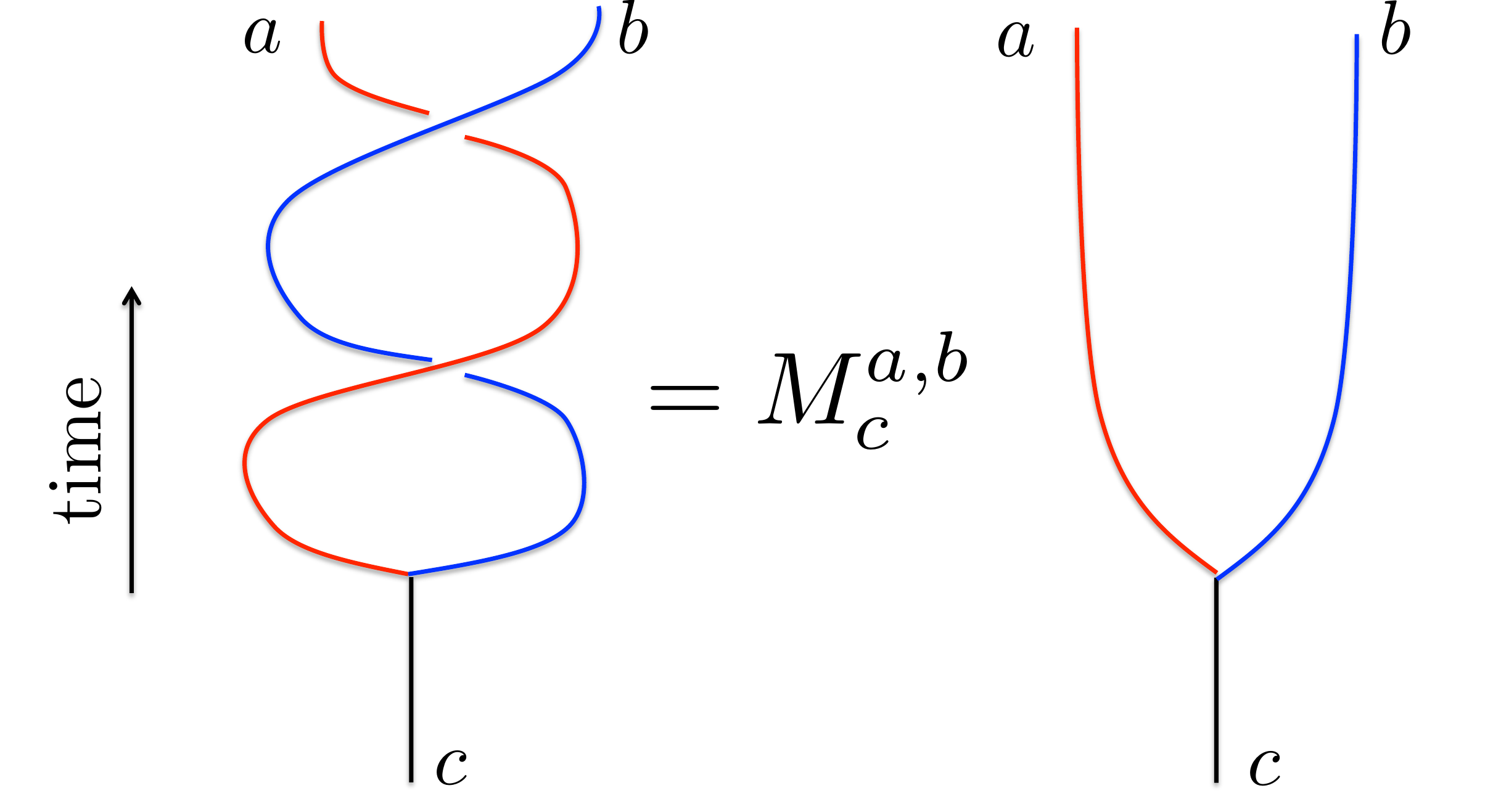}
\caption{\label{BraidFig}  A space-time depiction of a process in which anyon $a$ is braided around anyon $b$, where the fusion channel of $a \times b$ is fixed to  be $c$.  The red (blue) line labeled $a$ ($b)$denotes the trajectory of anyon $a$ $(b)$ in time; the vertex $(a,b,c)$ indicates that $a$ and $b$ were fused at some point in the past to $c$.}
\end{figure}
\end{enumerate}
%

Mathematically, these properties (among others) are described by a unitary modular tensor category (UMTC)\cite{KitaevHoney} -- a structure which is roughly analogous to the symmetry group in a conventional phase.\footnote{Technically the properties listed above, and indeed anyon condensation in general, do not require the anyon model to be modular; in fact a braided fusion category is sufficient.  However, here we will restrict our discussion to anyon condensation in the modular case.}

\subsubsection{Examples}
In what follows we will primarily focus on examples constructed from two well-known topological orders, which we describe here.  

{\bf Toric code:} The Toric code\cite{Wegner71,KitaevToric} 
 describes the topological order of an $s$ wave superconductor\cite{SondhiSC} (or equivalently, of a $\mz_2$ spin liquid\cite{pwa-rvb,WenSLTO,ReadSachdev,SachdevRead,IsingDual}).  It contains one fermion (the Bogolon) $\psi$, and two types of abelian bosons: a $\pi$-flux vortex which we call $m$, and a bound state of a fermion and a vortex, which we will call $e$.  
Since two Bogolons can form a Cooper pair,  two $\psi$ particles can annihilate to give the vacuum. 
Similarly a pair of $\pi$-flux vortices has trivial statistical interactions with the fermion, implying that 
two $m$ particles can also annihilate.  
Note that we do not require that combining the two $m$ particles actually returns the system locally to its ground state; the topological order is indifferent to questions of local energetics and depends only on the net Berry phase detected by a fermion encircling the region from a distance.

These properties are described by the following braiding and fusion rules:
\begin{align} \label{ToricData}
\psi \times \psi = 1 \ , \ \ \ m \times m = 1 \ , \ \ \  \psi \times m = e \\
s_e =s_m = 1 \ , \ \ \ s_\psi = -1 \\
M^{a a}_1 = 1 \ , \ \ \ 
M^{a b}_{a \times b} = -1 \ ,  a \neq b
\end{align}

{\bf Ising:} Our second example is the Ising topological order, which essentially describes a spinless chiral $p$ wave superconductor. \cite{VolovikPip,Ivanov,ReadGreen}
The superconductor has a fermion (the Bogolon) $\psi$ and a $\pi$ vortex.  In this case we will call the vortex $\sigma$, as it has a Majorana bound state.\cite{VolovikPip,Ivanov,ReadGreen} This implies that $\sigma$ is a non-abelian anyon, since a pair of vortices can have a net topological charge of either 1 (even fermion parity) or $\psi$ (odd fermion parity).  

This leads to the following braiding and fusion rules:
\begin{align}
\psi \times \psi = 1 \ , \ \ \ \psi \times \sigma = \sigma \ , \ \ \  \sigma \times \sigma = 1 + \psi \n
s_\psi =- 1 \ , \ \ \ s_\sigma = e^{ i \pi /8} \n
M^{\psi \psi}_1 = 1 \ , M^{\psi \sigma}_\sigma = -1 \n
 M^{\sigma \sigma}_1 =e^{ - i \pi/4}  \ , M^{\sigma \sigma}_\psi = e^{ 3 i \pi/4} 
\end{align}
In this case, there is an important feature of the statistical interaction not captured by $M$ and the fusion rules:  if a $\sigma$ anyon is braided around one member of a pair of $\sigma$ anyons, the fermion parity of this pair (encoded by whether the fusion channel of the two $\sigma$ anyons is $1$ or $\psi$) will be flipped after the braiding process.\cite{Ivanov}  This will be important to our analysis later.

One important difference between the true Toric code and Ising topological orders and their superconductor counterparts is that in the latter, vortices are created by externally applied electromagnetic fields, whereas in the former they are excitations intrinsic to the 2D system, as can be realized in somewhat less familiar models.\cite{MooreRead,KitaevHoney,KitaevToric,WenToric}   Readers should be aware that throughout this work, $m$ and $\sigma$ are intrinsic 2D excitations, rather than superconducting vortices.


{\bf G$_k$:} In addition, we will sometimes allude to other topological orders described by Chern-Simons gauge theories with gauge group $G$ in 2 spatial dimensions.    Because they are intimately related to Wess-Zumino-Witten models,\cite{WessWZW,WittenWZWa,WittenWZWb,NovikovWZW} 
exact results from 2D conformal field theory can be leveraged to understand the corresponding topological orders,\cite{WittenJones} as well as TSB in these systems.\cite{ConformalZoo,SlingerlandBais} 
Anyons in these models are associated with fluxes (a.k.a. Wilson lines) of the Chern-Simons gauge field; their statistical interactions realize a wide variety of anyon theories depending on the choice of the integer-valued Chern-Simons coupling $k$ and the gauge group $G$.  We will primarily take $G$= SU$(2)$, and denote the resulting topological order as SU$(2)_k$.   
Details of these topological orders, together with a much more thorough introduction to topological order than we give here, can be found in Refs. \cite{BondersonThesis,KitaevHoney}.

{\bf Combining and conjugating topological orders:}  Finally, it is useful to define the product of two topological orders $\mac{C} \times \mac{D}$, which simply denotes that we have the anyons of both $\mac{C}$ and $\mac{D}$, where anyons in $\mac{C}$ have trivial braiding (i.e. no statistical interactions) with anyons in $\mac{D}$.  In addition, the conjugate $\overline{\mac{C}}$ of a topological order $\mac{C}$ is a topological order with the same fusion rules, but where the topological spins and braiding phases are complex conjugated.

\subsubsection{Condensable bosons}

In order to discuss Bose condensation, we must first specify what is required for anyons to be bosons.  For non-abelian anyons, we require only that the particles be bosons under exchange (i.e. $M^{a \overline{a}}_c =1$) in at least one of the possible fusion channels $c$; this is sufficient to allow condensation to occur.\cite{SlingerlandBais}  For abelian anyons where the fusion outcome is fixed, they must be bosons under exchange in the usual sense.

  As discussed above, if $a = \overline{a}$, then by this definition $a$ is a boson if its topological spin is $1$.  If not, it turns out that it is both necessary and sufficient  for both $a$ and one of the fusion products $b \in a \times a$ to have topological spin $1$.\cite{TSBPRL,SlingerlandBais,Eliens}  Thus in general, in order for an anyon $a$ to be a boson, we require that $s_a = 1$ and that $N_{a,a}^b >0$ for some $b$ (which may be the vacuum) with $s_b = 1$.

One can also consider condensing multiple bosons $\gamma_1, ... \gamma_n$ simultaneously.  In this case, we additionally require that there is at least one fusion channel in which each pair of particles braid trivially with each other.  This will be the case if, for each $i$ and $j$ there is at least one fusion channel $c$ such that $M^{\gamma_i \gamma_j}_c =1$.  
If the condensing boson(s) are abelian, meaning that $\gamma_i \times \gamma_j = c$ (with no sum), this turns out to requires that $c$ is also a boson; if $c\neq 1$ then $c$ will also condense.\cite{Eliens,NeupertBernevig}

\subsection{Understanding the condensed phase}

To understand the effect of forming a Bose condensate of anyons, it is helpful to draw parallels with the more familiar setting 
 of spontaneous symmetry breaking.   Here Bose condensation reduces a microscopic symmetry group $G$ to a subgroup $H$ of $G$.  Before condensation quasiparticles can be classified according to the representations of $G$; afterwards we label them by the representations of $H$.  
This has two possible effects, both of which can occur in TSB.

First, distinct particles may become {\it identified} in the condensed phase.  For example, in an insulator we may (in the absence of screening) label excitations by their charge, which could be any integer multiple of $e$; in a superconductor, where Cooper pairs of charge $2e$ have condensed, we label them by their charge {\it modulo $2e$}.  Thus a region with any even number of electrons is now indistinguishable from the vacuum;  any odd number is equivalent to having a single unpaired electron (at least for experiments attempting to measure the local charge).  We have identified excitations of charge $q$ and $q+2e$.  

Second, if the symmetry is non-abelian, such that it has representations of dimension greater than 1, it is also possible that one representation in the unbroken phase {\it splits} into two or more distinct representations after the symmetry is broken.  For example, if the SU$(2)$ spin rotation symmetry of a magnet is broken to the group of rotations about the $z$ axis, the spin $S$ representation splits into $2s+1$ distinct representations with different eigenvalues of the now conserved quantity$S^z$.  

In addition to the possibility of identifying or splitting quasi-particles, we must account for the fact that the bosons we condense have non-trivial braiding with some of the other anyons.  These anyons will therefore cause destructive interference between different configuration histories of the condensate, and become {\it confined} in the condensed phase.   A similar phenomenon occurs for magnetic flux in a superconductor: general fluxes have non-trivial Aharonov Bohm phases with the condensed Cooper pairs, and flux (where permitted at all) is confined to vortex cores carrying an integer number of half-flux quanta.  By confined, we mean that subdividing the vortex core into two halves would incur an energy cost that increases at least logarithmically with their separation, such that the two halves are energetically bound together.  

The parallels between TSB and spontaneous symmetry breaking are summarized in Table \ref{CompTab}.

 \begin{widetext}
 
 \begin{table}[htp]
 \begin{tabular}{|c|c|c|}
 \hline
 Topological order & Symmetry breaking & Analogy \\
 \hline
 UMTC $T$ & Symmetry group $G$ & Underlying mathematical structure \\
 Topological charge & Irreducible representation (irreps) & specifies conserved properties of quasiparticles \\
 &&  (i.e. charge, braiding statistics, etc.) \\
 Fusion & Tensor product of irreps & Describes possible conserved properties of a region  \\
 && in space with two or more quasiparticles \\
 Braiding &   \footnote{ Braiding is present only in theories where the symmetry is gauged, where it corresponds to a Berry phase between particles and vortices.} & Long-ranged statistical interaction between quasiparticles \\
Condensed topological order $\tilde{T}$ & $H$ (a subgroup of $G$) & Mathematical structure of the condensed phase \\
Identification $a_1 \rightarrow \tilde{a}, a_2 \rightarrow \tilde{a} $ & 2 irreps of $G \rightarrow$ same irrep of $H$ & After condensation, two sets of conserved properties \\
&& are no longer physically distinguishable \\
Splitting $a \rightarrow \tilde{a}_1 + \tilde{a}_2 $ & 1 irrep of $G \rightarrow 2$ irreps of $H$ & After condensation, multiple internal states of the same\\
&&  anyon or irrep become physically distinct  \\
Confinement & &  Analogous to  confinement in Anderson-Higgs phases\\
& Goldstone mode & Not present in TSB (or Anderson-Higgs phases) \\
\hline
\end{tabular}
\caption{\label{CompTab}  Analogy between Bose condensation leading to spontaneous symmetry breaking, and Bose condensation leading to TSB.  Braiding, and consequently confinement, do not occur in theories with a global symmetry-- although they do arise when the symmetry is gauged (in which case spontaneous symmetry breaking describes the Anderson-Higgs mechanism).  Like Higgs phases, a phase obtained from TSB is fully gapped, and never contains Goldstone modes. }
\end{table}
 
 \end{widetext}

\subsubsection{Example} \label{CondSecEx}
The anyons in the condensed phase can be understood using these three principles. 
Before discussing the general case, let us illustrate this with an example.  We begin with Ising $\times \overline{\text{Ising}}$ -- i.e. two copies of the Ising topological order with opposite chirality.  Roughly speaking, this descripes a time-reversal invariant bilayer chiral $p$ wave superconductor, in which the two layers are initially decoupled.  
The non-trivial anyons are the fermions $\psi_1, \psi_2$ of layers $1$ and $2$ respectively, the two vortices $ \sigma_1,\sigma_2$, and the $4$ combinations  $\sigma_1 \psi_2, \psi_1 \sigma_2, \psi_1 \psi_2,$ and $\sigma_1 \sigma_2$.  
(Note that in a real superconductor  $\sigma_1$ and $\sigma_2$ would not typically be independent, since an external flux would thread both layers.  However in the Ising topological order $\sigma$ is an intrinsic 2D excitation, so that the two are independent).  Since the two layers to have opposite chirality, the topological spins of $\sigma_1$ and $\sigma_2$ are complex conjugates.  This ensures that $s_{\sigma_1 \sigma_2} =1$, and thus since $\sigma_1 \sigma_2$ is its own anti-particle (i.e. $\sigma_1 \sigma_2 \times \sigma_1 \sigma_2 = 1 + ... $) , it satisfies our criterion to be a boson. 
 
Let us now condense $\psi_1 \psi_2$ \cite{HormoziSlingerland,TSBShort}  -- an inter-layer bound pair of fermions.   What  effect does this have on the allowed topological charges?  First, after condensation the fermionic quasiparticle is a superposition of $\psi_1$ and $\psi_2$, in much the same way that the Bogolon itself is a superposition of particle and hole excitations.  Similarly the new vacuum $\tilde{1}$ is a superposition of the original vacuum and the inter-layer pairs.  Thus up to normalization the vacuum $\tilde{1}$ and fermion $\tilde{\psi}$ of the condensed phase are:
\be
\tilde{\psi} \sim \left(  \psi_1 + \psi_2 \right ) \ , \ \ \ \tilde{1} \sim \left( 1 + \psi_1 \psi_2 \right )
\ee
 We say that $\psi_1$ and $\psi_2$ have been identified, as have $1$ and $\psi_1 \psi_2$.

Since $\psi_1$ ($\psi_2$) has a non-trivial Berry phase only with $\sigma_1$ ($\sigma_2$), braiding an inter-layer pair around an isolated vortex in either layer gives the Berry phase $M^{\sigma \psi}_{\sigma}= -1$.   This means that the four anyons $\sigma_1,\sigma_2, \sigma_1 \psi_2$,  and $ \psi_1 \sigma_2$ all induce branch cuts in our condensate wave-function, whose energy cost grows with the length of the branch cut.  These anyons are therefore confined, and are not part of the topoloigcal order of the condensed phase.   

The bound state $\sigma_1 \sigma_2$, however, remains deconfined after the transition.
 Because $\sigma_1$ and $ \sigma_2$ each have an associated Majorana bound state, the resulting anyon $\tilde{\sigma}$  comes in two flavours-- one with even fermion parity, and one with odd.   
Before condensation, we have:
\be
(\sigma_1 \sigma_2) \times (\sigma_1 \sigma_2)  = 1+ \psi_1 \psi_2  + \psi_1 + \psi_2 
\ee
After condensation, this becomes:
\be
\tilde{\sigma} \times \tilde{ \sigma}   = \tilde{1}+ \tilde{1}+ \tilde{\psi}  + \tilde{\psi}  
\ee
The two copies of the vacuum state correspond to two orthogonal states that have trivial topological charge in the condensed phase.  However, if $\tilde{\sigma}$ represents a single anyon type, this multiplicity of the vacuum in the fusion product is not compatible with the mathematical structure required for a consistent topological order.  The resolution is that the two flavours of $\tilde{\sigma}$ (with even and odd fermion parity, respectively) split into two distinct anyon types, $\tilde{\sigma}_e$ and $\tilde{\sigma}_m$.

\begin{figure}[ht]
\includegraphics[width=0.4\textwidth]{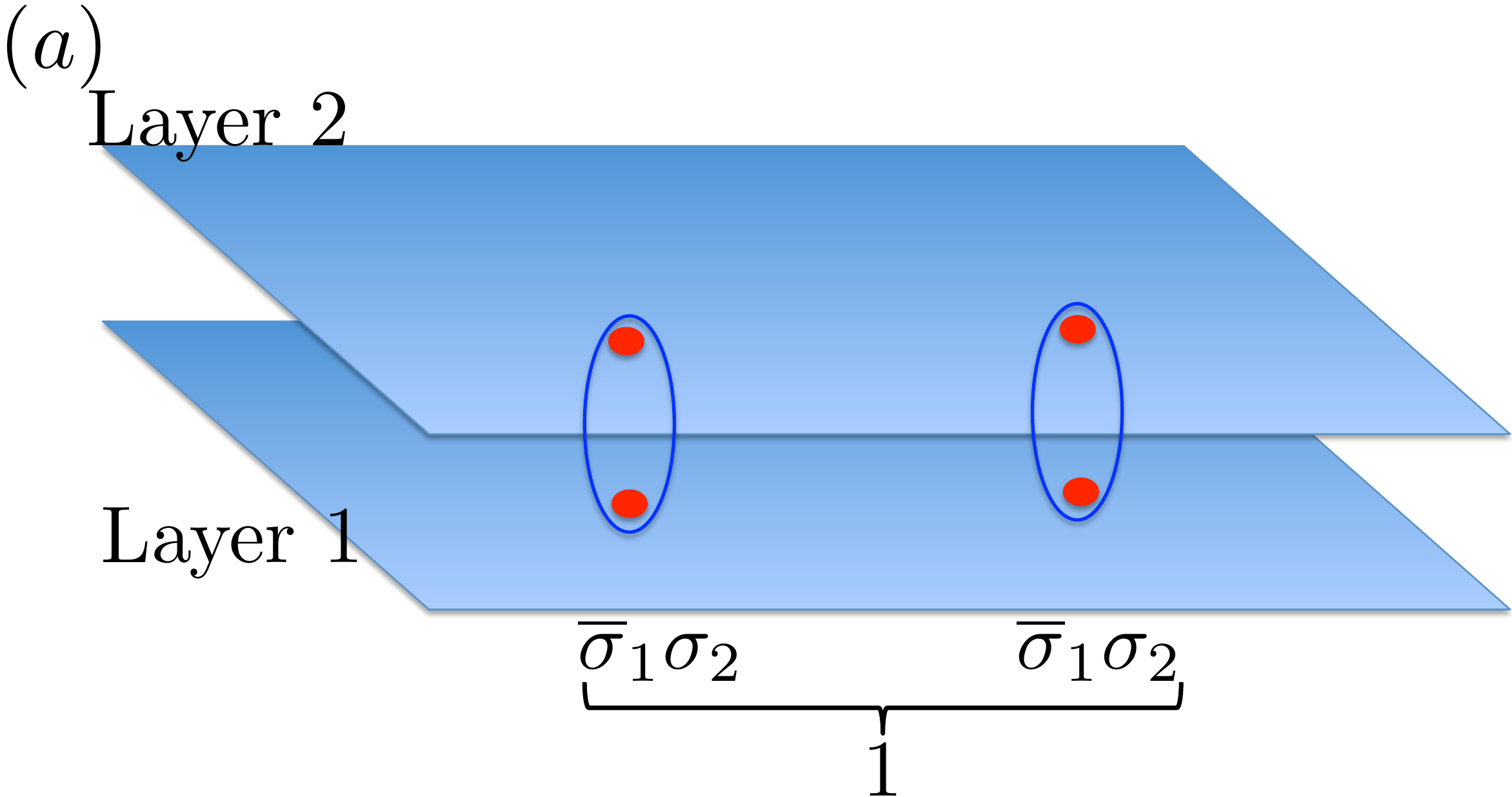} \\
\includegraphics[width=0.4\textwidth]{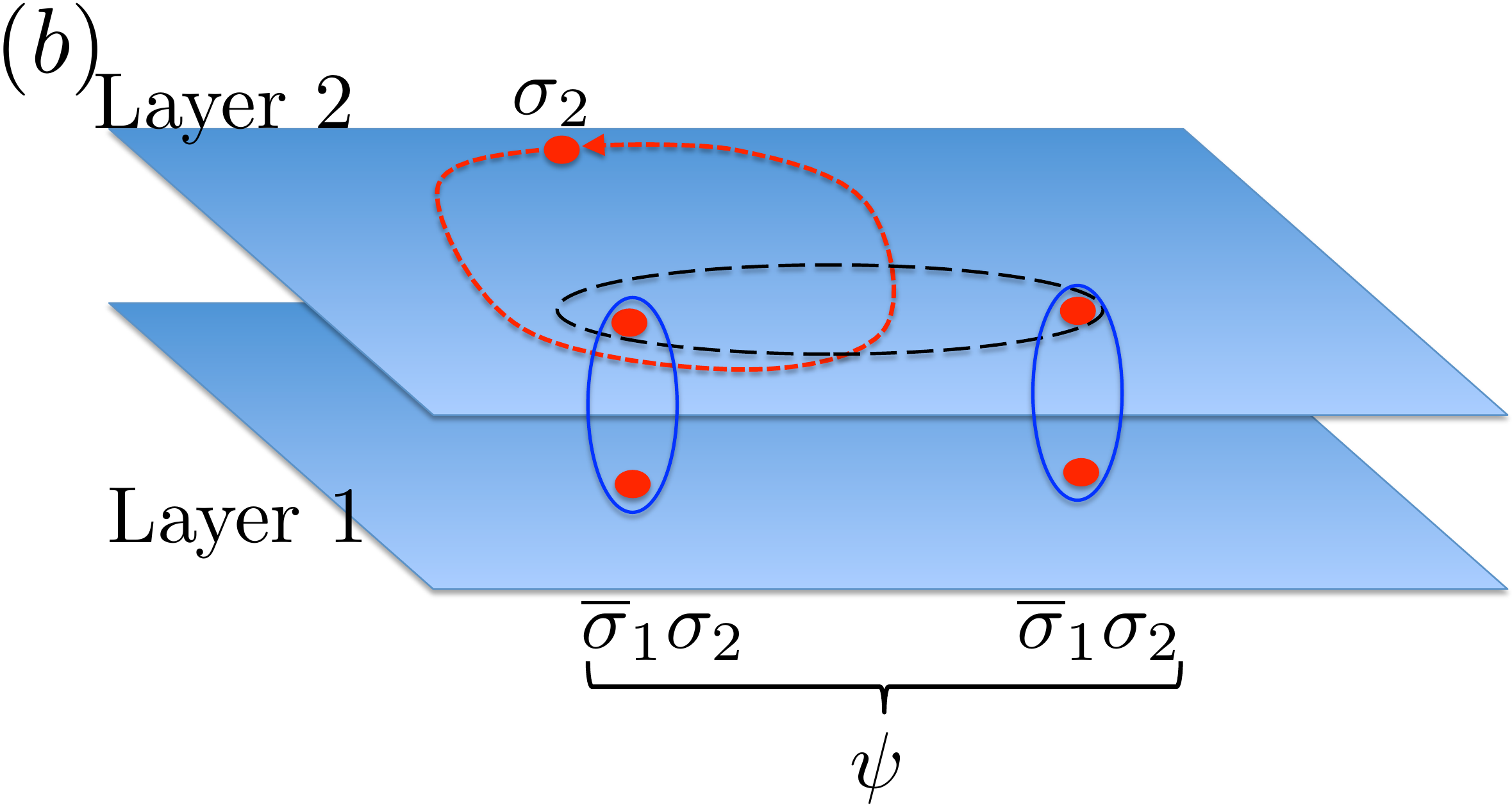} 
\caption{\label{SigBraidFig} A pair of $\sigma_1 \sigma_2$ bound states, initially with both $\sigma_1$ particles, and both $\sigma_2$ particles, fusing to the vacuum, is braided with a $\sigma_2$ particle.  (a) Before braiding, the net fermion parity of the two pairs is even.   (b)  After braiding a $\sigma_2$ anyon around one of the $\overline{\sigma}_1\sigma_2$ anyons, the fermion parity of the remaining two $\sigma$'s in layer 2 (represented by a black dashed line) has changed.  The net fermion parity of the two $\sigma_1\sigma_2$ pairs is now odd, meaning that one of these pairs has changed its fermion parity.  A similar argument holds in other initial fusion channels.
}
\end{figure}

The end resut of this process is a topological order with one species of fermion $\tilde{\psi}$, and two abelian anyons $\tilde{\sigma}_e$ and $\tilde{\sigma}_m$, both with $\pi$ Berry phase when braided with the fermion.  As $\sigma_1$ and $\sigma_2$ have opposite topological spins, these particles are both bosons.  The fusion rules are:
\begin{align}
\tilde{\psi} \times \tilde{\psi} = 1 \ ,\ \ \ \tilde{\psi} \times \tilde{\sigma}_e = \tilde{\sigma}_m   \ ,\ \ \ \tilde{\psi} \times \tilde{\sigma}_m = \tilde{\sigma}_e \n
 \tilde{\sigma}_e \times  \tilde{\sigma}_e =  \tilde{\sigma}_m \times  \tilde{\sigma}_m = 1 \ , \ \ \  \tilde{\sigma}_e \times  \tilde{\sigma}_m = \tilde{\psi}
\end{align}
Identifying $\tilde{\sigma}_m = m, \tilde{\sigma}_e= e$, we have recovered the topological order of the Toric code!  

One might rightly ask why, before condensation, $\sigma_1 \sigma_2$ could not be split into a fermion parity even and fermion parity odd particle.  
Before condensation, braiding a $\sigma_1 \sigma_2$ pair with a $\sigma_2$ anyon changes the pair's fermion parity.  (See Fig. \ref{SigBraidFig}).  Since topological charge must be conserved under braiding operations, these two internal states therefore cannot comprise different anyon types.  Once the single-layer vortices $\sigma_1$ and $\sigma_2$ are confined, however, the pair's fermion parity is conserved under braiding, and it splits into two distinct anyon types.

\subsubsection{General prescription}

This example illustrates the main phenomena that arise in TSB transitions.  
We now summarize these in the general case.  The main complications that can arise here are that (1) we may condense multiple boson types $\{ \gamma_i \}$, which is often necessary in order to obtain a consistent vacuum $\tilde{1}$ in the condensed phase; and (2) the condensing boson may be non-abelian. 

 Let $\{ \gamma_i \}$ be a set of bosons that we wish to condense.  Then:\cite{Gepner89,ConformalZoo,SlingerlandBais,Eliens}
\begin{enumerate}
\m {\it Identification}:
An anyon  $\tilde{t}$  in the condensed phase is often comprised of a superposition of the original anyons: up to normalization, 
\be  \label{IdqE}
\tilde{t} \sim \sum_a n_a a
\ee
As was the case for the two fermions $\psi_1$ and $\psi_2$ in our example above, in the condensed phase often two anyons $a$ and $b$ will only appear as the superposition $a + b$, and are therefore reduced to a single anyon type.  In this case we say  that $a$ and $b$ have been identified.  

The condensing bosons are always identified with the vacuum.  
If the condensing boson $\gamma_i$ is abelian, $
 a $ and $ \gamma_i \times a$ are always identified.  Thus fusing any anyon with the condensate returns the same anyon type.  (Clearly this must be the case if $\gamma_i$ is to be topologically equivalent to the vacuum after condensation!)

 \m {\it Splitting}:
 If 
 \be
 a \times \overline{a} = 1+ \gamma_i + ... \ ,
 \ee
 then in the condensed phase $ \tilde{a} \times \overline{\tilde{a} }$ contains orthogonal copies of the vacuum $\tilde{1}$.   This implies that $\tilde{a}$ cannot be a single anyon type (otherwise the topological order would be mathematically inconsistent).  Thus  $\tilde{a}$ (and consequently $\overline{\tilde{a}}$) split into multiple anyon types.  Clearly this can only happen if, before condensation, $a$ is non-abelian. 
 
  As in our example, where one excitation $\sigma_1 \sigma_2$ with two internal states split into two excitations $\sigma_e$ and $\sigma_m$, each with one internal state, splitting conserves the total number of internal states (called the total quantum dimension) of the $a$ anyons.  
 
\m {\it Confinement}:
If the condensing bosons are abelian, only anyons which obey:
\be
M^{ a, \gamma_i}_{a \times \gamma_i}  = 1
\ee
remain deconfined after condensation.  

If $\gamma_i$ is non-abelian, this condition is required to hold only in some of the fusion channels $a \times \gamma_i$.  This is because different superpositions (\ref{IdqE}) involving the anyon $a$ are associated with different fusion channels of $a \times \gamma_i$, and may therefore have different braidings with $\gamma_i$.  After condensation, this means that some linear combinations involving $a$ become confined while others do not. 
This also applies to the condensate itself: if $N_{\gamma_i,\gamma_i}^c >0$ and $c$ is not a boson, then $c$ 
is confined in the condensed phase.  Since $c$ is manifestly not topologically equivalent to the vacuum, this is necessary to ensure that the point particles that remain after condensation have conserved topological charge.
Ref. \cite{Eliens} gives an example of how this occurs in SU(2)$_{10}$.   
\end{enumerate}

In general, it can be somewhat technically involved to work out the details of the topological order that is obtained by condensation; systematic approaches to doing this are described in Refs. \cite{Eliens,NeupertBernevig} and will not be a focus of this review. 
Additionally, many interesting examples have been studied in the context of conformal field theory.\cite{GoddardCoset1,GoddardCoset2,ConformalZoo} 

At this point, the reader may be wondering how TSB relates to more familiar field-theoretic treatments of symmetry breaking, such as the Higgs mechanism.  If the topological order describes a discrete gauge theory,\cite{BaisDiscrete} and the condensing boson can be mapped to a charge (rather than flux) type excitation, then the two yield the same gapped phases.  
For topological orders described by Chern-Simons theories, one might wonder whether a Chern-Simons-Higgs description  similarly captures the phenomenology of TSB.  Though there are situations where this holds\cite{WenTCLGCS,ClarkeNayak}, it need not be the case: for example, a TSB transition relates the Chern-Simons theory with gauge group $G=SU(2)$ to that with gauge group $G=SO(3)$; however there is no Higgs field that reduces the gauge symmetry from $SU(2)$ to $SO(3)$, since these have the same Lie algebra.  
.

\section{Anyon condensation and physics at the edge}

One interesting feature of topologically ordered systems is that they can possess ungappable boundaries.  
By ungappable, we mean that they cannot be gapped even in the absence of any symmetry.  This is the case, for example, for the boundary between an integer quantum Hall system and the vacuum -- but {\it not} between an integer spin Hall system and the vacuum.
One may then wonder: given two phases with different topological orders $\mac{C}$ and $\mac{D}$, when can the boundary between them be fully gapped?  
This question turns out\cite{BaisBoundaries,KongCond,Beigi11, BaisChiralBoundaries,YanWanCond}  to be intimately related to the question, ``when can a set of non-local bosons in $\mac{C}$ be condensed to obtain $\mac{D}$?"  (Or vice versa).  Here we will review this relationship, and some of its interesting physical consequences.

In general,  two phases that cannot be separated by a gapped boundary must be {\it topologically distinct}: For example, an ordinary band insulator admits a gapped boundary with a Mott insulating state, but a quantum Hall insulator, whose band structure is topologically nontrivial, does not.
This stems
 from the fact that the two topologically distinct phases contribute differently to the flow of energy on the boundary, requiring an ungappable boundary mode to ensure energy conservation.\cite{RyuAnomaly,FurasakiAnomaly,WenAnomaly} 
For example,  if phases $\mac{C}$ and $\mac{D}$ have different thermal Hall conductivities\cite{KaneFisherThermalHall}, the boundary between them must be gapless.  
There are known examples,\cite{LevinBdy} however, where $\mac{C}$ and $\mac{D}$ have the same thermal Hall conductivity, but the boundary nonetheless cannot be gapped.

Given that the thermal Hall conductance is not sufficient, what features of the topological orders $\mac{C}$ and $\mac{D}$ determine whether their boundary can be gapped?  
Here, we review arguments\cite{LevinBdy,KongCond,KitaevKong} implying that boundaries between topological orders related by TSB are gappable, while other boundaries are not.  

\begin{figure}[ht]
\begin{tabular}{lc}
\includegraphics[width=0.25\textwidth]{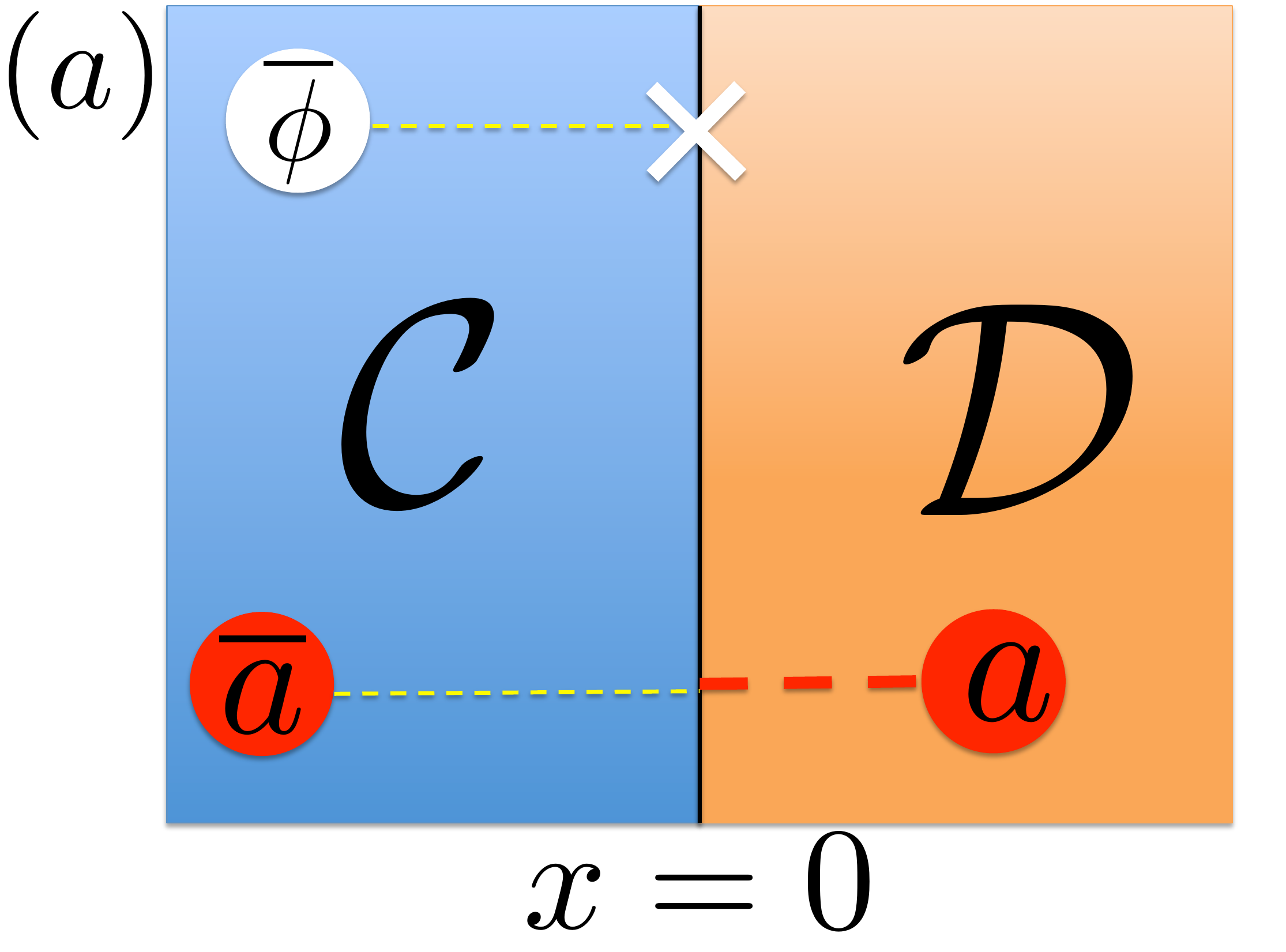} & \includegraphics[width=0.17\textwidth]{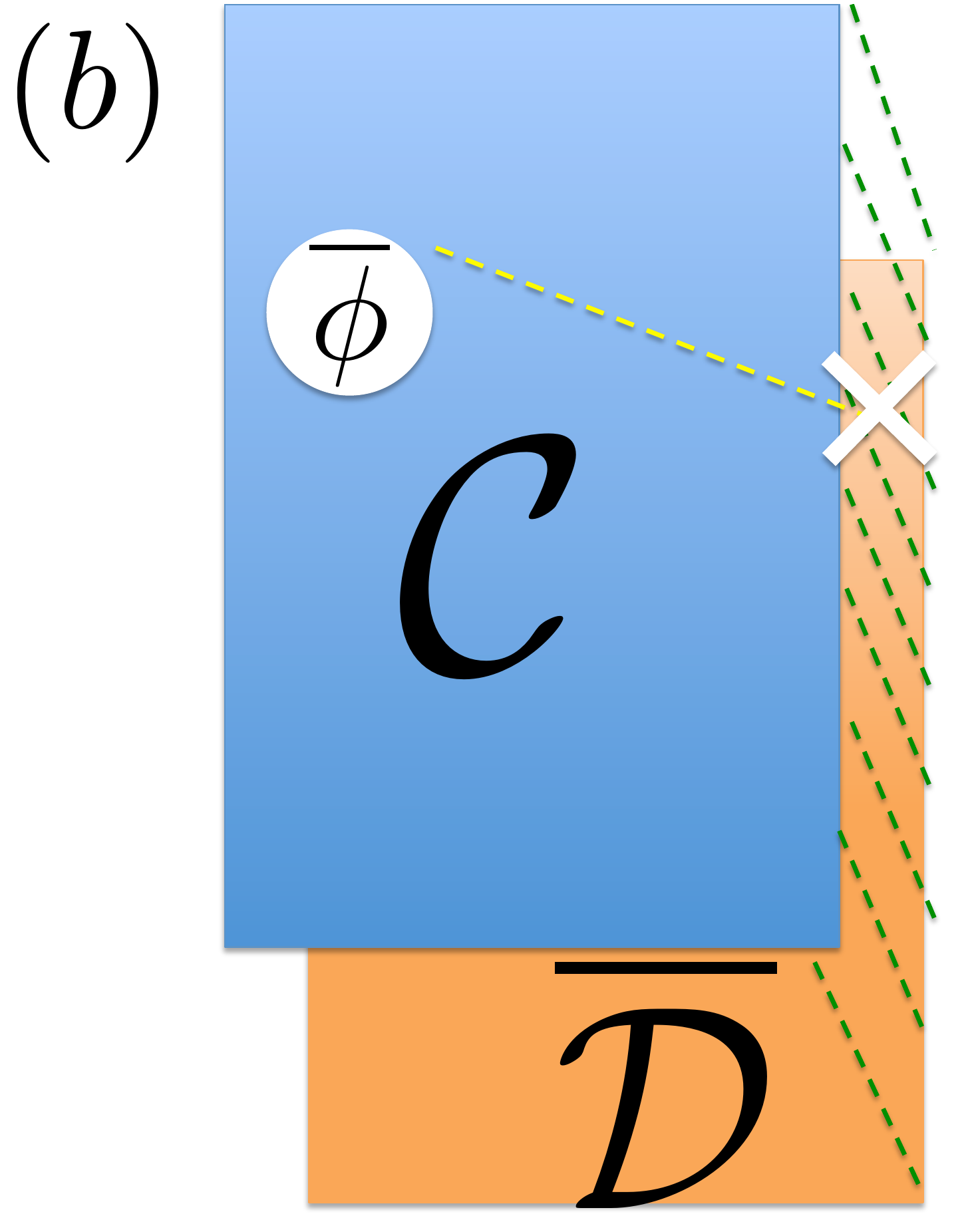} \\
\end{tabular}
\caption{\label{edgefig}(a) Creating a boundary between topological orders $\mac{C}$ and $\mac{D}$ by condensing the boson $\gamma$ in the region $x>0$.  $\gamma$ anyons can ``dissappear" (indicated by an x) at this boundary; a confined anyon $a$ brought to $x>0$ incurs an energy cost proportional to its separation from the boundary (indicated by the thick dashed red line). (b) By folding the plane along the line $x=0$, we see that this is equivalent to a boundary  $\mac{C} \times \overline{\mac{D}}$ with the vacuum.  
}
\end{figure}

To simplify our task, we note that a gapped boundary between $\mac{C}$ and $\mac{D}$ is equivalent to a gapped boundary between $\mac{C} \times \overline{\mac{D}}$ and the vacuum.   This is shown in Fig. \ref{edgefig}: ``folding" along the gapped boundary between $\mac{C}$ and $\mac{D}$ gives a gapped boundary between $\mac{C} \times \overline{\mac{D}}$ and the vacuum.  (Folding reverses the orientation of $\mac{D}$, and therefore also flips its chirality, giving $\overline{\mac{D}}$).    Therefore to understand in general when boundaries may be gapped, we need only to study gapped boundaries between a general topological order $\mac{T} \equiv \mac{C} \times \overline{\mac{D}}$ and the vacuum.  


To test whether $\mac{T}$ admits a gapped boundary with the vacuum, imagine a 1D system comprised of a thin strip of our topological phase, on which the lower boundary is gapped, but the upper boundary remains gapless.  
 At energy scales below the gap, this 1D system is described by the conformal field theory (CFT) corresponding to the upper gapless edge alone.  
Following Levin,\cite{LevinBdy} we can assess whether the lower edge of our strip really can be gapped by determining whether this CFT 
 is a well-defined 1D theory.  If not, we arrive at a contradiction, and will be forced to conclude that either the upper boundary is also gapped, or the lower boundary is gapless -- i.e. the upper and lower boundaries are not individually gappable.
 
 To see how this question is connected to TSB,
we consider 
 the fate of a quasiparticle $a$ created in the $\mac{T}$ region and brought to the gapped boundary.  If $a$ is among the anyons that become confined, then it cannot cross into the vacuum without paying a high energy cost.  Evidently, if the boundary is gapped, in this case $a$ anyons correspond to gapped excitations at the boundary.   If $a \in \{ \gamma_i \}$ is among the bosons condensed in the vacuum, then the $a$ quasiparticle ``dissappears" at the boundary; its energy cost falls to $0$ as it enters the condensed region, where it is indistinguishable from the vacuum.  This must hold even if the boundary is gapped.  Since all anyons must be either confined or condensed in the vacuum, these are the only two possibilities.\cite{BaisBoundaries}

At energies below the bulk gap, our strip model is therefore described by a 1D CFT whose local operators can carry the topological charges of the $\{ \gamma_i \}$. 
 since we can create a $\gamma_i \overline{\gamma}_i$ pair in the bulk, and bring one of them (say $\overline{\gamma}_i$) to the lower gapped boundary, and the other (say $\gamma_i$) to the gapless boundary.  
 Since $\overline{\gamma}_i$ dissappears into the vacuum at the gapped boundary, the result is  a local operator with topological charge $\gamma_i$.   For all other anyons, our only option is to bring both $a$ and $\overline{a}$ to the gapless boundary, corresponding to an operator with trivial topological charge.    

To determine whether the lower boundary is indeed gappable, we must ascertain whether this set of local operators corresponds to a legitimate 1D CFT.
It is believed that this requires\cite{CardyModular,FreidanShenker} the partition function constructed from these local operators to be invariant under a set of operations known as modular transformations on the torus.
As is well known in CFT\cite{MooreSeibergNatural,AlvarezGaume,DijkgraafVerlinde,CapelliADE} (where the construction analogous to TSB is known as ``extending the chiral algebra"\cite{ConformalZoo,SlingerlandBais}), this is guaranteed if $\{ \gamma_i \}$ are a set of bosons in $\mac{T}$ that can be condensed to produce the vacuum.  

In other words, the 1D CFT associated with the boundary between two phases related by TSB is  always invariant under modular transformations.  In this case the corresponding boundary is always gappable.
In fact, this can be proven by direct construction: if $\mac{T}$ contains a set of bosons (known as a Lagrangian algebra\cite{KongCond}) that can be condensed to give the vacuum, then $\mac{T}$ is a special type of modular tensor category (i.e. of topological order) known as a Drinfeld center;\cite{Davydov13}  for these topological orders Hamiltonian models with gapped boundaries can be constructed explicitly.\cite{KitaevKong,LW,KongCond,LevinLin}

 Conversely, we could ask whether a topological order $\mac{T}$ which is not related to the vacuum by TSB is always gapless.  
In this case, we arrive at a potential contradiction: the strip with 1 gapped boundary has a  partition function that is modular invariant {\it only} if  $\mac{T}$ and the vacuum are either isomorphic\cite{DijkgraafVerlinde} or related by extending the chiral algebra (e.g. by TSB).\cite{MooreSeibergNatural,Schellekens89,Fuchs11,KongModular,Fuchs13,Hung15}  This is believed to imply that the lower boundary of our strip cannot be gapped -- i.e. the boundary between $\mac{T}$ and the vacuum is ungappable.

In summary, results developed in the study of CFT strongly suggest that the boundary between $\mac{C}$ and $\mac{D}$ can be gapped if (and only if) the two topological orders are related by TSB.  Here ``related by" means that $\mac{C} \times \overline{\mac{D}}$ contains a set of bosons (known as a Lagrangian algebra) that can be condensed to give the vacuum.\cite{LevinBdy,KongCond}  Physically, this means that TSB -- like Bose condensation in conventional systems -- does not alter any interdependence of the bulk and boundary.  For example, it cannot alter the topological central charge,\cite{SlingerlandBais} which describes the thermal Hall conductivity modulo 8.\cite{ReadGreen}\footnote{The reason that this quantity is conserved only modulo $8$ is that there exists a 2D bulk state of interacting bosons, known as the Kitaev $E_8$ state, with no topological order and boundary central charge 8.\cite{KitaevHoney}  In principle we could add copies of this $E_8$ state to any 2D system without changing the bulk topological order; hence the statistics of the anyons can determine the boundary's central charge only modulo 8.  Indeed SU$(9)_1$ has central charge $8$, and can be reduced to the vacuum by condensing either of the two bosons.}

 \subsection{Multiple boundary types and topological bound states} \label{EdgeBS}

It often happens that there is more than one choice of bosons to condense in $\mac{C}$ to obtain $\mac{D}$; this can lead to multiple distinct types of gapped boundary between these two topological orders.
The simplest example of this phenomenon occurs in the Toric code model, for which either $e$ or $m$ (but not both) can be condensed to obtain the vacuum.  This leads to two distinct boundaries with the vacuum\cite{BravyiKitaev}: an $e$-edge, where $e$ particles may be locally annihilated, and an $m$ edge where $m$ particles can be locally annihilated.  

\begin{figure}[ht]
\includegraphics[width=0.3\textwidth]{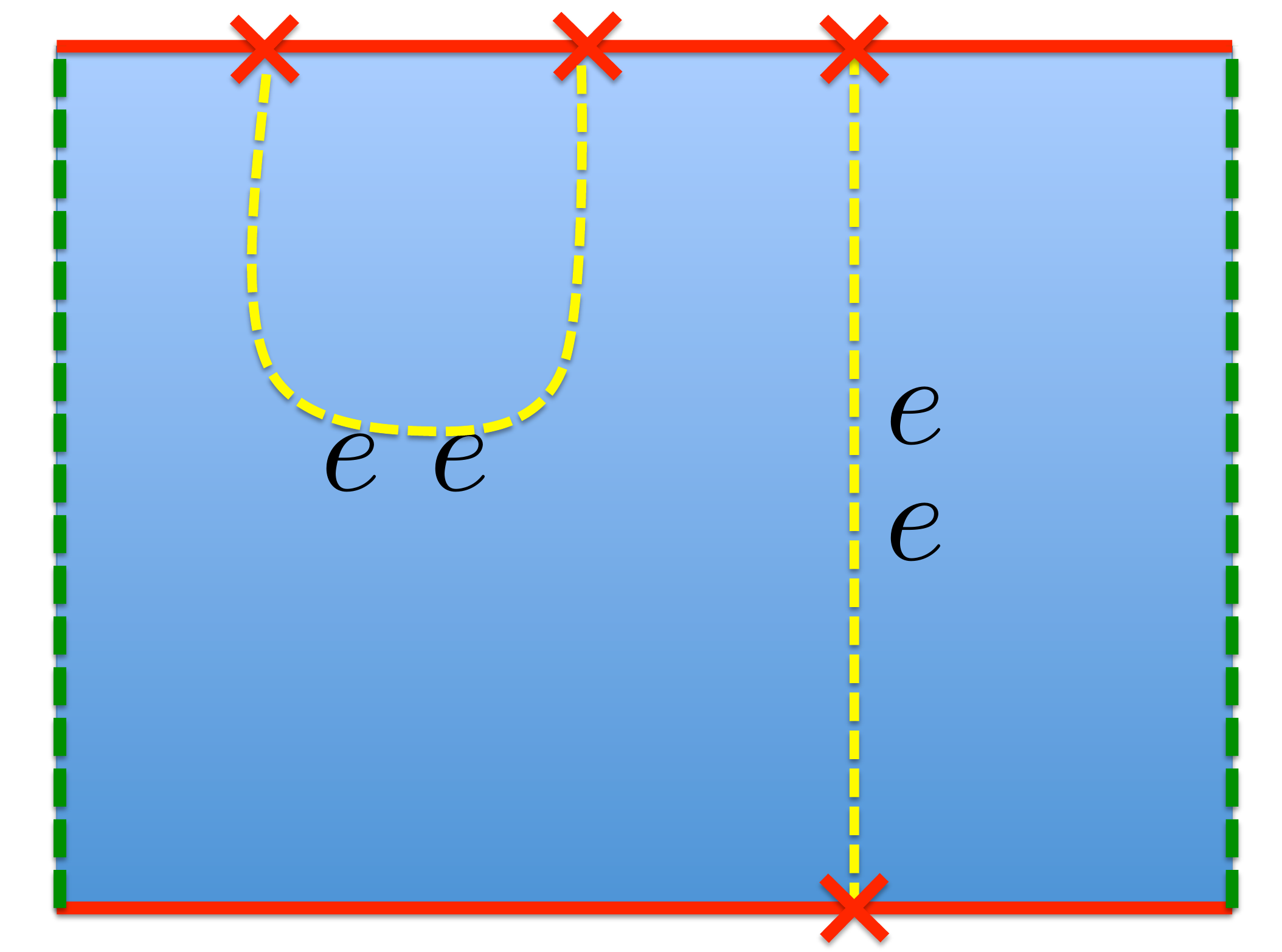} 
\caption{\label{TCBdiesFig}  Mixed boundary conditions between the Toric code and the vacuum: along the top and bottom edges of the square (solid red lines) $e$ anyons have condensed; on the left and right edges (dashed green lines) $m$ anyons have.  Dashed yellow lines show two distinct paths by which a pair of $e$ particles created in the bulk can be brought to the $e$ boundaries, where they can disappear.
}
\end{figure}

Given that there are two distinct types of boundary with the vacuum, it is natural to ask what happens when we consider a system with mixed boundary conditions (Fig. \ref{TCBdiesFig}), where some regions have $e$-type boundaries, and some have $m$-type boundaries.   Despite the fact that outside the square the system is uniformly in a topologically trivial phase (which is certainly possible here since the regions of condensed $e$ and of condensed $m$ can be adiabatically connected\cite{FradkinShenker}), the square with mixed boundary conditions has a larger number of ground states than one with uniform boundary conditions!  To see this, consider a process where a pair of $e$ quasiparticles are created in the bulk of the system and brought to an $e$ boundary.  If the boundary conditions are uniform, each $e$ particle may be brought to any point on the boundary, and all such choices are equivalent.  If the boundary conditions are mixed as in Fig. \ref{TCBdiesFig}, however, there are two different ways to bring these quasiparticles to the boundary: either we can bring them both to the top (or bottom) edge of the square, or we can send one to the top and one to the bottom.  These two choices are energetically degenerate (both return the system to its ground state), but are nonetheless physically distinct: an $m$ particle tunneling between the left and right boundaries of the square will give a different phase depending on whether the charges went to the same or to opposite edges.   

Numerous other examples of topological orders with multiple types of gapped boundary have been studied;\cite{BravyiKitaev,KapustinAbelianBdies,WangWenBdy,LanWangWen,YanWanCond} in all cases increasing the number of interfaces between the different boundary types increases the number of ground states.  In some cases a bosonized treatment of the edges in question can be used to derive this increased ground state degeneracy,\cite{GaneshanBdy} and show that it can be attributed to the presence of bound states at the interface between the two domain wall types.\cite{FuKane,LindnerPara,ClarkePara,ChengPara,HughesSantos}  These bound states encode a degenerate ground-state Hilbert space, and can be viewed as non-abelian anyons that have been pinned to a specific spatial location.  In the Toric code example given above, a Majorana bound state occurs at each interface between different domain wall types; thus each new segment of $m$-type domain wall after the first one increases the number of ground states by a factor of 2.

\section{TSB, defects, and anyon-permuting symmetries}

Another way to create interesting non-abelian bound states is by making defects in a topologically ordered phase with a symmetry that permutes anyon types.\cite{KitaevKong,BombinTwist,BarkeshliQi,WenGenon,BarkeshliAbelian1,BarkeshliAbelian2,BarkeshliBraiding}    This has generated considerable interest in such anyon permuting symmetries (APS).\cite{TeoTwist,LongQPaper,FradkinTwist,TarantinoTwist,ChengAPS,Heinrich16} Here, we describe how such symmetries are related to TSB.

One example of an APS occurs in the Toric code:\cite{KitaevKong,Beigi11}   from the fusion and braiding rules in Eq. (\ref{ToricData}), it is evident that exchanging the two bosons $e$ and $m$ leaves the fusion and braiding rules that define this topological order invariant.
(This symmetry at the level of fusion and braiding rules is not necessarily a symmetry of Hamiltonians that realize this phase.\cite{KitaevToric,WenToric,ChengAPS,Heinrich16})
To understand the origin of this symmetry, recall that we may obtain the Toric code from two copies of the Ising topological order 
(with opposite chirality) by
condensing the  boson 
$\psi_1 \psi_2$, as
described in Sec. \ref{CondSecEx}.  After condensation, the non-abelian anyon $\sigma_1 \sigma_2$ splits into two abelian anyons $ \tilde{\sigma}_e$ and $\tilde{\sigma}_m$, each with spin $1$ and identical braiding, which we identify with the $e$ and $m$ anyons of the Toric code.
Thus the anyon-permuting symmetry can be viewed as a consequence of the splitting of $\sigma_1 \sigma_2$ into $ \tilde{\sigma}_e$ and $ \tilde{\sigma}_m$.

One reason this symmetry is interesting is that, even if it is not a symmetry of the Hamiltonian realizing the Toric code phase, this phase always admits {\it defect lines}  with the property that an $e$ particle crossing the line becomes an $m$ particle, and vice versa.  The end-points of such a defect line, and their analogues for other APS,  host non-abelian bound states similar to those encountered in Sec. \ref{EdgeBS}.\cite{BombinTwist,KitaevKong,BarkeshliBraiding}  
These defect lines are also easy to understand in the condensation picture: recall that after condensation the four anyon types 
$\sigma_1, \sigma_2,\sigma_1 \psi_2, \psi_1 \sigma_2$ become confined.  Any attempt to create particle-antiparticle pairs of these will create a  defect line (i.e. the energy cost will be proportional to the separation between the particle and anti-particle).  
Before condensation, we showed (see Fig. \ref{SigBraidFig}) that braiding $\sigma_1 \sigma_2$ around a $\sigma_1$ particle (i.e. around one end of the defect line) interchanges its two internal states.  As can be shown using explicit Hamiltonians realizing this transition\cite{schulzIsi}, this aspect of the statistical interaction remains even in the condensed phase,\cite{LongQPaper} where these two internal states correspond to the two bosons $e \equiv \tilde{\sigma}_e, m \equiv \tilde{\sigma}_m$.  
Thus $e$ and $m$ are interchanged when crossing a line defect.

The relationship between APS -- or indeed any global symmetry acting on a system of anyons -- and condensation transitions is quite general.\cite{GuWan14,HungWan14,GarciaCond} 
A topological order with non-trivial APS-- which host multiple gapped boundaries {\it with itself}\cite{DijkgraafVerlinde,KitaevKong,Fuchs11,Fuchs13} -- will admit defect lines across which the anyons are permuted by the symmetry.  These lines terminate on  bound states which can be viewed as confined non-abelian anyons.\cite{BarkeshliQi,BarkeshliAbelian1,TeoTwist,LongQPaper,FradkinTwist} 
This suggests that there is a parent topological order in which these bound states become deconfined anyons, connected to our symmetric topological order by a TSB transition.
In this parent topological order, however, 
 excitations (such as $e$ and $m$ in our example) that are permuted when crossing a defect line must become a single anyon type, since this is the only way for the topological charge to be conserved under braiding.  
Thus the TSB transition in question always splits a single anyon type in this parent phase into all of the anyon types related by symmetry.
 It turns out that any topological order in which a subset of the anyons transform under a discrete symmetry group can be obtained by TSB from such a parent topological order, and vice versa.\cite{DrinfeldGauging,LongQPaper}
This connection between APS and TSB gives an alternate route to identifying examples where such symmetry exists, as well as to constructing symmetric Hamiltonians.\cite{Heinrich16,ChengAPS}


\section{TSB and criticality}

For continuous phase transitions in which a global symmetry is broken, there is a close relationship between the nature of the broken symmetry and the long-distance physics at the critical point. 
One might therefore hope that a similar result holds for phase transitions driven by TSB.  Though a general framework comparable to the Landau-Ginzburg-Wilson theory for symmetry-breaking transitions is not currently known, in certain cases significant progress has been made towards characterizing these critical points.  

To discuss critical points, we require some understanding of the underlying Hamiltonian that describes the transition itself.  
The more experimentally relevant Hamiltonian realizations of TSB involve bilayer quantum Hall systems,\cite{WenBarkeshli,WenBarkeshliLong,WB2,HormoziSlingerland,ClarkeNayak} since a number of candidate topological orders to describe fractional quantum Hall states contain bosons that can potentially condense.  Notable examples include the $\mz_4$ Read-Rezayi state,\cite{ReadRezayi,WenBarkeshli} and a bilayer of the Moore-Read state.\cite{TSBShort,ClarkeNayak}  
However, to study the critical point in a systematic way it is advantageous to use 
 more tractable (but less physically realistic) lattice models,\cite{FradkinShenker,Gilsetal,BaisMonteCarlo,TSBShort,TSBLong,KitaevHoney,LongQPaper,HughesKimChiral,MotrunichSET,MotrunichU1,Isakov11,ArdonneConformalQCP,BhattFQHETrans,ChengParafermionCrit,DSemToTC} which will be our focus here.
Tensor-network constructions\cite{GarciaCond,JiangRanTensors} provide an alternative, wave-function based, approach to TSB on the lattice.  

A powerful approach to studying 
transitions in which the condensing bosons are not local is to use a duality mapping.  One advantage of lattice models is that often this duality can be shown to be exact at the level of the lattice Hamiltonian, even when the appropriate field theory (and hence its dual) is not readily apparent.  This approach was originally used by Wegner\cite{Wegner71}, and later by Fradkin and Shenker\cite{FradkinShenker}, who showed that the two transitions which, in the notation of this paper, amount to condensing the $e$ or the $m$ boson in the Toric code, are described by the same 3D Ising critical point.  
Numerous subsequent studies have confirmed and expanded upon this result.\cite{IsingDual,castelnovo10,VidalToric,TrebstTC,KitaevPhase}
However, it can be generalized\cite{TSBShort,TSBLong,FengKitaevJW,schulzIsi} to a wide class of transitions in which the condensing boson is abelian (and non-chiral), even when the underlying topological order is non-abelian.  One example is the condensation of $\psi_1 \psi_2$ in the example of Sec. \ref{CondSecEx}, for which the low-energy degrees of freedom (but {\it not} the non-abelian anyons, which remain gapped at the transition) are dual to critical Ising spins.  
More generally, an abelian boson $\gamma$ for which $\gamma^p =1$ is dual to a $p$-state Potts spin; in this case the Hamiltonian can be chosen to drive transitions of the Potts or Clock type.\cite{TSBLong} 
Such duality mappings can also be used to characterize dynamical behaviour of time-dependent Hamiltonians near the phase transition in terms of the dynamics of their (much better understood) spin model counterparts.\cite{ChandranBurnell}

More generally, one may consider transitions where the condensing bosons have the fusion and braiding rules of a set of charges of a discrete gauge theory.  This would be the case, for example, in any situation where TSB results in an anyon model with a discrete global symmetry, {\it even if the underlying topological order is not, itself, a discrete gauge theory}.\cite{LongQPaper}  
In this case, one expects that the critical point will be identical to that of the relevant Higgs transition of the discrete gauge theory.
Models realizing such condensation transitions have been studied by several groups,\cite{VidalToric2,Bombin,BaisMonteCarlo}  and are in principle amenable to Monte Carlo techniques.

If we are in search of exotic second-order critical points, however, this leaves only transitions in which the condensing boson is non-abelian (and in fact, not equivalent to a charge of any discrete gauge theory).  In this case, although the condensing anyons are bosons in the sense described in Sec. \ref{Sec2}, they generally have long-ranged statistical interactions which rule out a dual spin description of the type described above. A simple example of this type is to condense the $\sigma_1 \sigma_2$ boson in our bilayer Ising example.  $\sigma_1 \sigma_2$ is a boson, in the sense that no phase is accrued upon exchanging a pair of $\sigma_1 \sigma_2$ particles fused to the vacuum.  However, $\sigma_1 \sigma_2$ has long-ranged statistical interactions with all other particles except $\psi_1 \psi_2$ -- including with itself, since braiding a $\sigma_1 \sigma_2$ particle around another can alter its fusion channel with the remaining $\sigma_1 \sigma_2$ particles.

These transitions are only poorly understood.
They can be described analytically in a 1D variant, where the condensing anyons may fuse but not braid; in this case they can be mapped onto so-called restricted solid on solid models for which the critical behaviour is known.\cite{Gilsetal,GilsIsing}  In 2D several examples have been studied by a combination of numerics and high-order perturbative expansions.\cite{SchulzFib,SchulzSU2,VidalBS}  The current accuracy of these combined methods is not sufficient to determine the order of the transition in these cases, and there is evidence that they may generically be first order.\cite{VidalMF,VidalBS}   However if they are second order, series expansion results suggest\cite{SchulzFib} that they may comprise novel universality classes, and as such are worthy of further study.

In addition to transitions which only affect the topological order, one can consider transitions in which a change in the topological order coincides with another type of change in the system, such as breaking an underlying symmetry.  A number of such  scenarios have been studied in the context of $\mz_2$ spin liquids, including transitions to magnetically ordered\cite{Jalabert91,Vojta00,ChubukovSenthilSachdev} and gapless\cite{ReadSachdev,SachdevRead,SenthilFisher} phases.  Transitions in which reduction of a non-abelian topological order coincides with spontaneous breaking of lattice symmetry are also known to exist.\cite{schulzIsi}  In general the set of possible critical points in this situation is richer than that described above, and little is known about the possibilities.

\section{Summary and outlook}

Anyon condensation is a cornerstone of our understanding of topological orders and the relationship between them.  
Topological symmetry breaking (TSB)-- i.e. condensing bosons in a topologically ordered system -- has had a particularly long reach in this respect, due in large part to the fact that it is relatively well understood, having been extensively studied in the context both of conformal field theory and of mathematics.  Our focus here has been to provide a non-technical review of the essence of this approach, and some of its most timely applications in the study of 2D topological phases.  

We have seen that a deep correspondence exists between TSB and gappable boundaries between differing topological orders, which follows from the relation of TSB to (``off-diagonal") modular invariants in CFT.   Of particular interest is the systematic classification\cite{KitaevKong,KongCond} of systems admitting multiple gapped boundary types, which admit interesting bound states at their interfaces.

We have also reviewed how global symmetries in anyon systems arise from TSB in of the topological order in which the symmetry has been gauged; the anyons that become confined as this transition is crossed become defect lines hosting non-abelian bound states.  Among other things, this perspective can be useful in constructing examples of anyonic symmetry, since a large number of examples of such transitions (including those in which an anyon type splits) exist in the literature.\cite{ConformalZoo,CapelliADE}

Though in some cases the critical points can be well-characterised, TSB provides many interesting examples of phase transitions that are qualitatively different from both ordinary symmetry-breaking transitions and Higgs transitions, even of the non-abelian type.  Studying lattice Hamiltonians with TSB  transitions has enabled a limited analysis of these poorly understood critical points.  
However, a general framework to reveal the relationship between phases separated by these more exotic transitions and the nature of the critical points remains an interesting open problem.

\section*{ACKNOWLEDGMENTS}
The author is grateful to Shivaji Sondhi, Michael Levin, Steve Simon, Anushya Chandran and Chris Laumann for helpful comments on this manuscript, as well as to NSF-DMR $1352271$ and the Sloan Foundation FG-$2015$-$65927$ for their financial support.

\bibliography{Reviewbib}

\begin{thebibliography}{149}%
\makeatletter
\providecommand \@ifxundefined [1]{%
 \@ifx{#1\undefined}
}%
\providecommand \@ifnum [1]{%
 \ifnum #1\expandafter \@firstoftwo
 \else \expandafter \@secondoftwo
 \fi
}%
\providecommand \@ifx [1]{%
 \ifx #1\expandafter \@firstoftwo
 \else \expandafter \@secondoftwo
 \fi
}%
\providecommand \natexlab [1]{#1}%
\providecommand \enquote  [1]{``#1''}%
\providecommand \bibnamefont  [1]{#1}%
\providecommand \bibfnamefont [1]{#1}%
\providecommand \citenamefont [1]{#1}%
\providecommand \href@noop [0]{\@secondoftwo}%
\providecommand \href [0]{\begingroup \@sanitize@url \@href}%
\providecommand \@href[1]{\@@startlink{#1}\@@href}%
\providecommand \@@href[1]{\endgroup#1\@@endlink}%
\providecommand \@sanitize@url [0]{\catcode `\\12\catcode `\$12\catcode
  `\&12\catcode `\#12\catcode `\^12\catcode `\_12\catcode `\%12\relax}%
\providecommand \@@startlink[1]{}%
\providecommand \@@endlink[0]{}%
\providecommand \url  [0]{\begingroup\@sanitize@url \@url }%
\providecommand \@url [1]{\endgroup\@href {#1}{\urlprefix }}%
\providecommand \urlprefix  [0]{URL }%
\providecommand \Eprint [0]{\href }%
\providecommand \doibase [0]{http://dx.doi.org/}%
\providecommand \selectlanguage [0]{\@gobble}%
\providecommand \bibinfo  [0]{\@secondoftwo}%
\providecommand \bibfield  [0]{\@secondoftwo}%
\providecommand \translation [1]{[#1]}%
\providecommand \BibitemOpen [0]{}%
\providecommand \bibitemStop [0]{}%
\providecommand \bibitemNoStop [0]{.\EOS\space}%
\providecommand \EOS [0]{\spacefactor3000\relax}%
\providecommand \BibitemShut  [1]{\csname bibitem#1\endcsname}%
\let\auto@bib@innerbib\@empty
\bibitem [{\citenamefont {Wen}(1989)}]{WenGSDeg}%
  \BibitemOpen
  \bibfield  {author} {\bibinfo {author} {\bibfnamefont {X.~G.}\ \bibnamefont
  {Wen}},\ }\href {\doibase 10.1103/PhysRevB.40.7387} {\bibfield  {journal}
  {\bibinfo  {journal} {Phys. Rev. B}\ }\textbf {\bibinfo {volume} {40}},\
  \bibinfo {pages} {7387} (\bibinfo {year} {1989})}\BibitemShut {NoStop}%
\bibitem [{\citenamefont {Wen}(1990)}]{WenTopOrd}%
  \BibitemOpen
  \bibfield  {author} {\bibinfo {author} {\bibfnamefont {X.~G.}\ \bibnamefont
  {Wen}},\ }\href {\doibase 10.1142/S0217979290000139} {\bibfield  {journal}
  {\bibinfo  {journal} {International Journal of Modern Physics B}\ }\textbf
  {\bibinfo {volume} {04}},\ \bibinfo {pages} {239} (\bibinfo {year}
  {1990})}\BibitemShut {NoStop}%
\bibitem [{\citenamefont {Wilczek}(1982)}]{WilczekAnyon}%
  \BibitemOpen
  \bibfield  {author} {\bibinfo {author} {\bibfnamefont {F.}~\bibnamefont
  {Wilczek}},\ }\href {\doibase 10.1103/PhysRevLett.49.957} {\bibfield
  {journal} {\bibinfo  {journal} {Phys. Rev. Lett.}\ }\textbf {\bibinfo
  {volume} {49}},\ \bibinfo {pages} {957} (\bibinfo {year} {1982})}\BibitemShut
  {NoStop}%
\bibitem [{\citenamefont {Kosterlitz}\ and\ \citenamefont
  {Thouless}(1973)}]{KT}%
  \BibitemOpen
  \bibfield  {author} {\bibinfo {author} {\bibfnamefont {J.~M.}\ \bibnamefont
  {Kosterlitz}}\ and\ \bibinfo {author} {\bibfnamefont {D.~J.}\ \bibnamefont
  {Thouless}},\ }\href {http://stacks.iop.org/0022-3719/6/i=7/a=010} {\bibfield
   {journal} {\bibinfo  {journal} {Journal of Physics C: Solid State Physics}\
  }\textbf {\bibinfo {volume} {6}},\ \bibinfo {pages} {1181} (\bibinfo {year}
  {1973})}\BibitemShut {NoStop}%
\bibitem [{\citenamefont {Kosterlitz}(1974)}]{KosterlitzRG}%
  \BibitemOpen
  \bibfield  {author} {\bibinfo {author} {\bibfnamefont {J.~M.}\ \bibnamefont
  {Kosterlitz}},\ }\href {http://stacks.iop.org/0022-3719/7/i=6/a=005}
  {\bibfield  {journal} {\bibinfo  {journal} {Journal of Physics C: Solid State
  Physics}\ }\textbf {\bibinfo {volume} {7}},\ \bibinfo {pages} {1046}
  (\bibinfo {year} {1974})}\BibitemShut {NoStop}%
\bibitem [{\citenamefont {Aharonov}\ and\ \citenamefont
  {Casher}(1984)}]{AharonovCasher}%
  \BibitemOpen
  \bibfield  {author} {\bibinfo {author} {\bibfnamefont {Y.}~\bibnamefont
  {Aharonov}}\ and\ \bibinfo {author} {\bibfnamefont {A.}~\bibnamefont
  {Casher}},\ }\href {\doibase 10.1103/PhysRevLett.53.319} {\bibfield
  {journal} {\bibinfo  {journal} {Phys. Rev. Lett.}\ }\textbf {\bibinfo
  {volume} {53}},\ \bibinfo {pages} {319} (\bibinfo {year} {1984})}\BibitemShut
  {NoStop}%
\bibitem [{\citenamefont {Hansson}\ \emph {et~al.}(2004)\citenamefont
  {Hansson}, \citenamefont {Oganesyan},\ and\ \citenamefont
  {Sondhi}}]{SondhiSC}%
  \BibitemOpen
  \bibfield  {author} {\bibinfo {author} {\bibfnamefont {T.~H.}\ \bibnamefont
  {Hansson}}, \bibinfo {author} {\bibfnamefont {V.}~\bibnamefont {Oganesyan}},
  \ and\ \bibinfo {author} {\bibfnamefont {S.~L.}\ \bibnamefont {Sondhi}},\
  }\href@noop {} {\bibfield  {journal} {\bibinfo  {journal} {Annals of
  Physics}\ }\textbf {\bibinfo {volume} {313}},\ \bibinfo {pages} {497 }
  (\bibinfo {year} {2004})}\BibitemShut {NoStop}%
\bibitem [{\citenamefont {Dasgupta}\ and\ \citenamefont
  {Halperin}(1981)}]{DasguptaHalperin}%
  \BibitemOpen
  \bibfield  {author} {\bibinfo {author} {\bibfnamefont {C.}~\bibnamefont
  {Dasgupta}}\ and\ \bibinfo {author} {\bibfnamefont {B.~I.}\ \bibnamefont
  {Halperin}},\ }\href {\doibase 10.1103/PhysRevLett.47.1556} {\bibfield
  {journal} {\bibinfo  {journal} {Phys. Rev. Lett.}\ }\textbf {\bibinfo
  {volume} {47}},\ \bibinfo {pages} {1556} (\bibinfo {year}
  {1981})}\BibitemShut {NoStop}%
\bibitem [{\citenamefont {Lee}\ and\ \citenamefont {Fisher}(1989)}]{LeeFisher}%
  \BibitemOpen
  \bibfield  {author} {\bibinfo {author} {\bibfnamefont {D.-H.}\ \bibnamefont
  {Lee}}\ and\ \bibinfo {author} {\bibfnamefont {M.}~\bibnamefont {Fisher}},\
  }\href@noop {} {\bibfield  {journal} {\bibinfo  {journal} {Physical Review
  Letters}\ }\textbf {\bibinfo {volume} {63}},\ \bibinfo {pages} {903}
  (\bibinfo {year} {1989})}\BibitemShut {NoStop}%
\bibitem [{\citenamefont {Lee}\ and\ \citenamefont {Fisher}(1991)}]{FisherLee}%
  \BibitemOpen
  \bibfield  {author} {\bibinfo {author} {\bibfnamefont {D.-H.}\ \bibnamefont
  {Lee}}\ and\ \bibinfo {author} {\bibfnamefont {M.}~\bibnamefont {Fisher}},\
  }\href@noop {} {\bibfield  {journal} {\bibinfo  {journal} {International
  Journal of Modern Physics B}\ }\textbf {\bibinfo {volume} {5}},\ \bibinfo
  {pages} {2675} (\bibinfo {year} {1991})}\BibitemShut {NoStop}%
\bibitem [{\citenamefont {Wen}\ and\ \citenamefont {Zee}(1992)}]{WenKMatrix}%
  \BibitemOpen
  \bibfield  {author} {\bibinfo {author} {\bibfnamefont {X.~G.}\ \bibnamefont
  {Wen}}\ and\ \bibinfo {author} {\bibfnamefont {A.}~\bibnamefont {Zee}},\
  }\href {\doibase 10.1103/PhysRevB.46.2290} {\bibfield  {journal} {\bibinfo
  {journal} {Phys. Rev. B}\ }\textbf {\bibinfo {volume} {46}},\ \bibinfo
  {pages} {2290} (\bibinfo {year} {1992})}\BibitemShut {NoStop}%
\bibitem [{\citenamefont {Barkeshli}\ \emph {et~al.}(2015)\citenamefont
  {Barkeshli}, \citenamefont {Yao},\ and\ \citenamefont
  {Laumann}}]{BarkeshliYaoLaumann}%
  \BibitemOpen
  \bibfield  {author} {\bibinfo {author} {\bibfnamefont {M.}~\bibnamefont
  {Barkeshli}}, \bibinfo {author} {\bibfnamefont {N.~Y.}\ \bibnamefont {Yao}},
  \ and\ \bibinfo {author} {\bibfnamefont {C.~R.}\ \bibnamefont {Laumann}},\
  }\href {\doibase 10.1103/PhysRevLett.115.026802} {\bibfield  {journal}
  {\bibinfo  {journal} {Phys. Rev. Lett.}\ }\textbf {\bibinfo {volume} {115}},\
  \bibinfo {pages} {026802} (\bibinfo {year} {2015})}\BibitemShut {NoStop}%
\bibitem [{\citenamefont {Haldane}(1983)}]{HaldaneHierarchy}%
  \BibitemOpen
  \bibfield  {author} {\bibinfo {author} {\bibfnamefont {F.~D.~M.}\
  \bibnamefont {Haldane}},\ }\href {\doibase 10.1103/PhysRevLett.51.605}
  {\bibfield  {journal} {\bibinfo  {journal} {Phys. Rev. Lett.}\ }\textbf
  {\bibinfo {volume} {51}},\ \bibinfo {pages} {605} (\bibinfo {year}
  {1983})}\BibitemShut {NoStop}%
\bibitem [{\citenamefont {Halperin}(1984)}]{HalperinHierarchy}%
  \BibitemOpen
  \bibfield  {author} {\bibinfo {author} {\bibfnamefont {B.~I.}\ \bibnamefont
  {Halperin}},\ }\href {\doibase 10.1103/PhysRevLett.52.1583} {\bibfield
  {journal} {\bibinfo  {journal} {Phys. Rev. Lett.}\ }\textbf {\bibinfo
  {volume} {52}},\ \bibinfo {pages} {1583} (\bibinfo {year}
  {1984})}\BibitemShut {NoStop}%
\bibitem [{\citenamefont {Read}(1990)}]{ReadHierarchy}%
  \BibitemOpen
  \bibfield  {author} {\bibinfo {author} {\bibfnamefont {N.}~\bibnamefont
  {Read}},\ }\href {\doibase 10.1103/PhysRevLett.65.1502} {\bibfield  {journal}
  {\bibinfo  {journal} {Phys. Rev. Lett.}\ }\textbf {\bibinfo {volume} {65}},\
  \bibinfo {pages} {1502} (\bibinfo {year} {1990})}\BibitemShut {NoStop}%
\bibitem [{\citenamefont {Frohlich}\ and\ \citenamefont
  {Zee}(1991)}]{FrohlichHierarcy}%
  \BibitemOpen
  \bibfield  {author} {\bibinfo {author} {\bibfnamefont {J.}~\bibnamefont
  {Frohlich}}\ and\ \bibinfo {author} {\bibfnamefont {A.}~\bibnamefont {Zee}},\
  }\href {\doibase http://dx.doi.org/10.1016/0550-3213(91)90275-3} {\bibfield
  {journal} {\bibinfo  {journal} {Nuclear Physics B}\ }\textbf {\bibinfo
  {volume} {364}},\ \bibinfo {pages} {517 } (\bibinfo {year}
  {1991})}\BibitemShut {NoStop}%
\bibitem [{\citenamefont {Wang}\ and\ \citenamefont
  {Senthil}(2015)}]{WangSenthil}%
  \BibitemOpen
  \bibfield  {author} {\bibinfo {author} {\bibfnamefont {C.}~\bibnamefont
  {Wang}}\ and\ \bibinfo {author} {\bibfnamefont {T.}~\bibnamefont {Senthil}},\
  }\href {\doibase 10.1103/PhysRevX.5.041031} {\bibfield  {journal} {\bibinfo
  {journal} {Phys. Rev. X}\ }\textbf {\bibinfo {volume} {5}},\ \bibinfo {pages}
  {041031} (\bibinfo {year} {2015})}\BibitemShut {NoStop}%
\bibitem [{\citenamefont {Metlitski}\ \emph {et~al.}(2015)\citenamefont
  {Metlitski}, \citenamefont {Kane},\ and\ \citenamefont
  {Fisher}}]{MetlitskiKaneFisher}%
  \BibitemOpen
  \bibfield  {author} {\bibinfo {author} {\bibfnamefont {M.~A.}\ \bibnamefont
  {Metlitski}}, \bibinfo {author} {\bibfnamefont {C.~L.}\ \bibnamefont {Kane}},
  \ and\ \bibinfo {author} {\bibfnamefont {M.~P.~A.}\ \bibnamefont {Fisher}},\
  }\href {\doibase 10.1103/PhysRevB.92.125111} {\bibfield  {journal} {\bibinfo
  {journal} {Phys. Rev. B}\ }\textbf {\bibinfo {volume} {92}},\ \bibinfo
  {pages} {125111} (\bibinfo {year} {2015})}\BibitemShut {NoStop}%
\bibitem [{\citenamefont {Vishwanath}\ and\ \citenamefont
  {Senthil}(2013)}]{SenthilVishwanath}%
  \BibitemOpen
  \bibfield  {author} {\bibinfo {author} {\bibfnamefont {A.}~\bibnamefont
  {Vishwanath}}\ and\ \bibinfo {author} {\bibfnamefont {T.}~\bibnamefont
  {Senthil}},\ }\href {\doibase 10.1103/PhysRevX.3.011016} {\bibfield
  {journal} {\bibinfo  {journal} {Phys. Rev. X}\ }\textbf {\bibinfo {volume}
  {3}},\ \bibinfo {pages} {011016} (\bibinfo {year} {2013})}\BibitemShut
  {NoStop}%
\bibitem [{\citenamefont {Moore}\ and\ \citenamefont
  {Seiberg}(1989{\natexlab{a}})}]{ConformalZoo}%
  \BibitemOpen
  \bibfield  {author} {\bibinfo {author} {\bibfnamefont {G.}~\bibnamefont
  {Moore}}\ and\ \bibinfo {author} {\bibfnamefont {N.}~\bibnamefont
  {Seiberg}},\ }\href {\doibase http://dx.doi.org/10.1016/0370-2693(89)90897-6}
  {\bibfield  {journal} {\bibinfo  {journal} {Physics Letters B}\ }\textbf
  {\bibinfo {volume} {220}},\ \bibinfo {pages} {422 } (\bibinfo {year}
  {1989}{\natexlab{a}})}\BibitemShut {NoStop}%
\bibitem [{\citenamefont {Moore}\ and\ \citenamefont
  {Seiberg}(1989{\natexlab{b}})}]{MooreSeibergNatural}%
  \BibitemOpen
  \bibfield  {author} {\bibinfo {author} {\bibfnamefont {G.}~\bibnamefont
  {Moore}}\ and\ \bibinfo {author} {\bibfnamefont {N.}~\bibnamefont
  {Seiberg}},\ }\href {\doibase http://dx.doi.org/10.1016/0550-3213(89)90511-7}
  {\bibfield  {journal} {\bibinfo  {journal} {Nuclear Physics B}\ }\textbf
  {\bibinfo {volume} {313}},\ \bibinfo {pages} {16 } (\bibinfo {year}
  {1989}{\natexlab{b}})}\BibitemShut {NoStop}%
\bibitem [{\citenamefont {Gepner}(1989)}]{Gepner89}%
  \BibitemOpen
  \bibfield  {author} {\bibinfo {author} {\bibfnamefont {D.}~\bibnamefont
  {Gepner}},\ }\href {\doibase http://dx.doi.org/10.1016/0370-2693(89)91253-7}
  {\bibfield  {journal} {\bibinfo  {journal} {Physics Letters B}\ }\textbf
  {\bibinfo {volume} {222}},\ \bibinfo {pages} {207 } (\bibinfo {year}
  {1989})}\BibitemShut {NoStop}%
\bibitem [{\citenamefont {Schellekens}\ and\ \citenamefont
  {Yankielowicz}(1989)}]{Schellekens89}%
  \BibitemOpen
  \bibfield  {author} {\bibinfo {author} {\bibfnamefont {A.~N.}\ \bibnamefont
  {Schellekens}}\ and\ \bibinfo {author} {\bibfnamefont {S.}~\bibnamefont
  {Yankielowicz}},\ }\href {\doibase DOI: 10.1016/0550-3213(89)90310-6}
  {\bibfield  {journal} {\bibinfo  {journal} {Nuclear Physics B}\ }\textbf
  {\bibinfo {volume} {327}},\ \bibinfo {pages} {673 } (\bibinfo {year}
  {1989})}\BibitemShut {NoStop}%
\bibitem [{\citenamefont {Goodard}\ \emph {et~al.}(1985)\citenamefont
  {Goodard}, \citenamefont {Kent},\ and\ \citenamefont
  {Olive}}]{GoddardCoset1}%
  \BibitemOpen
  \bibfield  {author} {\bibinfo {author} {\bibfnamefont {P.}~\bibnamefont
  {Goodard}}, \bibinfo {author} {\bibfnamefont {A.}~\bibnamefont {Kent}}, \
  and\ \bibinfo {author} {\bibfnamefont {D.}~\bibnamefont {Olive}},\ }\href
  {\doibase http://dx.doi.org/10.1016/0370-2693(85)91145-1} {\bibfield
  {journal} {\bibinfo  {journal} {Physics Letters B}\ }\textbf {\bibinfo
  {volume} {152}},\ \bibinfo {pages} {88 } (\bibinfo {year}
  {1985})}\BibitemShut {NoStop}%
\bibitem [{\citenamefont {Goddard}\ \emph {et~al.}(1986)\citenamefont
  {Goddard}, \citenamefont {Kent},\ and\ \citenamefont
  {Olive}}]{GoddardCoset2}%
  \BibitemOpen
  \bibfield  {author} {\bibinfo {author} {\bibfnamefont {P.}~\bibnamefont
  {Goddard}}, \bibinfo {author} {\bibfnamefont {A.}~\bibnamefont {Kent}}, \
  and\ \bibinfo {author} {\bibfnamefont {D.}~\bibnamefont {Olive}},\ }\href
  {\doibase 10.1007/BF01464283} {\bibfield  {journal} {\bibinfo  {journal}
  {Communications in Mathematical Physics}\ }\textbf {\bibinfo {volume}
  {103}},\ \bibinfo {pages} {105} (\bibinfo {year} {1986})}\BibitemShut
  {NoStop}%
\bibitem [{\citenamefont {Bais}\ \emph {et~al.}(2002)\citenamefont {Bais},
  \citenamefont {Schroers},\ and\ \citenamefont {Slingerland}}]{TSBPRL}%
  \BibitemOpen
  \bibfield  {author} {\bibinfo {author} {\bibfnamefont {F.~A.}\ \bibnamefont
  {Bais}}, \bibinfo {author} {\bibfnamefont {B.~J.}\ \bibnamefont {Schroers}},
  \ and\ \bibinfo {author} {\bibfnamefont {J.~K.}\ \bibnamefont
  {Slingerland}},\ }\href@noop {} {\bibfield  {journal} {\bibinfo  {journal}
  {Phys. Rev. Lett.}\ }\textbf {\bibinfo {volume} {89}},\ \bibinfo {pages}
  {181601} (\bibinfo {year} {2002})}\BibitemShut {NoStop}%
\bibitem [{\citenamefont {Bais}\ and\ \citenamefont
  {Slingerland}(2009)}]{SlingerlandBais}%
  \BibitemOpen
  \bibfield  {author} {\bibinfo {author} {\bibfnamefont {F.~A.}\ \bibnamefont
  {Bais}}\ and\ \bibinfo {author} {\bibfnamefont {J.~K.}\ \bibnamefont
  {Slingerland}},\ }\href {\doibase 10.1103/PhysRevB.79.045316} {\bibfield
  {journal} {\bibinfo  {journal} {Phys. Rev. B}\ }\textbf {\bibinfo {volume}
  {79}},\ \bibinfo {pages} {045316} (\bibinfo {year} {2009})}\BibitemShut
  {NoStop}%
\bibitem [{\citenamefont {Bais}\ and\ \citenamefont {Mathy}(2007)}]{BaisMathy}%
  \BibitemOpen
  \bibfield  {author} {\bibinfo {author} {\bibfnamefont {F.}~\bibnamefont
  {Bais}}\ and\ \bibinfo {author} {\bibfnamefont {C.}~\bibnamefont {Mathy}},\
  }\href {\doibase http://dx.doi.org/10.1016/j.aop.2006.05.010} {\bibfield
  {journal} {\bibinfo  {journal} {Annals of Physics}\ }\textbf {\bibinfo
  {volume} {322}},\ \bibinfo {pages} {552 } (\bibinfo {year}
  {2007})}\BibitemShut {NoStop}%
\bibitem [{\citenamefont {Eli\"ens}\ \emph {et~al.}(2014)\citenamefont
  {Eli\"ens}, \citenamefont {Romers},\ and\ \citenamefont {Bais}}]{Eliens}%
  \BibitemOpen
  \bibfield  {author} {\bibinfo {author} {\bibfnamefont {I.~S.}\ \bibnamefont
  {Eli\"ens}}, \bibinfo {author} {\bibfnamefont {J.~C.}\ \bibnamefont
  {Romers}}, \ and\ \bibinfo {author} {\bibfnamefont {F.~A.}\ \bibnamefont
  {Bais}},\ }\href {\doibase 10.1103/PhysRevB.90.195130} {\bibfield  {journal}
  {\bibinfo  {journal} {Phys. Rev. B}\ }\textbf {\bibinfo {volume} {90}},\
  \bibinfo {pages} {195130} (\bibinfo {year} {2014})}\BibitemShut {NoStop}%
\bibitem [{\citenamefont {Neupert}\ \emph
  {et~al.}(2016{\natexlab{a}})\citenamefont {Neupert}, \citenamefont {He},
  \citenamefont {von Keyserlingk}, \citenamefont {Sierra},\ and\ \citenamefont
  {Bernevig}}]{NeupertBernevig}%
  \BibitemOpen
  \bibfield  {author} {\bibinfo {author} {\bibfnamefont {T.}~\bibnamefont
  {Neupert}}, \bibinfo {author} {\bibfnamefont {H.}~\bibnamefont {He}},
  \bibinfo {author} {\bibfnamefont {C.}~\bibnamefont {von Keyserlingk}},
  \bibinfo {author} {\bibfnamefont {G.}~\bibnamefont {Sierra}}, \ and\ \bibinfo
  {author} {\bibfnamefont {B.~A.}\ \bibnamefont {Bernevig}},\ }\href {\doibase
  10.1103/PhysRevB.93.115103} {\bibfield  {journal} {\bibinfo  {journal} {Phys.
  Rev. B}\ }\textbf {\bibinfo {volume} {93}},\ \bibinfo {pages} {115103}
  (\bibinfo {year} {2016}{\natexlab{a}})}\BibitemShut {NoStop}%
\bibitem [{\citenamefont {Muger}(2000)}]{MugerTSB}%
  \BibitemOpen
  \bibfield  {author} {\bibinfo {author} {\bibfnamefont {M.}~\bibnamefont
  {Muger}},\ }\href {\doibase http://dx.doi.org/10.1006/aima.1999.1860}
  {\bibfield  {journal} {\bibinfo  {journal} {Advances in Mathematics}\
  }\textbf {\bibinfo {volume} {150}},\ \bibinfo {pages} {151 } (\bibinfo {year}
  {2000})}\BibitemShut {NoStop}%
\bibitem [{\citenamefont {Brugui{\`e}res}(2000)}]{BruguieresTSB}%
  \BibitemOpen
  \bibfield  {author} {\bibinfo {author} {\bibfnamefont {A.}~\bibnamefont
  {Brugui{\`e}res}},\ }\href {\doibase 10.1007/s002080050011} {\bibfield
  {journal} {\bibinfo  {journal} {Mathematische Annalen}\ }\textbf {\bibinfo
  {volume} {316}},\ \bibinfo {pages} {215} (\bibinfo {year}
  {2000})}\BibitemShut {NoStop}%
\bibitem [{\citenamefont {MŸger}(2003)}]{Muger}%
  \BibitemOpen
  \bibfield  {author} {\bibinfo {author} {\bibfnamefont {M.}~\bibnamefont
  {MŸger}},\ }\href@noop {} {\bibfield  {journal} {\bibinfo  {journal} {J. Pure
  Appl. Algebra}\ }\textbf {\bibinfo {volume} {180}},\ \bibinfo {pages}
  {159Ð219} (\bibinfo {year} {2003})}\BibitemShut {NoStop}%
\bibitem [{\citenamefont {Kirillov}\ and\ \citenamefont
  {Ostrik}(2002)}]{KIRILLOV2002183}%
  \BibitemOpen
  \bibfield  {author} {\bibinfo {author} {\bibfnamefont {A.}~\bibnamefont
  {Kirillov}}\ and\ \bibinfo {author} {\bibfnamefont {V.}~\bibnamefont
  {Ostrik}},\ }\href {\doibase http://dx.doi.org/10.1006/aima.2002.2072}
  {\bibfield  {journal} {\bibinfo  {journal} {Advances in Mathematics}\
  }\textbf {\bibinfo {volume} {171}},\ \bibinfo {pages} {183 } (\bibinfo {year}
  {2002})}\BibitemShut {NoStop}%
\bibitem [{\citenamefont {Davydov}\ \emph {et~al.}(2013)\citenamefont
  {Davydov}, \citenamefont {Muger}, \citenamefont {Nikshych},\ and\
  \citenamefont {Ostrik}}]{Davydov13}%
  \BibitemOpen
  \bibfield  {author} {\bibinfo {author} {\bibfnamefont {A.}~\bibnamefont
  {Davydov}}, \bibinfo {author} {\bibfnamefont {M.}~\bibnamefont {Muger}},
  \bibinfo {author} {\bibfnamefont {D.}~\bibnamefont {Nikshych}}, \ and\
  \bibinfo {author} {\bibfnamefont {V.}~\bibnamefont {Ostrik}},\ }\href@noop {}
  {\bibfield  {journal} {\bibinfo  {journal} {J. Reine Angew. Math.}\ }\textbf
  {\bibinfo {volume} {677}},\ \bibinfo {pages} {135?177} (\bibinfo {year}
  {2013})}\BibitemShut {NoStop}%
\bibitem [{\citenamefont {Hung}\ and\ \citenamefont
  {Wan}(2015{\natexlab{a}})}]{Hung15}%
  \BibitemOpen
  \bibfield  {author} {\bibinfo {author} {\bibfnamefont {L.-Y.}\ \bibnamefont
  {Hung}}\ and\ \bibinfo {author} {\bibfnamefont {Y.}~\bibnamefont {Wan}},\
  }\href {\doibase 10.1007/JHEP07(2015)120} {\bibfield  {journal} {\bibinfo
  {journal} {Journal of High Energy Physics}\ }\textbf {\bibinfo {volume}
  {2015}},\ \bibinfo {pages} {120} (\bibinfo {year}
  {2015}{\natexlab{a}})}\BibitemShut {NoStop}%
\bibitem [{\citenamefont {Neupert}\ \emph
  {et~al.}(2016{\natexlab{b}})\citenamefont {Neupert}, \citenamefont {He},
  \citenamefont {von Keyserlingk}, \citenamefont {Sierra},\ and\ \citenamefont
  {Bernevig}}]{NeupertNoGo}%
  \BibitemOpen
  \bibfield  {author} {\bibinfo {author} {\bibfnamefont {T.}~\bibnamefont
  {Neupert}}, \bibinfo {author} {\bibfnamefont {H.}~\bibnamefont {He}},
  \bibinfo {author} {\bibfnamefont {C.}~\bibnamefont {von Keyserlingk}},
  \bibinfo {author} {\bibfnamefont {G.}~\bibnamefont {Sierra}}, \ and\ \bibinfo
  {author} {\bibfnamefont {B.~A.}\ \bibnamefont {Bernevig}},\ }\href
  {http://stacks.iop.org/1367-2630/18/i=12/a=123009} {\bibfield  {journal}
  {\bibinfo  {journal} {New Journal of Physics}\ }\textbf {\bibinfo {volume}
  {18}},\ \bibinfo {pages} {123009} (\bibinfo {year}
  {2016}{\natexlab{b}})}\BibitemShut {NoStop}%
\bibitem [{\citenamefont {Bais}\ \emph {et~al.}(2003)\citenamefont {Bais},
  \citenamefont {Schroers},\ and\ \citenamefont
  {Slingerland}}]{BaisGaugeTheories}%
  \BibitemOpen
  \bibfield  {author} {\bibinfo {author} {\bibfnamefont {F.~A.}\ \bibnamefont
  {Bais}}, \bibinfo {author} {\bibfnamefont {B.~J.}\ \bibnamefont {Schroers}},
  \ and\ \bibinfo {author} {\bibfnamefont {J.~K.}\ \bibnamefont
  {Slingerland}},\ }\href@noop {} {\bibfield  {journal} {\bibinfo  {journal}
  {J. H. E. P.}\ }\textbf {\bibinfo {volume} {05}},\ \bibinfo {pages} {068}
  (\bibinfo {year} {2003})}\BibitemShut {NoStop}%
\bibitem [{\citenamefont {Schellekens}\ and\ \citenamefont
  {Yankielowicz}(1990)}]{SCHELLEKENSFixed}%
  \BibitemOpen
  \bibfield  {author} {\bibinfo {author} {\bibfnamefont {A.}~\bibnamefont
  {Schellekens}}\ and\ \bibinfo {author} {\bibfnamefont {S.}~\bibnamefont
  {Yankielowicz}},\ }\href {\doibase
  http://dx.doi.org/10.1016/0550-3213(90)90657-Y} {\bibfield  {journal}
  {\bibinfo  {journal} {Nuclear Physics B}\ }\textbf {\bibinfo {volume}
  {334}},\ \bibinfo {pages} {67 } (\bibinfo {year} {1990})}\BibitemShut
  {NoStop}%
\bibitem [{\citenamefont {Fuchs}\ \emph {et~al.}(1996)\citenamefont {Fuchs},
  \citenamefont {Schellekens},\ and\ \citenamefont {Schweigert}}]{FUCHSFixed}%
  \BibitemOpen
  \bibfield  {author} {\bibinfo {author} {\bibfnamefont {J.}~\bibnamefont
  {Fuchs}}, \bibinfo {author} {\bibfnamefont {B.}~\bibnamefont {Schellekens}},
  \ and\ \bibinfo {author} {\bibfnamefont {C.}~\bibnamefont {Schweigert}},\
  }\href {\doibase http://dx.doi.org/10.1016/0550-3213(95)00623-0} {\bibfield
  {journal} {\bibinfo  {journal} {Nuclear Physics B}\ }\textbf {\bibinfo
  {volume} {461}},\ \bibinfo {pages} {371 } (\bibinfo {year}
  {1996})}\BibitemShut {NoStop}%
\bibitem [{\citenamefont {Fršhlich}\ \emph {et~al.}(2004)\citenamefont
  {Fršhlich}, \citenamefont {Fuchs}, \citenamefont {Runkel},\ and\
  \citenamefont {Schweigert}}]{FuchsCoset}%
  \BibitemOpen
  \bibfield  {author} {\bibinfo {author} {\bibfnamefont {J.}~\bibnamefont
  {Fršhlich}}, \bibinfo {author} {\bibfnamefont {J.}~\bibnamefont {Fuchs}},
  \bibinfo {author} {\bibfnamefont {I.}~\bibnamefont {Runkel}}, \ and\ \bibinfo
  {author} {\bibfnamefont {C.}~\bibnamefont {Schweigert}},\ }\href {\doibase
  10.1002/prop.200310162} {\bibfield  {journal} {\bibinfo  {journal}
  {Fortschritte der Physik}\ }\textbf {\bibinfo {volume} {52}},\ \bibinfo
  {pages} {672} (\bibinfo {year} {2004})}\BibitemShut {NoStop}%
\bibitem [{\citenamefont {Kitaev}(2006)}]{KitaevHoney}%
  \BibitemOpen
  \bibfield  {author} {\bibinfo {author} {\bibfnamefont {A.~Y.}\ \bibnamefont
  {Kitaev}},\ }\href@noop {} {\bibfield  {journal} {\bibinfo  {journal} {Annals
  of Physics}\ }\textbf {\bibinfo {volume} {321}},\ \bibinfo {pages} {2}
  (\bibinfo {year} {2006})},\ \bibinfo {note} {cond-mat/0506438}\BibitemShut
  {NoStop}%
\bibitem [{\citenamefont {Bravyi}\ and\ \citenamefont
  {Kitaev}(1998)}]{BravyiKitaev}%
  \BibitemOpen
  \bibfield  {author} {\bibinfo {author} {\bibfnamefont {S.~B.}\ \bibnamefont
  {Bravyi}}\ and\ \bibinfo {author} {\bibfnamefont {A.~Y.}\ \bibnamefont
  {Kitaev}},\ }\href@noop {} {\bibfield  {journal} {\bibinfo  {journal}
  {quant-ph/9811052}\ } (\bibinfo {year} {1998})}\BibitemShut {NoStop}%
\bibitem [{\citenamefont {{Kitaev}}\ and\ \citenamefont
  {{Kong}}(2012)}]{KitaevKong}%
  \BibitemOpen
  \bibfield  {author} {\bibinfo {author} {\bibfnamefont {A.}~\bibnamefont
  {{Kitaev}}}\ and\ \bibinfo {author} {\bibfnamefont {L.}~\bibnamefont
  {{Kong}}},\ }\href@noop {} {\bibfield  {journal} {\bibinfo  {journal}
  {Communications in Mathematical Physics}\ }\textbf {\bibinfo {volume}
  {313}},\ \bibinfo {pages} {351} (\bibinfo {year} {2012})}\BibitemShut
  {NoStop}%
\bibitem [{\citenamefont {Levin}(2013)}]{LevinBdy}%
  \BibitemOpen
  \bibfield  {author} {\bibinfo {author} {\bibfnamefont {M.}~\bibnamefont
  {Levin}},\ }\href {\doibase 10.1103/PhysRevX.3.021009} {\bibfield  {journal}
  {\bibinfo  {journal} {Phys. Rev. X}\ }\textbf {\bibinfo {volume} {3}},\
  \bibinfo {pages} {021009} (\bibinfo {year} {2013})}\BibitemShut {NoStop}%
\bibitem [{\citenamefont {Kong}(2014)}]{KongCond}%
  \BibitemOpen
  \bibfield  {author} {\bibinfo {author} {\bibfnamefont {L.}~\bibnamefont
  {Kong}},\ }\href {\doibase http://dx.doi.org/10.1016/j.nuclphysb.2014.07.003}
  {\bibfield  {journal} {\bibinfo  {journal} {Nuclear Physics B}\ }\textbf
  {\bibinfo {volume} {886}},\ \bibinfo {pages} {436 } (\bibinfo {year}
  {2014})}\BibitemShut {NoStop}%
\bibitem [{\citenamefont {Bais}\ \emph {et~al.}(2009)\citenamefont {Bais},
  \citenamefont {Slingerland},\ and\ \citenamefont {Haaker}}]{BaisBoundaries}%
  \BibitemOpen
  \bibfield  {author} {\bibinfo {author} {\bibfnamefont {F.~A.}\ \bibnamefont
  {Bais}}, \bibinfo {author} {\bibfnamefont {J.~K.}\ \bibnamefont
  {Slingerland}}, \ and\ \bibinfo {author} {\bibfnamefont {S.~M.}\ \bibnamefont
  {Haaker}},\ }\href {\doibase 10.1103/PhysRevLett.102.220403} {\bibfield
  {journal} {\bibinfo  {journal} {Phys. Rev. Lett.}\ }\textbf {\bibinfo
  {volume} {102}},\ \bibinfo {pages} {220403} (\bibinfo {year}
  {2009})}\BibitemShut {NoStop}%
\bibitem [{\citenamefont {Beigi}\ \emph {et~al.}(2011)\citenamefont {Beigi},
  \citenamefont {Shor},\ and\ \citenamefont {Whalen}}]{Beigi11}%
  \BibitemOpen
  \bibfield  {author} {\bibinfo {author} {\bibfnamefont {S.}~\bibnamefont
  {Beigi}}, \bibinfo {author} {\bibfnamefont {P.~W.}\ \bibnamefont {Shor}}, \
  and\ \bibinfo {author} {\bibfnamefont {D.}~\bibnamefont {Whalen}},\ }\href
  {\doibase 10.1007/s00220-011-1294-x} {\bibfield  {journal} {\bibinfo
  {journal} {Communications in Mathematical Physics}\ }\textbf {\bibinfo
  {volume} {306}},\ \bibinfo {pages} {663} (\bibinfo {year}
  {2011})}\BibitemShut {NoStop}%
\bibitem [{\citenamefont {Bais}\ and\ \citenamefont
  {Haaker}(2015)}]{BaisChiralBoundaries}%
  \BibitemOpen
  \bibfield  {author} {\bibinfo {author} {\bibfnamefont {F.~A.}\ \bibnamefont
  {Bais}}\ and\ \bibinfo {author} {\bibfnamefont {S.~M.}\ \bibnamefont
  {Haaker}},\ }\href {\doibase 10.1103/PhysRevB.92.075427} {\bibfield
  {journal} {\bibinfo  {journal} {Phys. Rev. B}\ }\textbf {\bibinfo {volume}
  {92}},\ \bibinfo {pages} {075427} (\bibinfo {year} {2015})}\BibitemShut
  {NoStop}%
\bibitem [{\citenamefont {Hung}\ and\ \citenamefont
  {Wan}(2015{\natexlab{b}})}]{YanWanCond}%
  \BibitemOpen
  \bibfield  {author} {\bibinfo {author} {\bibfnamefont {L.-Y.}\ \bibnamefont
  {Hung}}\ and\ \bibinfo {author} {\bibfnamefont {Y.}~\bibnamefont {Wan}},\
  }\href {\doibase 10.1103/PhysRevLett.114.076401} {\bibfield  {journal}
  {\bibinfo  {journal} {Phys. Rev. Lett.}\ }\textbf {\bibinfo {volume} {114}},\
  \bibinfo {pages} {076401} (\bibinfo {year} {2015}{\natexlab{b}})}\BibitemShut
  {NoStop}%
\bibitem [{\citenamefont {Kapustin}(2014)}]{KapustinAbelianBdies}%
  \BibitemOpen
  \bibfield  {author} {\bibinfo {author} {\bibfnamefont {A.}~\bibnamefont
  {Kapustin}},\ }\href {\doibase 10.1103/PhysRevB.89.125307} {\bibfield
  {journal} {\bibinfo  {journal} {Phys. Rev. B}\ }\textbf {\bibinfo {volume}
  {89}},\ \bibinfo {pages} {125307} (\bibinfo {year} {2014})}\BibitemShut
  {NoStop}%
\bibitem [{\citenamefont {Wang}\ and\ \citenamefont {Wen}(2015)}]{WangWenBdy}%
  \BibitemOpen
  \bibfield  {author} {\bibinfo {author} {\bibfnamefont {J.~C.}\ \bibnamefont
  {Wang}}\ and\ \bibinfo {author} {\bibfnamefont {X.-G.}\ \bibnamefont {Wen}},\
  }\href {\doibase 10.1103/PhysRevB.91.125124} {\bibfield  {journal} {\bibinfo
  {journal} {Phys. Rev. B}\ }\textbf {\bibinfo {volume} {91}},\ \bibinfo
  {pages} {125124} (\bibinfo {year} {2015})}\BibitemShut {NoStop}%
\bibitem [{\citenamefont {Lan}\ \emph {et~al.}(2015)\citenamefont {Lan},
  \citenamefont {Wang},\ and\ \citenamefont {Wen}}]{LanWangWen}%
  \BibitemOpen
  \bibfield  {author} {\bibinfo {author} {\bibfnamefont {T.}~\bibnamefont
  {Lan}}, \bibinfo {author} {\bibfnamefont {J.~C.}\ \bibnamefont {Wang}}, \
  and\ \bibinfo {author} {\bibfnamefont {X.-G.}\ \bibnamefont {Wen}},\ }\href
  {\doibase 10.1103/PhysRevLett.114.076402} {\bibfield  {journal} {\bibinfo
  {journal} {Phys. Rev. Lett.}\ }\textbf {\bibinfo {volume} {114}},\ \bibinfo
  {pages} {076402} (\bibinfo {year} {2015})}\BibitemShut {NoStop}%
\bibitem [{\citenamefont {Ganeshan}\ \emph {et~al.}(2017)\citenamefont
  {Ganeshan}, \citenamefont {Gorshkov}, \citenamefont {Gurarie},\ and\
  \citenamefont {Galitski}}]{GaneshanBdy}%
  \BibitemOpen
  \bibfield  {author} {\bibinfo {author} {\bibfnamefont {S.}~\bibnamefont
  {Ganeshan}}, \bibinfo {author} {\bibfnamefont {A.~V.}\ \bibnamefont
  {Gorshkov}}, \bibinfo {author} {\bibfnamefont {V.}~\bibnamefont {Gurarie}}, \
  and\ \bibinfo {author} {\bibfnamefont {V.~M.}\ \bibnamefont {Galitski}},\
  }\href {\doibase 10.1103/PhysRevB.95.045309} {\bibfield  {journal} {\bibinfo
  {journal} {Phys. Rev. B}\ }\textbf {\bibinfo {volume} {95}},\ \bibinfo
  {pages} {045309} (\bibinfo {year} {2017})}\BibitemShut {NoStop}%
\bibitem [{\citenamefont {Drinfeld}\ \emph {et~al.}(2010)\citenamefont
  {Drinfeld}, \citenamefont {Gelaki}, \citenamefont {Nikshych},\ and\
  \citenamefont {Ostrik}}]{DrinfeldGauging}%
  \BibitemOpen
  \bibfield  {author} {\bibinfo {author} {\bibfnamefont {V.~G.}\ \bibnamefont
  {Drinfeld}}, \bibinfo {author} {\bibfnamefont {S.}~\bibnamefont {Gelaki}},
  \bibinfo {author} {\bibfnamefont {D.}~\bibnamefont {Nikshych}}, \ and\
  \bibinfo {author} {\bibfnamefont {V.}~\bibnamefont {Ostrik}},\ }\href@noop {}
  {\bibfield  {journal} {\bibinfo  {journal} {Selecta Mathematica}\ }\textbf
  {\bibinfo {volume} {16}},\ \bibinfo {pages} {1119} (\bibinfo {year}
  {2010})}\BibitemShut {NoStop}%
\bibitem [{\citenamefont {Gu}\ \emph {et~al.}(2014)\citenamefont {Gu},
  \citenamefont {Hung},\ and\ \citenamefont {Wan}}]{GuWan14}%
  \BibitemOpen
  \bibfield  {author} {\bibinfo {author} {\bibfnamefont {Y.}~\bibnamefont
  {Gu}}, \bibinfo {author} {\bibfnamefont {L.-Y.}\ \bibnamefont {Hung}}, \ and\
  \bibinfo {author} {\bibfnamefont {Y.}~\bibnamefont {Wan}},\ }\href {\doibase
  10.1103/PhysRevB.90.245125} {\bibfield  {journal} {\bibinfo  {journal} {Phys.
  Rev. B}\ }\textbf {\bibinfo {volume} {90}},\ \bibinfo {pages} {245125}
  (\bibinfo {year} {2014})}\BibitemShut {NoStop}%
\bibitem [{\citenamefont {Hung}\ and\ \citenamefont {Wan}(2014)}]{HungWan14}%
  \BibitemOpen
  \bibfield  {author} {\bibinfo {author} {\bibfnamefont {L.-Y.}\ \bibnamefont
  {Hung}}\ and\ \bibinfo {author} {\bibfnamefont {Y.}~\bibnamefont {Wan}},\
  }\href {\doibase 10.1142/S0217979214501720} {\bibfield  {journal} {\bibinfo
  {journal} {International Journal of Modern Physics B}\ }\textbf {\bibinfo
  {volume} {28}},\ \bibinfo {pages} {1450172} (\bibinfo {year}
  {2014})}\BibitemShut {NoStop}%
\bibitem [{\citenamefont {Garre-Rubio}\ \emph {et~al.}(2017)\citenamefont
  {Garre-Rubio}, \citenamefont {Iblisdir},\ and\ \citenamefont
  {PŽrez-Garc'a}}]{GarciaCond}%
  \BibitemOpen
  \bibfield  {author} {\bibinfo {author} {\bibfnamefont {J.}~\bibnamefont
  {Garre-Rubio}}, \bibinfo {author} {\bibfnamefont {S.}~\bibnamefont
  {Iblisdir}}, \ and\ \bibinfo {author} {\bibfnamefont {D.}~\bibnamefont
  {PŽrez-Garc'a}},\ }\href@noop {} {\bibfield  {journal} {\bibinfo  {journal}
  {arXiv}\ } (\bibinfo {year} {2017})}\BibitemShut {NoStop}%
\bibitem [{\citenamefont {Teo}\ \emph {et~al.}(2014)\citenamefont {Teo},
  \citenamefont {Roy},\ and\ \citenamefont {Chen}}]{TeoTwist}%
  \BibitemOpen
  \bibfield  {author} {\bibinfo {author} {\bibfnamefont {J.~C.~Y.}\
  \bibnamefont {Teo}}, \bibinfo {author} {\bibfnamefont {A.}~\bibnamefont
  {Roy}}, \ and\ \bibinfo {author} {\bibfnamefont {X.}~\bibnamefont {Chen}},\
  }\href {\doibase 10.1103/PhysRevB.90.115118} {\bibfield  {journal} {\bibinfo
  {journal} {Phys. Rev. B}\ }\textbf {\bibinfo {volume} {90}},\ \bibinfo
  {pages} {115118} (\bibinfo {year} {2014})}\BibitemShut {NoStop}%
\bibitem [{\citenamefont {Barkeshli}\ \emph {et~al.}(2014)\citenamefont
  {Barkeshli}, \citenamefont {Bonderson}, \citenamefont {Cheng},\ and\
  \citenamefont {Wang}}]{LongQPaper}%
  \BibitemOpen
  \bibfield  {author} {\bibinfo {author} {\bibfnamefont {M.}~\bibnamefont
  {Barkeshli}}, \bibinfo {author} {\bibfnamefont {P.}~\bibnamefont
  {Bonderson}}, \bibinfo {author} {\bibfnamefont {M.}~\bibnamefont {Cheng}}, \
  and\ \bibinfo {author} {\bibfnamefont {Z.}~\bibnamefont {Wang}},\ }\href@noop
  {} {\bibfield  {journal} {\bibinfo  {journal} {arXiv preprint
  arXiv:1410.4540}\ } (\bibinfo {year} {2014})}\BibitemShut {NoStop}%
\bibitem [{\citenamefont {Cheng}\ \emph {et~al.}(2016)\citenamefont {Cheng},
  \citenamefont {Gu}, \citenamefont {Jiang},\ and\ \citenamefont
  {Qi}}]{ChengAPS}%
  \BibitemOpen
  \bibfield  {author} {\bibinfo {author} {\bibfnamefont {M.}~\bibnamefont
  {Cheng}}, \bibinfo {author} {\bibfnamefont {Z.-C.}\ \bibnamefont {Gu}},
  \bibinfo {author} {\bibfnamefont {S.}~\bibnamefont {Jiang}}, \ and\ \bibinfo
  {author} {\bibfnamefont {Y.}~\bibnamefont {Qi}},\ }\href@noop {} {\bibfield
  {journal} {\bibinfo  {journal} {arXiv:}\ ,\ \bibinfo {pages} {1606.08482}}
  (\bibinfo {year} {2016})}\BibitemShut {NoStop}%
\bibitem [{\citenamefont {Heinrich}\ \emph {et~al.}(2016)\citenamefont
  {Heinrich}, \citenamefont {Burnell}, \citenamefont {Fidkowski},\ and\
  \citenamefont {Levin}}]{Heinrich16}%
  \BibitemOpen
  \bibfield  {author} {\bibinfo {author} {\bibfnamefont {C.}~\bibnamefont
  {Heinrich}}, \bibinfo {author} {\bibfnamefont {F.}~\bibnamefont {Burnell}},
  \bibinfo {author} {\bibfnamefont {L.}~\bibnamefont {Fidkowski}}, \ and\
  \bibinfo {author} {\bibfnamefont {M.}~\bibnamefont {Levin}},\ }\href
  {\doibase 10.1103/PhysRevB.94.235136} {\bibfield  {journal} {\bibinfo
  {journal} {Phys. Rev. B}\ }\textbf {\bibinfo {volume} {94}},\ \bibinfo
  {pages} {235136} (\bibinfo {year} {2016})}\BibitemShut {NoStop}%
\bibitem [{\citenamefont {{Teo}}\ \emph {et~al.}(2015)\citenamefont {{Teo}},
  \citenamefont {{Hughes}},\ and\ \citenamefont {{Fradkin}}}]{FradkinTwist}%
  \BibitemOpen
  \bibfield  {author} {\bibinfo {author} {\bibfnamefont {J.~C.~Y.}\
  \bibnamefont {{Teo}}}, \bibinfo {author} {\bibfnamefont {T.~L.}\ \bibnamefont
  {{Hughes}}}, \ and\ \bibinfo {author} {\bibfnamefont {E.}~\bibnamefont
  {{Fradkin}}},\ }\href@noop {} {\bibfield  {journal} {\bibinfo  {journal}
  {ArXiv e-prints}\ } (\bibinfo {year} {2015})},\ \Eprint
  {http://arxiv.org/abs/1503.06812} {arXiv:1503.06812} \BibitemShut {NoStop}%
\bibitem [{\citenamefont {Tarantino}\ \emph {et~al.}(2016)\citenamefont
  {Tarantino}, \citenamefont {Lindner},\ and\ \citenamefont
  {Fidkowski}}]{TarantinoTwist}%
  \BibitemOpen
  \bibfield  {author} {\bibinfo {author} {\bibfnamefont {N.}~\bibnamefont
  {Tarantino}}, \bibinfo {author} {\bibfnamefont {N.~H.}\ \bibnamefont
  {Lindner}}, \ and\ \bibinfo {author} {\bibfnamefont {L.}~\bibnamefont
  {Fidkowski}},\ }\href {http://stacks.iop.org/1367-2630/18/i=3/a=035006}
  {\bibfield  {journal} {\bibinfo  {journal} {New Journal of Physics}\ }\textbf
  {\bibinfo {volume} {18}},\ \bibinfo {pages} {035006} (\bibinfo {year}
  {2016})}\BibitemShut {NoStop}%
\bibitem [{\citenamefont {Fu}\ and\ \citenamefont {Kane}(2008)}]{FuKane}%
  \BibitemOpen
  \bibfield  {author} {\bibinfo {author} {\bibfnamefont {L.}~\bibnamefont
  {Fu}}\ and\ \bibinfo {author} {\bibfnamefont {C.~L.}\ \bibnamefont {Kane}},\
  }\href@noop {} {\bibfield  {journal} {\bibinfo  {journal} {Phys. Rev. Lett.}\
  }\textbf {\bibinfo {volume} {100}},\ \bibinfo {pages} {096407} (\bibinfo
  {year} {2008})}\BibitemShut {NoStop}%
\bibitem [{\citenamefont {Lindner}\ \emph {et~al.}(2012)\citenamefont
  {Lindner}, \citenamefont {Berg}, \citenamefont {Refael},\ and\ \citenamefont
  {Stern}}]{LindnerPara}%
  \BibitemOpen
  \bibfield  {author} {\bibinfo {author} {\bibfnamefont {N.~H.}\ \bibnamefont
  {Lindner}}, \bibinfo {author} {\bibfnamefont {E.}~\bibnamefont {Berg}},
  \bibinfo {author} {\bibfnamefont {G.}~\bibnamefont {Refael}}, \ and\ \bibinfo
  {author} {\bibfnamefont {A.}~\bibnamefont {Stern}},\ }\href {\doibase
  10.1103/PhysRevX.2.041002} {\bibfield  {journal} {\bibinfo  {journal} {Phys.
  Rev. X}\ }\textbf {\bibinfo {volume} {2}},\ \bibinfo {pages} {041002}
  (\bibinfo {year} {2012})}\BibitemShut {NoStop}%
\bibitem [{\citenamefont {Clarke}\ \emph {et~al.}(2013)\citenamefont {Clarke},
  \citenamefont {Alicea},\ and\ \citenamefont {Shtengel}}]{ClarkePara}%
  \BibitemOpen
  \bibfield  {author} {\bibinfo {author} {\bibfnamefont {D.~J.}\ \bibnamefont
  {Clarke}}, \bibinfo {author} {\bibfnamefont {J.}~\bibnamefont {Alicea}}, \
  and\ \bibinfo {author} {\bibfnamefont {K.}~\bibnamefont {Shtengel}},\ }\href
  {\doibase 10.1038/ncomms2340} {\bibfield  {journal} {\bibinfo  {journal}
  {Nat. Commun.}\ }\textbf {\bibinfo {volume} {4}},\ \bibinfo {pages} {1348}
  (\bibinfo {year} {2013})}\BibitemShut {NoStop}%
\bibitem [{\citenamefont {Cheng}(2012)}]{ChengPara}%
  \BibitemOpen
  \bibfield  {author} {\bibinfo {author} {\bibfnamefont {M.}~\bibnamefont
  {Cheng}},\ }\href {\doibase 10.1103/PhysRevB.86.195126} {\bibfield  {journal}
  {\bibinfo  {journal} {Phys. Rev. B}\ }\textbf {\bibinfo {volume} {86}},\
  \bibinfo {pages} {195126} (\bibinfo {year} {2012})}\BibitemShut {NoStop}%
\bibitem [{\citenamefont {Barkeshli}\ and\ \citenamefont
  {Qi}(2012)}]{BarkeshliQi}%
  \BibitemOpen
  \bibfield  {author} {\bibinfo {author} {\bibfnamefont {M.}~\bibnamefont
  {Barkeshli}}\ and\ \bibinfo {author} {\bibfnamefont {X.-L.}\ \bibnamefont
  {Qi}},\ }\href {\doibase 10.1103/PhysRevX.2.031013} {\bibfield  {journal}
  {\bibinfo  {journal} {Phys. Rev. X}\ }\textbf {\bibinfo {volume} {2}},\
  \bibinfo {pages} {031013} (\bibinfo {year} {2012})}\BibitemShut {NoStop}%
\bibitem [{\citenamefont {Bombin}(2010)}]{BombinTwist}%
  \BibitemOpen
  \bibfield  {author} {\bibinfo {author} {\bibfnamefont {H.}~\bibnamefont
  {Bombin}},\ }\href {\doibase 10.1103/PhysRevLett.105.030403} {\bibfield
  {journal} {\bibinfo  {journal} {Phys. Rev. Lett.}\ }\textbf {\bibinfo
  {volume} {105}},\ \bibinfo {pages} {030403} (\bibinfo {year}
  {2010})}\BibitemShut {NoStop}%
\bibitem [{\citenamefont {You}\ and\ \citenamefont {Wen}(2012)}]{WenGenon}%
  \BibitemOpen
  \bibfield  {author} {\bibinfo {author} {\bibfnamefont {Y.-Z.}\ \bibnamefont
  {You}}\ and\ \bibinfo {author} {\bibfnamefont {X.-G.}\ \bibnamefont {Wen}},\
  }\href {\doibase 10.1103/PhysRevB.86.161107} {\bibfield  {journal} {\bibinfo
  {journal} {Phys. Rev. B}\ }\textbf {\bibinfo {volume} {86}},\ \bibinfo
  {pages} {161107} (\bibinfo {year} {2012})}\BibitemShut {NoStop}%
\bibitem [{\citenamefont {Barkeshli}\ \emph
  {et~al.}(2013{\natexlab{a}})\citenamefont {Barkeshli}, \citenamefont {Jian},\
  and\ \citenamefont {Qi}}]{BarkeshliAbelian1}%
  \BibitemOpen
  \bibfield  {author} {\bibinfo {author} {\bibfnamefont {M.}~\bibnamefont
  {Barkeshli}}, \bibinfo {author} {\bibfnamefont {C.-M.}\ \bibnamefont {Jian}},
  \ and\ \bibinfo {author} {\bibfnamefont {X.-L.}\ \bibnamefont {Qi}},\ }\href
  {\doibase 10.1103/PhysRevB.88.235103} {\bibfield  {journal} {\bibinfo
  {journal} {Phys. Rev. B}\ }\textbf {\bibinfo {volume} {88}},\ \bibinfo
  {pages} {235103} (\bibinfo {year} {2013}{\natexlab{a}})}\BibitemShut
  {NoStop}%
\bibitem [{\citenamefont {Barkeshli}\ \emph
  {et~al.}(2013{\natexlab{b}})\citenamefont {Barkeshli}, \citenamefont {Jian},\
  and\ \citenamefont {Qi}}]{BarkeshliAbelian2}%
  \BibitemOpen
  \bibfield  {author} {\bibinfo {author} {\bibfnamefont {M.}~\bibnamefont
  {Barkeshli}}, \bibinfo {author} {\bibfnamefont {C.-M.}\ \bibnamefont {Jian}},
  \ and\ \bibinfo {author} {\bibfnamefont {X.-L.}\ \bibnamefont {Qi}},\ }\href
  {\doibase 10.1103/PhysRevB.88.241103} {\bibfield  {journal} {\bibinfo
  {journal} {Phys. Rev. B}\ }\textbf {\bibinfo {volume} {88}},\ \bibinfo
  {pages} {241103} (\bibinfo {year} {2013}{\natexlab{b}})}\BibitemShut
  {NoStop}%
\bibitem [{\citenamefont {Barkeshli}\ \emph
  {et~al.}(2013{\natexlab{c}})\citenamefont {Barkeshli}, \citenamefont {Jian},\
  and\ \citenamefont {Qi}}]{BarkeshliBraiding}%
  \BibitemOpen
  \bibfield  {author} {\bibinfo {author} {\bibfnamefont {M.}~\bibnamefont
  {Barkeshli}}, \bibinfo {author} {\bibfnamefont {C.-M.}\ \bibnamefont {Jian}},
  \ and\ \bibinfo {author} {\bibfnamefont {X.-L.}\ \bibnamefont {Qi}},\ }\href
  {\doibase 10.1103/PhysRevB.87.045130} {\bibfield  {journal} {\bibinfo
  {journal} {Phys. Rev. B}\ }\textbf {\bibinfo {volume} {87}},\ \bibinfo
  {pages} {045130} (\bibinfo {year} {2013}{\natexlab{c}})}\BibitemShut
  {NoStop}%
\bibitem [{\citenamefont {Santos}\ and\ \citenamefont
  {Hughes}(2016)}]{HughesSantos}%
  \BibitemOpen
  \bibfield  {author} {\bibinfo {author} {\bibfnamefont {L.}~\bibnamefont
  {Santos}}\ and\ \bibinfo {author} {\bibfnamefont {T.}~\bibnamefont
  {Hughes}},\ }\href@noop {} {\bibfield  {journal} {\bibinfo  {journal}
  {arXiv:1609.06714}\ } (\bibinfo {year} {2016})}\BibitemShut {NoStop}%
\bibitem [{\citenamefont {Burnell}\ \emph {et~al.}(2012)\citenamefont
  {Burnell}, \citenamefont {Slingerland},\ and\ \citenamefont
  {Simon}}]{TSBShort}%
  \BibitemOpen
  \bibfield  {author} {\bibinfo {author} {\bibfnamefont {F.~J.}\ \bibnamefont
  {Burnell}}, \bibinfo {author} {\bibfnamefont {J.}~\bibnamefont
  {Slingerland}}, \ and\ \bibinfo {author} {\bibfnamefont {S.~H.}\ \bibnamefont
  {Simon}},\ }\href@noop {} {\bibfield  {journal} {\bibinfo  {journal} {New
  Journal of Physics}\ }\textbf {\bibinfo {volume} {14}},\ \bibinfo {pages}
  {015004} (\bibinfo {year} {2012})}\BibitemShut {NoStop}%
\bibitem [{\citenamefont {Burnell}\ \emph {et~al.}(2011)\citenamefont
  {Burnell}, \citenamefont {Simon},\ and\ \citenamefont
  {Slingerland}}]{TSBLong}%
  \BibitemOpen
  \bibfield  {author} {\bibinfo {author} {\bibfnamefont {F.~J.}\ \bibnamefont
  {Burnell}}, \bibinfo {author} {\bibfnamefont {S.~H.}\ \bibnamefont {Simon}},
  \ and\ \bibinfo {author} {\bibfnamefont {J.}~\bibnamefont {Slingerland}},\
  }\href@noop {} {\bibfield  {journal} {\bibinfo  {journal} {Phys. Rev. B}\
  }\textbf {\bibinfo {volume} {84}},\ \bibinfo {pages} {125434} (\bibinfo
  {year} {2011})}\BibitemShut {NoStop}%
\bibitem [{\citenamefont {Barkeshli}\ and\ \citenamefont
  {Wen}(2010)}]{WenBarkeshli}%
  \BibitemOpen
  \bibfield  {author} {\bibinfo {author} {\bibfnamefont {M.}~\bibnamefont
  {Barkeshli}}\ and\ \bibinfo {author} {\bibfnamefont {X.-G.}\ \bibnamefont
  {Wen}},\ }\href {\doibase 10.1103/PhysRevLett.105.216804} {\bibfield
  {journal} {\bibinfo  {journal} {Phys. Rev. Lett.}\ }\textbf {\bibinfo
  {volume} {105}},\ \bibinfo {pages} {216804} (\bibinfo {year}
  {2010})}\BibitemShut {NoStop}%
\bibitem [{\citenamefont {Barkeshli}\ and\ \citenamefont
  {Wen}(2011)}]{WenBarkeshliLong}%
  \BibitemOpen
  \bibfield  {author} {\bibinfo {author} {\bibfnamefont {M.}~\bibnamefont
  {Barkeshli}}\ and\ \bibinfo {author} {\bibfnamefont {X.-G.}\ \bibnamefont
  {Wen}},\ }\href {\doibase 10.1103/PhysRevB.84.115121} {\bibfield  {journal}
  {\bibinfo  {journal} {Phys. Rev. B}\ }\textbf {\bibinfo {volume} {84}},\
  \bibinfo {pages} {115121} (\bibinfo {year} {2011})}\BibitemShut {NoStop}%
\bibitem [{\citenamefont {Barkeshli}\ and\ \citenamefont {Wen}(2012)}]{WB2}%
  \BibitemOpen
  \bibfield  {author} {\bibinfo {author} {\bibfnamefont {M.}~\bibnamefont
  {Barkeshli}}\ and\ \bibinfo {author} {\bibfnamefont {X.-G.}\ \bibnamefont
  {Wen}},\ }\href {\doibase 10.1103/PhysRevB.86.085114} {\bibfield  {journal}
  {\bibinfo  {journal} {Phys. Rev. B}\ }\textbf {\bibinfo {volume} {86}},\
  \bibinfo {pages} {085114} (\bibinfo {year} {2012})}\BibitemShut {NoStop}%
\bibitem [{\citenamefont {Bombin}\ and\ \citenamefont
  {Martin-Delgado}(2008)}]{Bombin}%
  \BibitemOpen
  \bibfield  {author} {\bibinfo {author} {\bibfnamefont {H.}~\bibnamefont
  {Bombin}}\ and\ \bibinfo {author} {\bibfnamefont {M.~A.}\ \bibnamefont
  {Martin-Delgado}},\ }\href {\doibase 10.1103/PhysRevB.78.115421} {\bibfield
  {journal} {\bibinfo  {journal} {Phys. Rev. B}\ }\textbf {\bibinfo {volume}
  {78}},\ \bibinfo {pages} {115421} (\bibinfo {year} {2008})}\BibitemShut
  {NoStop}%
\bibitem [{\citenamefont {M\"oller}\ \emph {et~al.}(2014)\citenamefont
  {M\"oller}, \citenamefont {Hormozi}, \citenamefont {Slingerland},\ and\
  \citenamefont {Simon}}]{HormoziSlingerland}%
  \BibitemOpen
  \bibfield  {author} {\bibinfo {author} {\bibfnamefont {G.}~\bibnamefont
  {M\"oller}}, \bibinfo {author} {\bibfnamefont {L.}~\bibnamefont {Hormozi}},
  \bibinfo {author} {\bibfnamefont {J.}~\bibnamefont {Slingerland}}, \ and\
  \bibinfo {author} {\bibfnamefont {S.~H.}\ \bibnamefont {Simon}},\ }\href
  {\doibase 10.1103/PhysRevB.90.235101} {\bibfield  {journal} {\bibinfo
  {journal} {Phys. Rev. B}\ }\textbf {\bibinfo {volume} {90}},\ \bibinfo
  {pages} {235101} (\bibinfo {year} {2014})}\BibitemShut {NoStop}%
\bibitem [{\citenamefont {Feng}\ \emph {et~al.}(2007)\citenamefont {Feng},
  \citenamefont {Zhang},\ and\ \citenamefont {Xiang}}]{FengKitaevJW}%
  \BibitemOpen
  \bibfield  {author} {\bibinfo {author} {\bibfnamefont {X.-Y.}\ \bibnamefont
  {Feng}}, \bibinfo {author} {\bibfnamefont {G.-M.}\ \bibnamefont {Zhang}}, \
  and\ \bibinfo {author} {\bibfnamefont {T.}~\bibnamefont {Xiang}},\ }\href
  {\doibase 10.1103/PhysRevLett.98.087204} {\bibfield  {journal} {\bibinfo
  {journal} {Phys. Rev. Lett.}\ }\textbf {\bibinfo {volume} {98}},\ \bibinfo
  {pages} {087204} (\bibinfo {year} {2007})}\BibitemShut {NoStop}%
\bibitem [{\citenamefont {Gils}\ \emph {et~al.}(2009)\citenamefont {Gils},
  \citenamefont {Trebst}, \citenamefont {Kitaev}, \citenamefont {Ludwig},
  \citenamefont {Troyer},\ and\ \citenamefont {Wang}}]{Gilsetal}%
  \BibitemOpen
  \bibfield  {author} {\bibinfo {author} {\bibfnamefont {C.}~\bibnamefont
  {Gils}}, \bibinfo {author} {\bibfnamefont {S.}~\bibnamefont {Trebst}},
  \bibinfo {author} {\bibfnamefont {A.}~\bibnamefont {Kitaev}}, \bibinfo
  {author} {\bibfnamefont {A.~W.~W.}\ \bibnamefont {Ludwig}}, \bibinfo {author}
  {\bibfnamefont {M.}~\bibnamefont {Troyer}}, \ and\ \bibinfo {author}
  {\bibfnamefont {Z.}~\bibnamefont {Wang}},\ }\href@noop {} {\bibfield
  {journal} {\bibinfo  {journal} {Nature Physics}\ }\textbf {\bibinfo {volume}
  {5}},\ \bibinfo {pages} {834} (\bibinfo {year} {2009.})}\BibitemShut
  {NoStop}%
\bibitem [{\citenamefont {Gils}(2009)}]{GilsIsing}%
  \BibitemOpen
  \bibfield  {author} {\bibinfo {author} {\bibfnamefont {C.}~\bibnamefont
  {Gils}},\ }\href {http://stacks.iop.org/1742-5468/2009/i=07/a=P07019}
  {\bibfield  {journal} {\bibinfo  {journal} {Journal of Statistical Mechanics:
  Theory and Experiment}\ }\textbf {\bibinfo {volume} {2009}},\ \bibinfo
  {pages} {P07019} (\bibinfo {year} {2009})}\BibitemShut {NoStop}%
\bibitem [{\citenamefont {Bais}\ and\ \citenamefont
  {Romers}(2012)}]{BaisMonteCarlo}%
  \BibitemOpen
  \bibfield  {author} {\bibinfo {author} {\bibfnamefont {F.~A.}\ \bibnamefont
  {Bais}}\ and\ \bibinfo {author} {\bibfnamefont {J.~C.}\ \bibnamefont
  {Romers}},\ }\href {http://stacks.iop.org/1367-2630/14/i=3/a=035024}
  {\bibfield  {journal} {\bibinfo  {journal} {New Journal of Physics}\ }\textbf
  {\bibinfo {volume} {14}},\ \bibinfo {pages} {035024} (\bibinfo {year}
  {2012})}\BibitemShut {NoStop}%
\bibitem [{\citenamefont {Schulz}\ and\ \citenamefont
  {Burnell}(2016)}]{schulzIsi}%
  \BibitemOpen
  \bibfield  {author} {\bibinfo {author} {\bibfnamefont {M.~D.}\ \bibnamefont
  {Schulz}}\ and\ \bibinfo {author} {\bibfnamefont {F.~J.}\ \bibnamefont
  {Burnell}},\ }\href {\doibase 10.1103/PhysRevB.94.165110} {\bibfield
  {journal} {\bibinfo  {journal} {Phys. Rev. B}\ }\textbf {\bibinfo {volume}
  {94}},\ \bibinfo {pages} {165110} (\bibinfo {year} {2016})}\BibitemShut
  {NoStop}%
\bibitem [{\citenamefont {Dusuel}\ and\ \citenamefont {Vidal}(2015)}]{VidalMF}%
  \BibitemOpen
  \bibfield  {author} {\bibinfo {author} {\bibfnamefont {S.}~\bibnamefont
  {Dusuel}}\ and\ \bibinfo {author} {\bibfnamefont {J.}~\bibnamefont {Vidal}},\
  }\href {\doibase 10.1103/PhysRevB.92.125150} {\bibfield  {journal} {\bibinfo
  {journal} {Phys. Rev. B}\ }\textbf {\bibinfo {volume} {92}},\ \bibinfo
  {pages} {125150} (\bibinfo {year} {2015})}\BibitemShut {NoStop}%
\bibitem [{\citenamefont {Schulz}\ \emph {et~al.}(2013)\citenamefont {Schulz},
  \citenamefont {Dusuel}, \citenamefont {Schmidt},\ and\ \citenamefont
  {Vidal}}]{SchulzFib}%
  \BibitemOpen
  \bibfield  {author} {\bibinfo {author} {\bibfnamefont {M.~D.}\ \bibnamefont
  {Schulz}}, \bibinfo {author} {\bibfnamefont {S.}~\bibnamefont {Dusuel}},
  \bibinfo {author} {\bibfnamefont {K.~P.}\ \bibnamefont {Schmidt}}, \ and\
  \bibinfo {author} {\bibfnamefont {J.}~\bibnamefont {Vidal}},\ }\href
  {\doibase 10.1103/PhysRevLett.110.147203} {\bibfield  {journal} {\bibinfo
  {journal} {Phys. Rev. Lett.}\ }\textbf {\bibinfo {volume} {110}},\ \bibinfo
  {pages} {147203} (\bibinfo {year} {2013})}\BibitemShut {NoStop}%
\bibitem [{\citenamefont {Schulz}\ \emph {et~al.}(2014)\citenamefont {Schulz},
  \citenamefont {Dusuel}, \citenamefont {Misguich}, \citenamefont {Schmidt},\
  and\ \citenamefont {Vidal}}]{SchulzSU2}%
  \BibitemOpen
  \bibfield  {author} {\bibinfo {author} {\bibfnamefont {M.~D.}\ \bibnamefont
  {Schulz}}, \bibinfo {author} {\bibfnamefont {S.}~\bibnamefont {Dusuel}},
  \bibinfo {author} {\bibfnamefont {G.}~\bibnamefont {Misguich}}, \bibinfo
  {author} {\bibfnamefont {K.~P.}\ \bibnamefont {Schmidt}}, \ and\ \bibinfo
  {author} {\bibfnamefont {J.}~\bibnamefont {Vidal}},\ }\href {\doibase
  10.1103/PhysRevB.89.201103} {\bibfield  {journal} {\bibinfo  {journal} {Phys.
  Rev. B}\ }\textbf {\bibinfo {volume} {89}},\ \bibinfo {pages} {201103}
  (\bibinfo {year} {2014})}\BibitemShut {NoStop}%
\bibitem [{\citenamefont {Schulz}\ \emph {et~al.}(2016)\citenamefont {Schulz},
  \citenamefont {Dusuel},\ and\ \citenamefont {Vidal}}]{VidalBS}%
  \BibitemOpen
  \bibfield  {author} {\bibinfo {author} {\bibfnamefont {M.~D.}\ \bibnamefont
  {Schulz}}, \bibinfo {author} {\bibfnamefont {S.}~\bibnamefont {Dusuel}}, \
  and\ \bibinfo {author} {\bibfnamefont {J.}~\bibnamefont {Vidal}},\ }\href
  {\doibase 10.1103/PhysRevB.94.205102} {\bibfield  {journal} {\bibinfo
  {journal} {Phys. Rev. B}\ }\textbf {\bibinfo {volume} {94}},\ \bibinfo
  {pages} {205102} (\bibinfo {year} {2016})}\BibitemShut {NoStop}%
\bibitem [{\citenamefont {Bais}\ \emph {et~al.}(1992)\citenamefont {Bais},
  \citenamefont {van Driel},\ and\ \citenamefont
  {de~Wild~Propitius}}]{BaisDiscrete}%
  \BibitemOpen
  \bibfield  {author} {\bibinfo {author} {\bibfnamefont {F.~A.}\ \bibnamefont
  {Bais}}, \bibinfo {author} {\bibfnamefont {P.}~\bibnamefont {van Driel}}, \
  and\ \bibinfo {author} {\bibfnamefont {M.}~\bibnamefont
  {de~Wild~Propitius}},\ }\href {\doibase
  http://dx.doi.org/10.1016/0370-2693(92)90773-W} {\bibfield  {journal}
  {\bibinfo  {journal} {Physics Letters B}\ }\textbf {\bibinfo {volume}
  {280}},\ \bibinfo {pages} {63 } (\bibinfo {year} {1992})}\BibitemShut
  {NoStop}%
\bibitem [{\citenamefont {Bonderson}(2007)}]{BondersonThesis}%
  \BibitemOpen
  \bibfield  {author} {\bibinfo {author} {\bibfnamefont {P.}~\bibnamefont
  {Bonderson}},\ }\href@noop {} {\bibinfo {type} {{PhD} thesis}},\ \bibinfo
  {school} {Caltech} (\bibinfo {year} {2007})\BibitemShut {NoStop}%
\bibitem [{Note1()}]{Note1}%
  \BibitemOpen
  \bibinfo {note} {Technically the properties listed above, and indeed anyon
  condensation in general, do not require the anyon model to be modular; in
  fact a braided fusion category is sufficient. However, here we will restrict
  our discussion to anyon condensation in the modular case.}\BibitemShut
  {Stop}%
\bibitem [{\citenamefont {Wegner}(1971)}]{Wegner71}%
  \BibitemOpen
  \bibfield  {author} {\bibinfo {author} {\bibfnamefont {F.~J.}\ \bibnamefont
  {Wegner}},\ }\href {\doibase http://dx.doi.org/10.1063/1.1665530} {\bibfield
  {journal} {\bibinfo  {journal} {Journal of Mathematical Physics}\ }\textbf
  {\bibinfo {volume} {12}},\ \bibinfo {pages} {2259} (\bibinfo {year}
  {1971})}\BibitemShut {NoStop}%
\bibitem [{\citenamefont {Kitaev}(2003)}]{KitaevToric}%
  \BibitemOpen
  \bibfield  {author} {\bibinfo {author} {\bibfnamefont {A.~Y.}\ \bibnamefont
  {Kitaev}},\ }\href@noop {} {\bibfield  {journal} {\bibinfo  {journal} {Annals
  of Physics}\ }\textbf {\bibinfo {volume} {303}},\ \bibinfo {pages} {2}
  (\bibinfo {year} {2003})}\BibitemShut {NoStop}%
\bibitem [{\citenamefont {anderson}(1973)}]{pwa-rvb}%
  \BibitemOpen
  \bibfield  {author} {\bibinfo {author} {\bibfnamefont {P.~W.}\ \bibnamefont
  {anderson}},\ }\href@noop {} {\bibfield  {journal} {\bibinfo  {journal} {Mat.
  Res. Bull.}\ }\textbf {\bibinfo {volume} {8}},\ \bibinfo {pages} {153}
  (\bibinfo {year} {1973})}\BibitemShut {NoStop}%
\bibitem [{\citenamefont {Wen}(1991)}]{WenSLTO}%
  \BibitemOpen
  \bibfield  {author} {\bibinfo {author} {\bibfnamefont {X.~G.}\ \bibnamefont
  {Wen}},\ }\href {\doibase 10.1103/PhysRevB.44.2664} {\bibfield  {journal}
  {\bibinfo  {journal} {Phys. Rev. B}\ }\textbf {\bibinfo {volume} {44}},\
  \bibinfo {pages} {2664} (\bibinfo {year} {1991})}\BibitemShut {NoStop}%
\bibitem [{\citenamefont {Read}\ and\ \citenamefont
  {Sachdev}(1991)}]{ReadSachdev}%
  \BibitemOpen
  \bibfield  {author} {\bibinfo {author} {\bibfnamefont {N.}~\bibnamefont
  {Read}}\ and\ \bibinfo {author} {\bibfnamefont {S.}~\bibnamefont {Sachdev}},\
  }\href {\doibase 10.1103/PhysRevLett.66.1773} {\bibfield  {journal} {\bibinfo
   {journal} {Phys. Rev. Lett.}\ }\textbf {\bibinfo {volume} {66}},\ \bibinfo
  {pages} {1773} (\bibinfo {year} {1991})}\BibitemShut {NoStop}%
\bibitem [{\citenamefont {Sachdev}\ and\ \citenamefont
  {Read}(1991)}]{SachdevRead}%
  \BibitemOpen
  \bibfield  {author} {\bibinfo {author} {\bibfnamefont {S.}~\bibnamefont
  {Sachdev}}\ and\ \bibinfo {author} {\bibfnamefont {N.}~\bibnamefont {Read}},\
  }\href {\doibase 10.1142/S0217979291000158} {\bibfield  {journal} {\bibinfo
  {journal} {International Journal of Modern Physics B}\ }\textbf {\bibinfo
  {volume} {05}},\ \bibinfo {pages} {219} (\bibinfo {year} {1991})}\BibitemShut
  {NoStop}%
\bibitem [{\citenamefont {Moessner}\ \emph {et~al.}(2001)\citenamefont
  {Moessner}, \citenamefont {Sondhi},\ and\ \citenamefont
  {Fradkin}}]{IsingDual}%
  \BibitemOpen
  \bibfield  {author} {\bibinfo {author} {\bibfnamefont {R.}~\bibnamefont
  {Moessner}}, \bibinfo {author} {\bibfnamefont {S.~L.}\ \bibnamefont
  {Sondhi}}, \ and\ \bibinfo {author} {\bibfnamefont {E.}~\bibnamefont
  {Fradkin}},\ }\href {\doibase 10.1103/PhysRevB.65.024504} {\bibfield
  {journal} {\bibinfo  {journal} {Phys. Rev. B}\ }\textbf {\bibinfo {volume}
  {65}},\ \bibinfo {pages} {024504} (\bibinfo {year} {2001})}\BibitemShut
  {NoStop}%
\bibitem [{\citenamefont {Volovik}(1999)}]{VolovikPip}%
  \BibitemOpen
  \bibfield  {author} {\bibinfo {author} {\bibfnamefont {G.~E.}\ \bibnamefont
  {Volovik}},\ }\href@noop {} {\bibfield  {journal} {\bibinfo  {journal} {JETP
  Lett.}\ }\textbf {\bibinfo {volume} {70}},\ \bibinfo {pages} {609} (\bibinfo
  {year} {1999})}\BibitemShut {NoStop}%
\bibitem [{\citenamefont {Ivanov}(2001)}]{Ivanov}%
  \BibitemOpen
  \bibfield  {author} {\bibinfo {author} {\bibfnamefont {D.~A.}\ \bibnamefont
  {Ivanov}},\ }\href {\doibase 10.1103/PhysRevLett.86.268} {\bibfield
  {journal} {\bibinfo  {journal} {Phys. Rev. Lett.}\ }\textbf {\bibinfo
  {volume} {86}},\ \bibinfo {pages} {268} (\bibinfo {year} {2001})}\BibitemShut
  {NoStop}%
\bibitem [{\citenamefont {Read}\ and\ \citenamefont {Green}(2000)}]{ReadGreen}%
  \BibitemOpen
  \bibfield  {author} {\bibinfo {author} {\bibfnamefont {N.}~\bibnamefont
  {Read}}\ and\ \bibinfo {author} {\bibfnamefont {D.}~\bibnamefont {Green}},\
  }\href {\doibase 10.1103/PhysRevB.61.10267} {\bibfield  {journal} {\bibinfo
  {journal} {Phys. Rev. B}\ }\textbf {\bibinfo {volume} {61}},\ \bibinfo
  {pages} {10267} (\bibinfo {year} {2000})}\BibitemShut {NoStop}%
\bibitem [{\citenamefont {Moore}\ and\ \citenamefont {Read}(1991)}]{MooreRead}%
  \BibitemOpen
  \bibfield  {author} {\bibinfo {author} {\bibfnamefont {G.}~\bibnamefont
  {Moore}}\ and\ \bibinfo {author} {\bibfnamefont {N.}~\bibnamefont {Read}},\
  }\href@noop {} {\bibfield  {journal} {\bibinfo  {journal} {Nucl. Phys. B}\
  }\textbf {\bibinfo {volume} {360}},\ \bibinfo {pages} {362} (\bibinfo {year}
  {1991})}\BibitemShut {NoStop}%
\bibitem [{\citenamefont {Wen}(2003)}]{WenToric}%
  \BibitemOpen
  \bibfield  {author} {\bibinfo {author} {\bibfnamefont {X.-G.}\ \bibnamefont
  {Wen}},\ }\href {\doibase 10.1103/PhysRevLett.90.016803} {\bibfield
  {journal} {\bibinfo  {journal} {Phys. Rev. Lett.}\ }\textbf {\bibinfo
  {volume} {90}},\ \bibinfo {pages} {016803} (\bibinfo {year}
  {2003})}\BibitemShut {NoStop}%
\bibitem [{\citenamefont {Wess}\ and\ \citenamefont {Zumino}(1971)}]{WessWZW}%
  \BibitemOpen
  \bibfield  {author} {\bibinfo {author} {\bibfnamefont {J.}~\bibnamefont
  {Wess}}\ and\ \bibinfo {author} {\bibfnamefont {B.}~\bibnamefont {Zumino}},\
  }\href {\doibase http://dx.doi.org/10.1016/0370-2693(71)90582-X} {\bibfield
  {journal} {\bibinfo  {journal} {Physics Letters B}\ }\textbf {\bibinfo
  {volume} {37}},\ \bibinfo {pages} {95 } (\bibinfo {year} {1971})}\BibitemShut
  {NoStop}%
\bibitem [{\citenamefont {Witten}(1983)}]{WittenWZWa}%
  \BibitemOpen
  \bibfield  {author} {\bibinfo {author} {\bibfnamefont {E.}~\bibnamefont
  {Witten}},\ }\href {\doibase http://dx.doi.org/10.1016/0550-3213(83)90063-9}
  {\bibfield  {journal} {\bibinfo  {journal} {Nuclear Physics B}\ }\textbf
  {\bibinfo {volume} {223}},\ \bibinfo {pages} {422 } (\bibinfo {year}
  {1983})}\BibitemShut {NoStop}%
\bibitem [{\citenamefont {Witten}(1984)}]{WittenWZWb}%
  \BibitemOpen
  \bibfield  {author} {\bibinfo {author} {\bibfnamefont {E.}~\bibnamefont
  {Witten}},\ }\href {\doibase 10.1007/BF01215276} {\bibfield  {journal}
  {\bibinfo  {journal} {Communications in Mathematical Physics}\ }\textbf
  {\bibinfo {volume} {92}},\ \bibinfo {pages} {455} (\bibinfo {year}
  {1984})}\BibitemShut {NoStop}%
\bibitem [{\citenamefont {Novikov}(1982)}]{NovikovWZW}%
  \BibitemOpen
  \bibfield  {author} {\bibinfo {author} {\bibfnamefont {S.~P.}\ \bibnamefont
  {Novikov}},\ }\href {http://stacks.iop.org/0036-0279/37/i=5/a=R01} {\bibfield
   {journal} {\bibinfo  {journal} {Russian Mathematical Surveys}\ }\textbf
  {\bibinfo {volume} {37}},\ \bibinfo {pages} {1} (\bibinfo {year}
  {1982})}\BibitemShut {NoStop}%
\bibitem [{\citenamefont {Witten}(1989)}]{WittenJones}%
  \BibitemOpen
  \bibfield  {author} {\bibinfo {author} {\bibfnamefont {E.}~\bibnamefont
  {Witten}},\ }\href {http://projecteuclid.org/euclid.cmp/1104178138}
  {\bibfield  {journal} {\bibinfo  {journal} {Comm. Math. Phys.}\ }\textbf
  {\bibinfo {volume} {121}},\ \bibinfo {pages} {351} (\bibinfo {year}
  {1989})}\BibitemShut {NoStop}%
\bibitem [{\citenamefont {Kou}\ \emph {et~al.}(2009)\citenamefont {Kou},
  \citenamefont {Yu},\ and\ \citenamefont {Wen}}]{WenTCLGCS}%
  \BibitemOpen
  \bibfield  {author} {\bibinfo {author} {\bibfnamefont {S.-P.}\ \bibnamefont
  {Kou}}, \bibinfo {author} {\bibfnamefont {J.}~\bibnamefont {Yu}}, \ and\
  \bibinfo {author} {\bibfnamefont {X.-G.}\ \bibnamefont {Wen}},\ }\href
  {\doibase 10.1103/PhysRevB.80.125101} {\bibfield  {journal} {\bibinfo
  {journal} {Phys. Rev. B}\ }\textbf {\bibinfo {volume} {80}},\ \bibinfo
  {pages} {125101} (\bibinfo {year} {2009})}\BibitemShut {NoStop}%
\bibitem [{\citenamefont {Clarke}\ and\ \citenamefont
  {Nayak}(2015)}]{ClarkeNayak}%
  \BibitemOpen
  \bibfield  {author} {\bibinfo {author} {\bibfnamefont {D.~J.}\ \bibnamefont
  {Clarke}}\ and\ \bibinfo {author} {\bibfnamefont {C.}~\bibnamefont {Nayak}},\
  }\href {\doibase 10.1103/PhysRevB.92.155110} {\bibfield  {journal} {\bibinfo
  {journal} {Phys. Rev. B}\ }\textbf {\bibinfo {volume} {92}},\ \bibinfo
  {pages} {155110} (\bibinfo {year} {2015})}\BibitemShut {NoStop}%
\bibitem [{\citenamefont {Ryu}\ \emph {et~al.}(2012)\citenamefont {Ryu},
  \citenamefont {Moore},\ and\ \citenamefont {Ludwig}}]{RyuAnomaly}%
  \BibitemOpen
  \bibfield  {author} {\bibinfo {author} {\bibfnamefont {S.}~\bibnamefont
  {Ryu}}, \bibinfo {author} {\bibfnamefont {J.~E.}\ \bibnamefont {Moore}}, \
  and\ \bibinfo {author} {\bibfnamefont {A.~W.~W.}\ \bibnamefont {Ludwig}},\
  }\href {\doibase 10.1103/PhysRevB.85.045104} {\bibfield  {journal} {\bibinfo
  {journal} {Phys. Rev. B}\ }\textbf {\bibinfo {volume} {85}},\ \bibinfo
  {pages} {045104} (\bibinfo {year} {2012})}\BibitemShut {NoStop}%
\bibitem [{\citenamefont {Furusaki}\ \emph {et~al.}(2013)\citenamefont
  {Furusaki}, \citenamefont {Nagaosa}, \citenamefont {Nomura}, \citenamefont
  {Ryu},\ and\ \citenamefont {Takayanagi}}]{FurasakiAnomaly}%
  \BibitemOpen
  \bibfield  {author} {\bibinfo {author} {\bibfnamefont {A.}~\bibnamefont
  {Furusaki}}, \bibinfo {author} {\bibfnamefont {N.}~\bibnamefont {Nagaosa}},
  \bibinfo {author} {\bibfnamefont {K.}~\bibnamefont {Nomura}}, \bibinfo
  {author} {\bibfnamefont {S.}~\bibnamefont {Ryu}}, \ and\ \bibinfo {author}
  {\bibfnamefont {T.}~\bibnamefont {Takayanagi}},\ }\href {\doibase
  http://dx.doi.org/10.1016/j.crhy.2013.03.002} {\bibfield  {journal} {\bibinfo
   {journal} {Comptes Rendus Physique}\ }\textbf {\bibinfo {volume} {14}},\
  \bibinfo {pages} {871 } (\bibinfo {year} {2013})}\BibitemShut {NoStop}%
\bibitem [{\citenamefont {Wen}(2013)}]{WenAnomaly}%
  \BibitemOpen
  \bibfield  {author} {\bibinfo {author} {\bibfnamefont {X.-G.}\ \bibnamefont
  {Wen}},\ }\href {\doibase 10.1103/PhysRevD.88.045013} {\bibfield  {journal}
  {\bibinfo  {journal} {Phys. Rev. D}\ }\textbf {\bibinfo {volume} {88}},\
  \bibinfo {pages} {045013} (\bibinfo {year} {2013})}\BibitemShut {NoStop}%
\bibitem [{\citenamefont {Kane}\ and\ \citenamefont
  {Fisher}(1997)}]{KaneFisherThermalHall}%
  \BibitemOpen
  \bibfield  {author} {\bibinfo {author} {\bibfnamefont {C.~L.}\ \bibnamefont
  {Kane}}\ and\ \bibinfo {author} {\bibfnamefont {M.~P.~A.}\ \bibnamefont
  {Fisher}},\ }\href {\doibase 10.1103/PhysRevB.55.15832} {\bibfield  {journal}
  {\bibinfo  {journal} {Phys. Rev. B}\ }\textbf {\bibinfo {volume} {55}},\
  \bibinfo {pages} {15832} (\bibinfo {year} {1997})}\BibitemShut {NoStop}%
\bibitem [{\citenamefont {Cardy}(1986)}]{CardyModular}%
  \BibitemOpen
  \bibfield  {author} {\bibinfo {author} {\bibfnamefont {J.~L.}\ \bibnamefont
  {Cardy}},\ }\href {\doibase http://dx.doi.org/10.1016/0550-3213(86)90552-3}
  {\bibfield  {journal} {\bibinfo  {journal} {Nuclear Physics B}\ }\textbf
  {\bibinfo {volume} {270}},\ \bibinfo {pages} {186 } (\bibinfo {year}
  {1986})}\BibitemShut {NoStop}%
\bibitem [{\citenamefont {Friedan}\ and\ \citenamefont
  {Shenker}(1987)}]{FreidanShenker}%
  \BibitemOpen
  \bibfield  {author} {\bibinfo {author} {\bibfnamefont {D.}~\bibnamefont
  {Friedan}}\ and\ \bibinfo {author} {\bibfnamefont {S.}~\bibnamefont
  {Shenker}},\ }\href {\doibase http://dx.doi.org/10.1016/0550-3213(87)90418-4}
  {\bibfield  {journal} {\bibinfo  {journal} {Nuclear Physics B}\ }\textbf
  {\bibinfo {volume} {281}},\ \bibinfo {pages} {509 } (\bibinfo {year}
  {1987})}\BibitemShut {NoStop}%
\bibitem [{\citenamefont {Alvarez-Gaume}\ \emph {et~al.}(1989)\citenamefont
  {Alvarez-Gaume}, \citenamefont {Sierra},\ and\ \citenamefont
  {Gomez}}]{AlvarezGaume}%
  \BibitemOpen
  \bibfield  {author} {\bibinfo {author} {\bibfnamefont {L.}~\bibnamefont
  {Alvarez-Gaume}}, \bibinfo {author} {\bibfnamefont {G.}~\bibnamefont
  {Sierra}}, \ and\ \bibinfo {author} {\bibfnamefont {C.}~\bibnamefont
  {Gomez}},\ }\href@noop {} {\  (\bibinfo {year} {1989})}\BibitemShut {NoStop}%
\bibitem [{\citenamefont {Dijkgraaf}\ and\ \citenamefont
  {Verlinde}(1988)}]{DijkgraafVerlinde}%
  \BibitemOpen
  \bibfield  {author} {\bibinfo {author} {\bibfnamefont {R.}~\bibnamefont
  {Dijkgraaf}}\ and\ \bibinfo {author} {\bibfnamefont {E.}~\bibnamefont
  {Verlinde}},\ }\href@noop {} {\bibfield  {journal} {\bibinfo  {journal}
  {Nucl. Phys. B}\ ,\ \bibinfo {pages} {87}} (\bibinfo {year}
  {1988})}\BibitemShut {NoStop}%
\bibitem [{\citenamefont {Cappelli}\ \emph {et~al.}(1987)\citenamefont
  {Cappelli}, \citenamefont {Otzykson},\ and\ \citenamefont
  {Zuber}}]{CapelliADE}%
  \BibitemOpen
  \bibfield  {author} {\bibinfo {author} {\bibfnamefont {A.}~\bibnamefont
  {Cappelli}}, \bibinfo {author} {\bibfnamefont {C.}~\bibnamefont {Otzykson}},
  \ and\ \bibinfo {author} {\bibfnamefont {J.-B.}\ \bibnamefont {Zuber}},\
  }\href {\doibase http://dx.doi.org/10.1016/0550-3213(87)90155-6} {\bibfield
  {journal} {\bibinfo  {journal} {Nuclear Physics B}\ }\textbf {\bibinfo
  {volume} {280}},\ \bibinfo {pages} {445 } (\bibinfo {year}
  {1987})}\BibitemShut {NoStop}%
\bibitem [{\citenamefont {Levin}\ and\ \citenamefont {Wen}(2005)}]{LW}%
  \BibitemOpen
  \bibfield  {author} {\bibinfo {author} {\bibfnamefont {M.~A.}\ \bibnamefont
  {Levin}}\ and\ \bibinfo {author} {\bibfnamefont {X.-G.}\ \bibnamefont
  {Wen}},\ }\href@noop {} {\bibfield  {journal} {\bibinfo  {journal} {Phys.
  Rev. B}\ }\textbf {\bibinfo {volume} {71}},\ \bibinfo {pages} {045110}
  (\bibinfo {year} {2005})}\BibitemShut {NoStop}%
\bibitem [{\citenamefont {Lin}\ and\ \citenamefont {Levin}(2014)}]{LevinLin}%
  \BibitemOpen
  \bibfield  {author} {\bibinfo {author} {\bibfnamefont {C.-H.}\ \bibnamefont
  {Lin}}\ and\ \bibinfo {author} {\bibfnamefont {M.}~\bibnamefont {Levin}},\
  }\href {\doibase 10.1103/PhysRevB.89.195130} {\bibfield  {journal} {\bibinfo
  {journal} {Phys. Rev. B}\ }\textbf {\bibinfo {volume} {89}},\ \bibinfo
  {pages} {195130} (\bibinfo {year} {2014})}\BibitemShut {NoStop}%
\bibitem [{\citenamefont {Fuchs}\ \emph {et~al.}(2011)\citenamefont {Fuchs},
  \citenamefont {Schweigert},\ and\ \citenamefont {Stigner}}]{Fuchs11}%
  \BibitemOpen
  \bibfield  {author} {\bibinfo {author} {\bibfnamefont {J.}~\bibnamefont
  {Fuchs}}, \bibinfo {author} {\bibfnamefont {C.}~\bibnamefont {Schweigert}}, \
  and\ \bibinfo {author} {\bibfnamefont {C.}~\bibnamefont {Stigner}},\ }\href
  {\doibase http://dx.doi.org/10.1016/j.nuclphysb.2010.10.008} {\bibfield
  {journal} {\bibinfo  {journal} {Nuclear Physics B}\ }\textbf {\bibinfo
  {volume} {843}},\ \bibinfo {pages} {673 } (\bibinfo {year}
  {2011})}\BibitemShut {NoStop}%
\bibitem [{\citenamefont {L.~Kong}(2009)}]{KongModular}%
  \BibitemOpen
  \bibfield  {author} {\bibinfo {author} {\bibfnamefont {I.~R.}\ \bibnamefont
  {L.~Kong}},\ }\href@noop {} {\bibfield  {journal} {\bibinfo  {journal}
  {Commun. Math. Phys}\ }\textbf {\bibinfo {volume} {292}},\ \bibinfo {pages}
  {871} (\bibinfo {year} {2009})}\BibitemShut {NoStop}%
\bibitem [{\citenamefont {Fuchs}\ \emph {et~al.}(2013)\citenamefont {Fuchs},
  \citenamefont {Schweigert},\ and\ \citenamefont {Valentino}}]{Fuchs13}%
  \BibitemOpen
  \bibfield  {author} {\bibinfo {author} {\bibfnamefont {J.}~\bibnamefont
  {Fuchs}}, \bibinfo {author} {\bibfnamefont {C.}~\bibnamefont {Schweigert}}, \
  and\ \bibinfo {author} {\bibfnamefont {A.}~\bibnamefont {Valentino}},\ }\href
  {\doibase 10.1007/s00220-013-1723-0} {\bibfield  {journal} {\bibinfo
  {journal} {Communications in Mathematical Physics}\ }\textbf {\bibinfo
  {volume} {321}},\ \bibinfo {pages} {543} (\bibinfo {year}
  {2013})}\BibitemShut {NoStop}%
\bibitem [{Note2()}]{Note2}%
  \BibitemOpen
  \bibinfo {note} {The reason that this quantity is conserved only modulo $8$
  is that there exists a 2D bulk state of interacting bosons, known as the
  Kitaev $E_8$ state, with no topological order and boundary central charge
  8.\cite {KitaevHoney} In principle we could add copies of this $E_8$ state to
  any 2D system without changing the bulk topological order; hence the
  statistics of the anyons can determine the boundary's central charge only
  modulo 8. Indeed SU$(9)_1$ has central charge $8$, and can be reduced to the
  vacuum by condensing either of the two bosons.}\BibitemShut {Stop}%
\bibitem [{\citenamefont {Fradkin}\ and\ \citenamefont
  {Shenker}(1979)}]{FradkinShenker}%
  \BibitemOpen
  \bibfield  {author} {\bibinfo {author} {\bibfnamefont {E.}~\bibnamefont
  {Fradkin}}\ and\ \bibinfo {author} {\bibfnamefont {S.~H.}\ \bibnamefont
  {Shenker}},\ }\href {\doibase 10.1103/PhysRevD.19.3682} {\bibfield  {journal}
  {\bibinfo  {journal} {Phys. Rev. D}\ }\textbf {\bibinfo {volume} {19}},\
  \bibinfo {pages} {3682} (\bibinfo {year} {1979})}\BibitemShut {NoStop}%
\bibitem [{\citenamefont {Read}\ and\ \citenamefont
  {Rezayi}(1999)}]{ReadRezayi}%
  \BibitemOpen
  \bibfield  {author} {\bibinfo {author} {\bibfnamefont {N.}~\bibnamefont
  {Read}}\ and\ \bibinfo {author} {\bibfnamefont {E.}~\bibnamefont {Rezayi}},\
  }\href {\doibase 10.1103/PhysRevB.59.8084} {\bibfield  {journal} {\bibinfo
  {journal} {Phys. Rev. B}\ }\textbf {\bibinfo {volume} {59}},\ \bibinfo
  {pages} {8084} (\bibinfo {year} {1999})}\BibitemShut {NoStop}%
\bibitem [{\citenamefont {Chung}\ \emph {et~al.}(2010)\citenamefont {Chung},
  \citenamefont {Yao}, \citenamefont {Hughes},\ and\ \citenamefont
  {Kim}}]{HughesKimChiral}%
  \BibitemOpen
  \bibfield  {author} {\bibinfo {author} {\bibfnamefont {S.~B.}\ \bibnamefont
  {Chung}}, \bibinfo {author} {\bibfnamefont {H.}~\bibnamefont {Yao}}, \bibinfo
  {author} {\bibfnamefont {T.~L.}\ \bibnamefont {Hughes}}, \ and\ \bibinfo
  {author} {\bibfnamefont {E.-A.}\ \bibnamefont {Kim}},\ }\href {\doibase
  10.1103/PhysRevB.81.060403} {\bibfield  {journal} {\bibinfo  {journal} {Phys.
  Rev. B}\ }\textbf {\bibinfo {volume} {81}},\ \bibinfo {pages} {060403}
  (\bibinfo {year} {2010})}\BibitemShut {NoStop}%
\bibitem [{\citenamefont {Lee}\ \emph {et~al.}(2016)\citenamefont {Lee},
  \citenamefont {Geraedts},\ and\ \citenamefont {Motrunich}}]{MotrunichSET}%
  \BibitemOpen
  \bibfield  {author} {\bibinfo {author} {\bibfnamefont {J.~Y.}\ \bibnamefont
  {Lee}}, \bibinfo {author} {\bibfnamefont {S.}~\bibnamefont {Geraedts}}, \
  and\ \bibinfo {author} {\bibfnamefont {O.~I.}\ \bibnamefont {Motrunich}},\
  }\href {\doibase 10.1103/PhysRevB.93.035103} {\bibfield  {journal} {\bibinfo
  {journal} {Phys. Rev. B}\ }\textbf {\bibinfo {volume} {93}},\ \bibinfo
  {pages} {035103} (\bibinfo {year} {2016})}\BibitemShut {NoStop}%
\bibitem [{\citenamefont {Geraedts}\ and\ \citenamefont
  {Motrunich}(2012)}]{MotrunichU1}%
  \BibitemOpen
  \bibfield  {author} {\bibinfo {author} {\bibfnamefont {S.~D.}\ \bibnamefont
  {Geraedts}}\ and\ \bibinfo {author} {\bibfnamefont {O.~I.}\ \bibnamefont
  {Motrunich}},\ }\href {\doibase 10.1103/PhysRevB.86.045106} {\bibfield
  {journal} {\bibinfo  {journal} {Phys. Rev. B}\ }\textbf {\bibinfo {volume}
  {86}},\ \bibinfo {pages} {045106} (\bibinfo {year} {2012})}\BibitemShut
  {NoStop}%
\bibitem [{\citenamefont {Isakov}\ \emph {et~al.}(2011)\citenamefont {Isakov},
  \citenamefont {Fendley}, \citenamefont {Ludwig}, \citenamefont {Trebst},\
  and\ \citenamefont {Troyer}}]{Isakov11}%
  \BibitemOpen
  \bibfield  {author} {\bibinfo {author} {\bibfnamefont {S.~V.}\ \bibnamefont
  {Isakov}}, \bibinfo {author} {\bibfnamefont {P.}~\bibnamefont {Fendley}},
  \bibinfo {author} {\bibfnamefont {A.~W.~W.}\ \bibnamefont {Ludwig}}, \bibinfo
  {author} {\bibfnamefont {S.}~\bibnamefont {Trebst}}, \ and\ \bibinfo {author}
  {\bibfnamefont {M.}~\bibnamefont {Troyer}},\ }\href {\doibase
  10.1103/PhysRevB.83.125114} {\bibfield  {journal} {\bibinfo  {journal} {Phys.
  Rev. B}\ }\textbf {\bibinfo {volume} {83}},\ \bibinfo {pages} {125114}
  (\bibinfo {year} {2011})}\BibitemShut {NoStop}%
\bibitem [{\citenamefont {Ardonne}\ \emph {et~al.}(2004)\citenamefont
  {Ardonne}, \citenamefont {Fendley},\ and\ \citenamefont
  {Fradkin}}]{ArdonneConformalQCP}%
  \BibitemOpen
  \bibfield  {author} {\bibinfo {author} {\bibfnamefont {E.}~\bibnamefont
  {Ardonne}}, \bibinfo {author} {\bibfnamefont {P.}~\bibnamefont {Fendley}}, \
  and\ \bibinfo {author} {\bibfnamefont {E.}~\bibnamefont {Fradkin}},\ }\href
  {\doibase http://dx.doi.org/10.1016/j.aop.2004.01.004} {\bibfield  {journal}
  {\bibinfo  {journal} {Annals of Physics}\ }\textbf {\bibinfo {volume}
  {310}},\ \bibinfo {pages} {493 } (\bibinfo {year} {2004})}\BibitemShut
  {NoStop}%
\bibitem [{\citenamefont {Liu}\ and\ \citenamefont
  {Bhatt}(2016)}]{BhattFQHETrans}%
  \BibitemOpen
  \bibfield  {author} {\bibinfo {author} {\bibfnamefont {Z.}~\bibnamefont
  {Liu}}\ and\ \bibinfo {author} {\bibfnamefont {R.~N.}\ \bibnamefont
  {Bhatt}},\ }\href {\doibase 10.1103/PhysRevLett.117.206801} {\bibfield
  {journal} {\bibinfo  {journal} {Phys. Rev. Lett.}\ }\textbf {\bibinfo
  {volume} {117}},\ \bibinfo {pages} {206801} (\bibinfo {year}
  {2016})}\BibitemShut {NoStop}%
\bibitem [{\citenamefont {Li}\ \emph {et~al.}(2015)\citenamefont {Li},
  \citenamefont {Yang}, \citenamefont {Tu},\ and\ \citenamefont
  {Cheng}}]{ChengParafermionCrit}%
  \BibitemOpen
  \bibfield  {author} {\bibinfo {author} {\bibfnamefont {W.}~\bibnamefont
  {Li}}, \bibinfo {author} {\bibfnamefont {S.}~\bibnamefont {Yang}}, \bibinfo
  {author} {\bibfnamefont {H.-H.}\ \bibnamefont {Tu}}, \ and\ \bibinfo {author}
  {\bibfnamefont {M.}~\bibnamefont {Cheng}},\ }\href {\doibase
  10.1103/PhysRevB.91.115133} {\bibfield  {journal} {\bibinfo  {journal} {Phys.
  Rev. B}\ }\textbf {\bibinfo {volume} {91}},\ \bibinfo {pages} {115133}
  (\bibinfo {year} {2015})}\BibitemShut {NoStop}%
\bibitem [{\citenamefont {Morampudi}\ \emph {et~al.}(2014)\citenamefont
  {Morampudi}, \citenamefont {von Keyserlingk},\ and\ \citenamefont
  {Pollmann}}]{DSemToTC}%
  \BibitemOpen
  \bibfield  {author} {\bibinfo {author} {\bibfnamefont {S.~C.}\ \bibnamefont
  {Morampudi}}, \bibinfo {author} {\bibfnamefont {C.}~\bibnamefont {von
  Keyserlingk}}, \ and\ \bibinfo {author} {\bibfnamefont {F.}~\bibnamefont
  {Pollmann}},\ }\href {\doibase 10.1103/PhysRevB.90.035117} {\bibfield
  {journal} {\bibinfo  {journal} {Phys. Rev. B}\ }\textbf {\bibinfo {volume}
  {90}},\ \bibinfo {pages} {035117} (\bibinfo {year} {2014})}\BibitemShut
  {NoStop}%
\bibitem [{\citenamefont {Jiang}\ and\ \citenamefont
  {Ran}(2016)}]{JiangRanTensors}%
  \BibitemOpen
  \bibfield  {author} {\bibinfo {author} {\bibfnamefont {S.}~\bibnamefont
  {Jiang}}\ and\ \bibinfo {author} {\bibfnamefont {Y.}~\bibnamefont {Ran}},\
  }\href@noop {} {\bibfield  {journal} {\bibinfo  {journal} {arXiv}\ }
  (\bibinfo {year} {2016})}\BibitemShut {NoStop}%
\bibitem [{\citenamefont {Castelnovo}\ \emph {et~al.}(2010)\citenamefont
  {Castelnovo}, \citenamefont {Troyer},\ and\ \citenamefont
  {Trebst}}]{castelnovo10}%
  \BibitemOpen
  \bibfield  {author} {\bibinfo {author} {\bibfnamefont {C.}~\bibnamefont
  {Castelnovo}}, \bibinfo {author} {\bibfnamefont {M.}~\bibnamefont {Troyer}},
  \ and\ \bibinfo {author} {\bibfnamefont {S.}~\bibnamefont {Trebst}},\
  }\enquote {\bibinfo {title} {{U}nderstanding {Q}uantum {P}hase
  {T}ransitions},}\ in\ \href {\doibase 10.1201/b10273-10} {\emph {\bibinfo
  {booktitle} {Understanding Quantum Phase Transitions}}}\ (\bibinfo
  {publisher} {CRC Press},\ \bibinfo {year} {2010})\ Chap.\ \bibinfo {chapter}
  {Topological {O}rder and {Q}uantum {C}riticality}, pp.\ \bibinfo {pages}
  {167--192}\BibitemShut {NoStop}%
\bibitem [{\citenamefont {Vidal}\ \emph {et~al.}(2009)\citenamefont {Vidal},
  \citenamefont {Dusuel},\ and\ \citenamefont {Schmidt}}]{VidalToric}%
  \BibitemOpen
  \bibfield  {author} {\bibinfo {author} {\bibfnamefont {J.}~\bibnamefont
  {Vidal}}, \bibinfo {author} {\bibfnamefont {S.}~\bibnamefont {Dusuel}}, \
  and\ \bibinfo {author} {\bibfnamefont {K.~P.}\ \bibnamefont {Schmidt}},\
  }\href {\doibase 10.1103/PhysRevB.79.033109} {\bibfield  {journal} {\bibinfo
  {journal} {Phys. Rev. B}\ }\textbf {\bibinfo {volume} {79}},\ \bibinfo
  {pages} {033109} (\bibinfo {year} {2009})}\BibitemShut {NoStop}%
\bibitem [{\citenamefont {Trebst}\ \emph {et~al.}(2007)\citenamefont {Trebst},
  \citenamefont {Werner}, \citenamefont {Troyer}, \citenamefont {Shtengel},\
  and\ \citenamefont {Nayak}}]{TrebstTC}%
  \BibitemOpen
  \bibfield  {author} {\bibinfo {author} {\bibfnamefont {S.}~\bibnamefont
  {Trebst}}, \bibinfo {author} {\bibfnamefont {P.}~\bibnamefont {Werner}},
  \bibinfo {author} {\bibfnamefont {M.}~\bibnamefont {Troyer}}, \bibinfo
  {author} {\bibfnamefont {K.}~\bibnamefont {Shtengel}}, \ and\ \bibinfo
  {author} {\bibfnamefont {C.}~\bibnamefont {Nayak}},\ }\href {\doibase
  10.1103/PhysRevLett.98.070602} {\bibfield  {journal} {\bibinfo  {journal}
  {Phys. Rev. Lett.}\ }\textbf {\bibinfo {volume} {98}},\ \bibinfo {pages}
  {070602} (\bibinfo {year} {2007})}\BibitemShut {NoStop}%
\bibitem [{\citenamefont {Tupitsyn}\ \emph {et~al.}(2010)\citenamefont
  {Tupitsyn}, \citenamefont {Kitaev}, \citenamefont {Prokof'ev},\ and\
  \citenamefont {Stamp}}]{KitaevPhase}%
  \BibitemOpen
  \bibfield  {author} {\bibinfo {author} {\bibfnamefont {I.~S.}\ \bibnamefont
  {Tupitsyn}}, \bibinfo {author} {\bibfnamefont {A.}~\bibnamefont {Kitaev}},
  \bibinfo {author} {\bibfnamefont {N.~V.}\ \bibnamefont {Prokof'ev}}, \ and\
  \bibinfo {author} {\bibfnamefont {P.~C.~E.}\ \bibnamefont {Stamp}},\ }\href
  {\doibase 10.1103/PhysRevB.82.085114} {\bibfield  {journal} {\bibinfo
  {journal} {Phys. Rev. B}\ }\textbf {\bibinfo {volume} {82}},\ \bibinfo
  {pages} {085114} (\bibinfo {year} {2010})}\BibitemShut {NoStop}%
\bibitem [{\citenamefont {Chandran}\ \emph {et~al.}(2013)\citenamefont
  {Chandran}, \citenamefont {Burnell}, \citenamefont {Khemani},\ and\
  \citenamefont {Sondhi}}]{ChandranBurnell}%
  \BibitemOpen
  \bibfield  {author} {\bibinfo {author} {\bibfnamefont {A.}~\bibnamefont
  {Chandran}}, \bibinfo {author} {\bibfnamefont {F.~J.}\ \bibnamefont
  {Burnell}}, \bibinfo {author} {\bibfnamefont {V.}~\bibnamefont {Khemani}}, \
  and\ \bibinfo {author} {\bibfnamefont {S.~L.}\ \bibnamefont {Sondhi}},\
  }\href {http://stacks.iop.org/0953-8984/25/i=40/a=404214} {\bibfield
  {journal} {\bibinfo  {journal} {Journal of Physics: Condensed Matter}\
  }\textbf {\bibinfo {volume} {25}},\ \bibinfo {pages} {404214} (\bibinfo
  {year} {2013})}\BibitemShut {NoStop}%
\bibitem [{\citenamefont {Dusuel}\ \emph {et~al.}(2011)\citenamefont {Dusuel},
  \citenamefont {Kamfor}, \citenamefont {Orus}, \citenamefont {Schmidt},\ and\
  \citenamefont {Vidal}}]{VidalToric2}%
  \BibitemOpen
  \bibfield  {author} {\bibinfo {author} {\bibfnamefont {S.}~\bibnamefont
  {Dusuel}}, \bibinfo {author} {\bibfnamefont {M.}~\bibnamefont {Kamfor}},
  \bibinfo {author} {\bibfnamefont {R.}~\bibnamefont {Orus}}, \bibinfo {author}
  {\bibfnamefont {K.~P.}\ \bibnamefont {Schmidt}}, \ and\ \bibinfo {author}
  {\bibfnamefont {J.}~\bibnamefont {Vidal}},\ }\href@noop {} {\bibfield
  {journal} {\bibinfo  {journal} {PRL}\ }\textbf {\bibinfo {volume} {106}},\
  \bibinfo {pages} {107203} (\bibinfo {year} {2011})}\BibitemShut {NoStop}%
\bibitem [{\citenamefont {Jalabert}\ and\ \citenamefont
  {Sachdev}(1991)}]{Jalabert91}%
  \BibitemOpen
  \bibfield  {author} {\bibinfo {author} {\bibfnamefont {R.~A.}\ \bibnamefont
  {Jalabert}}\ and\ \bibinfo {author} {\bibfnamefont {S.}~\bibnamefont
  {Sachdev}},\ }\href {\doibase 10.1103/PhysRevB.44.686} {\bibfield  {journal}
  {\bibinfo  {journal} {Phys. Rev. B}\ }\textbf {\bibinfo {volume} {44}},\
  \bibinfo {pages} {686} (\bibinfo {year} {1991})}\BibitemShut {NoStop}%
\bibitem [{\citenamefont {Vojta}\ and\ \citenamefont
  {Sachdev}(2000)}]{Vojta00}%
  \BibitemOpen
  \bibfield  {author} {\bibinfo {author} {\bibfnamefont {M.}~\bibnamefont
  {Vojta}}\ and\ \bibinfo {author} {\bibfnamefont {S.}~\bibnamefont
  {Sachdev}},\ }\href@noop {} {\bibfield  {journal} {\bibinfo  {journal}
  {Journal of the Physical Society of Japan}\ }\textbf {\bibinfo {volume}
  {69}},\ \bibinfo {pages} {1} (\bibinfo {year} {2000})}\BibitemShut {NoStop}%
\bibitem [{\citenamefont {Chubukov}\ \emph {et~al.}(1994)\citenamefont
  {Chubukov}, \citenamefont {Senthil},\ and\ \citenamefont
  {Sachdev}}]{ChubukovSenthilSachdev}%
  \BibitemOpen
  \bibfield  {author} {\bibinfo {author} {\bibfnamefont {A.~V.}\ \bibnamefont
  {Chubukov}}, \bibinfo {author} {\bibfnamefont {T.}~\bibnamefont {Senthil}}, \
  and\ \bibinfo {author} {\bibfnamefont {S.}~\bibnamefont {Sachdev}},\ }\href
  {\doibase 10.1103/PhysRevLett.72.2089} {\bibfield  {journal} {\bibinfo
  {journal} {Phys. Rev. Lett.}\ }\textbf {\bibinfo {volume} {72}},\ \bibinfo
  {pages} {2089} (\bibinfo {year} {1994})}\BibitemShut {NoStop}%
\bibitem [{\citenamefont {Senthil}\ and\ \citenamefont
  {Fisher}(2000)}]{SenthilFisher}%
  \BibitemOpen
  \bibfield  {author} {\bibinfo {author} {\bibfnamefont {T.}~\bibnamefont
  {Senthil}}\ and\ \bibinfo {author} {\bibfnamefont {M.~P.~A.}\ \bibnamefont
  {Fisher}},\ }\href {\doibase 10.1103/PhysRevB.62.7850} {\bibfield  {journal}
  {\bibinfo  {journal} {Phys. Rev. B}\ }\textbf {\bibinfo {volume} {62}},\
  \bibinfo {pages} {7850} (\bibinfo {year} {2000})}\BibitemShut {NoStop}%
\end{thebibliography}%

\end{document}